\documentclass[reprint,twocolumn,aps,amsmath,amssymb,pra,superscriptaddress,floatfix,tightenlines,nofootinbib,nobibnotes]{revtex4-2}

\usepackage{graphicx}
\usepackage[colorlinks,linkcolor=blue,urlcolor=blue, citecolor=blue]{hyperref}
\usepackage{xurl}
\usepackage[whole]{bxcjkjatype}
\usepackage{booktabs}
\usepackage{graphicx}
\usepackage{comment}
\usepackage{nicematrix}

\usepackage{dsfont}
\usepackage{physics}
\usepackage{nicefrac}
\usepackage{mathtools}
\usepackage{amsmath, amssymb, amsfonts, amsthm}
\usepackage{thmtools, thm-restate}

\usepackage{algorithmic}
\usepackage{algorithm}

\newtheorem{theorem}{Theorem}
\newtheorem*{theorem*}{Theorem}

\DeclareMathOperator{\polylog}{polylog}
\DeclareMathOperator{\poly}{poly}
\DeclareMathOperator{\adj}{adj}

\DeclareMathOperator{\cD}{\overline{\mathcal{V}}}
\DeclareMathOperator{\cE}{\mathcal{E}}
\DeclareMathOperator{\cG}{\mathcal{G}}

\allowdisplaybreaks

\makeatletter
\renewcommand{\@fnsymbol}[1]{%
  \ensuremath{%
    \ifcase#1
      \ast
    \or
      \dagger
    \or
      \ddagger
    \or
      \P
    \else
      \@ctrerr
    \fi
  }%
}
\makeatother

\begin{document}

\title{Doubly-polylog-time-overhead fault-tolerant quantum computation\\
by a polylog-time parallel minimum-weight perfect matching decoder}

\author{Yugo Takada$^{*,}$}
\email{yugo.takada1@gmail.com}
\affiliation{Graduate School of Engineering Science, The University of Osaka, 1--3 Machikaneyama, Toyonaka, Osaka 560--8531, Japan}
\author{Hayata Yamasaki$^{*,}$}
\email{hayata.yamasaki@gmail.com} 
\affiliation{Department of Computer Science, Graduate School of Information Science and Technology, The University of Tokyo, 7-3-1 Hongo, Bunkyo-ku, Tokyo 113-8656, Japan}
\affiliation{Department of Physics, Graduate School of Science, The University of Tokyo, 7--3--1 Hongo, Bunkyo-ku, Tokyo, 113--0033, Japan}
\begingroup
\renewcommand{\thefootnote}{$*$}
\footnotetext{The authors contributed equally to this work.}
\endgroup
\setcounter{footnote}{0}

\begin{abstract}
Reducing space and time overheads of fault-tolerant quantum computation (FTQC) has been receiving increasing attention as it is crucial for the development of quantum computers and also plays a fundamental role in understanding the feasibility and limitations of realizing quantum advantages. Shorter time overheads are particularly essential for demonstrating quantum computational speedups without compromising runtime advantages. However, surpassing the conventional polylogarithmic (polylog) scaling of time overheads has remained a significant challenge, since it requires addressing all potential bottlenecks, including the nonzero runtime of classical computation for decoding in practical implementations. In this work, we construct a protocol that achieves FTQC with doubly polylog time overhead while maintaining the conventional polylog space overhead. The key to our approach is the development of a highly parallelizable minimum-weight perfect matching (MWPM) decoder, which achieves a polylog parallel runtime in terms of the code size while providing theoretical guarantees on threshold existence and overhead bounds. Our protocol integrates this decoder with a topological-code protocol that incorporates single-shot decoding for efficient syndrome extraction; furthermore, we concatenate this with the concatenated Steane codes to guarantee the existence of the threshold while avoiding a backlog problem, enabling us to achieve doubly polylog time overheads even when accounting for the decoder's runtime. These results suggest the feasibility of surpassing the conventional polylog-time-overhead barrier, opening a new frontier in low-overhead FTQC\@.
\end{abstract}

\maketitle

\paragraph*{Teaser}
Developing a highly parallelizable method for widely used decoding drastically shortens the time overhead in fault-tolerant quantum computation.

\section*{Introduction}

Topological quantum codes, such as surface codes (also known as toric codes)~\cite{K2,bravyi1998quantumcodeslatticeboundary}\footnote{We refer to surface codes as those defined on lattices with open boundary conditions, while toric codes with periodic boundary conditions.} and color codes~\cite{PhysRevLett.97.180501}, constitute fundamental families of quantum error-correcting codes in quantum information theory and many-body quantum physics~\cite{doi:10.1063/1.1499754,Bombin_2013}.
A particularly notable feature of certain families of topological codes is their single-shot decodability~\cite{PhysRevX.5.031043}, a property that may arise in codes defined in three or more spatial dimensions~\cite{bombin2015gauge,kubica2022single,PRXQuantum.5.020310,Campbell_2019,RevModPhys.88.045005,PRXQuantum.2.020340}.
This property is closely related to self-correcting quantum memories and, potentially, finite-temperature topological order~\cite{Campbell_2019,RevModPhys.88.045005}; at the same time, it is expected to be useful for speeding up the implementation of fault-tolerant quantum computation (FTQC)~\cite{10.1145/258533.258579,10.1137/S0097539799359385}, a key milestone in quantum technologies~\cite{bluvstein2024logical,sunami2024scalablenetworkingneutralatomqubits,google2023suppressing}. 
Without the single-shot property, fault-tolerant protocols with two-dimensional (2D) topological codes typically require multiple rounds of measurements for syndrome extraction~\cite{doi:10.1063/1.1499754}.
In contrast, the single-shot property allows for syndrome extraction in just a single round~\cite{PhysRevX.5.031043,Campbell_2019}.
This capability has the potential to address the problem of reducing the overall overheads associated with FTQC, an increasingly critical challenge in the field of quantum information.

Reducing the overheads of FTQC is inherently difficult in general as it requires addressing all possible bottlenecks in its implementation, not just the number of syndrome extraction rounds.
Quantum computation~\cite{N4} aims to solve a family of computational problems, which we index by $m\in\{1,2,\ldots\}$.
We represent quantum computation by a polynomial-size quantum circuit of width $W(m)$ and depth $D(m)$, where $W(m)\to\infty$ and $D(m)=O\qty(\poly\qty(W(m)))\coloneqq O\qty(W(m)^\alpha)$ for some $\alpha>0$ as $m\to\infty$.
This circuit, called an original circuit, starts with state preparation, ends with measurements, and is composed of a finite, universal gate set.
If executed directly on noisy quantum devices without quantum error correction, the original circuit would fail to produce correct computational results due to the effect of noise, which is conventionally modeled as the local stochastic Pauli error model~\cite{gottesman2014faulttolerant} (see Methods for details).
The only established solution to this problem is to employ FTQC\@.
Given a target error $\epsilon(m)>0$ satisfying $\epsilon(m)=1/O(\poly(W(m)))$\footnote{It is also possible to choose the target error $\epsilon$ as a fixed constant as in a conventional setting~\cite{G,gottesman2014faulttolerant,yamasaki2022timeefficient,tamiya2024polylogtimeconstantspaceoverheadfaulttolerantquantum}. Here, however, we consider a more general setting to allow for asympotically vanishing $\epsilon$, which includes this conventional setting as a special case.}, the task of FTQC is to execute a fault-tolerant circuit under the error model in such a way that the computational output should be sampled from a probability distribution close to the original circuit's output distribution, up to a total variation distance of at most $\epsilon(m)$~\cite{gottesman2014faulttolerant,G,yamasaki2022timeefficient}.
To achieve this, a fault-tolerant protocol compiles the original circuit into a fault-tolerant circuit by encoding each qubit of the original circuit as a logical qubit of a quantum error-correcting code and implementing each operation as a logical operation.
With this compilation, we can suppress the error rate of each logical operation arbitrarily via quantum error correction.
A crucial component of quantum error correction is a classical decoder, which estimates the faulty locations in the noisy circuit based on syndrome measurement outcomes and outputs an appropriate recovery operation.
Since practical implementations must account for the nonzero runtime of such classical decoding, we explicitly consider such runtime in our analysis.
The size of the original circuit is given by $W(m)D(m)$.
To ensure the overall error within $\epsilon(m)$, following the union bound, fault-tolerant protocols will suppress the logical error rate below $\lessapprox \frac{\epsilon(m)}{W(m)D(m)}$.
This error suppression typically increases the width and depth of the fault-tolerant circuit, denoted by $W_\mathrm{FT}(m)$ and $D_\mathrm{FT}(m)$, respectively, compared to the original circuit.
The time overhead of a fault-tolerant protocol is defined as $\frac{D_\mathrm{FT}(m)}{D(m)}$ while the space overhead is defined as $\frac{W_\mathrm{FT}(m)}{W(m)}$~\cite{G,gottesman2014faulttolerant,yamasaki2022timeefficient}.\footnote{Following the convention of previous works~\cite{gottesman2014faulttolerant,G,yamasaki2022timeefficient}, we do not impose geometrical constraints on interactions in original and fault-tolerant circuits.}

Reducing overheads in FTQC is fundamental to understanding the true complexity of demonstrating quantum computational advantages and is equally crucial for the practical development of quantum computers; consequently, this topic has garnered significant attention in recent years.
Conventional fault-tolerant protocols, such as those based on 2D surface codes and concatenated Steane codes, implement FTQC with polylogarithmic (polylog) space and time overheads, scaling as $\Theta\qty(\polylog\qty(\frac{W(m)D(m)}{\epsilon(m)}))$~\cite{G,doi:10.1063/1.1499754}.
In contrast, between 2013 and 2019, a series of works~\cite{gottesman2014faulttolerant,PhysRevA.87.020304,8555154,Grospellier} demonstrated that the space overhead of FTQC can be reduced to a constant order $O(1)$; however, this improvement came at the cost of increasing the time overhead to a polynomial scaling $\Theta\qty(\poly\qty(\frac{W(m)D(m)}{\epsilon(m)}))$.
In 2024, Ref.~\cite{yamasaki2022timeefficient} introduced an alternative protocol achieving a constant space overhead $O(1)$ and the quasi-polylog time overhead $\Theta\qty(\text{quasi-polylog}\qty(\frac{W(m)D(m)}{\epsilon(m)}))$, where $\text{quasi-polylog}(x)\coloneqq\exp[\polylog(\polylog(x))]$ is substantially smaller than the polynomial scaling while slightly larger than the conventional polylog scaling.
More recently, Ref.~\cite{tamiya2024polylogtimeconstantspaceoverheadfaulttolerantquantum} proved that it is possible to simultaneously achieve the constant space overhead $O(1)$ and the polylog time overhead $\Theta\qty(\polylog\qty(\frac{W(m)D(m)}{\epsilon(m)}))$.\footnote{
Reference~\cite{nguyen2024quantumfaulttoleranceconstantspace} also analyzes a protocol with constant space overhead and polylog time overhead; however, its analysis does not apply to our setting since it does not fully account for the decoder's runtime. In contrast, Ref.~\cite{tamiya2024polylogtimeconstantspaceoverheadfaulttolerantquantum} provides a complete proof of the feasibility of constant-space-overhead and polylog-time-overhead FTQC by explicitly incorporating the decoder’s runtime into the analysis, without relying on unpublished results.
}
However, these theoretical advancements have primarily focused on reducing space overheads.
Meanwhile, breaking through the polylog-time-overhead barrier without incurring substantial increases in space overhead has remained a significant challenge, requiring fundamentally new techniques.

In this work, we address this challenge by constructing a fault-tolerant protocol that combines single-shot-decodable topological codes with concatenated codes to achieve doubly polylog time overhead while accounting for the nonzero runtime of decoders:
\begin{align}
\label{eq:doubly_polylog_time_overhead}
    \frac{D_\mathrm{FT}(m)}{D(m)}=O\qty(\polylog\qty(\polylog\qty(\frac{W(m)D(m)}{\epsilon(m)}))),
\end{align}
which also maintains the conventional polylog space overhead.
To bound the time overhead of fault-tolerant protocols based on topological codes, it is crucial to employ decoders with provable error-suppression guarantees---such as those based on finding a minimum-weight perfect matching (MWPM) in a given graph~\cite{doi:10.1063/1.1499754,gottesman2014faulttolerant,PhysRevLett.109.180502}---rather than relying on heuristic decoders that lack formal guarantees on threshold existence and overhead bounds.
A key challenge in this approach is the computational bottleneck introduced by classical decoding.
Under an assumption of instantaneous classical decoding, Ref.~\cite{PhysRevX.5.031043} has argued that a fault-tolerant protocol based on certain topological codes with single-shot features could achieve $O(1)$ time overhead.
However, as discussed in more detail below, in practical settings that account for nonzero runtime of classical decoders, a combination of such protocols~\cite{PhysRevX.5.031043,bombin2015gauge,kubica2022single} with existing MWPM decoding algorithms~\cite{doi:10.1063/1.1499754,10.5555/2011362.2011368,10.5555/2011383.2011385,PhysRevA.83.020302,PhysRevA.89.022326,DBLP:journals/qic/Fowler15a,higgott2023sparse,wu2023fusion,PhysRevA.89.012317,PRXQuantum.3.010310,Kubica2023efficientcolorcode} indeed results in polylog time overheads, due to the polynomial runtime of the blossom algorithm conventionally used for finding an MWPM~\cite{Edmonds_1965,edmonds1965maximum}.
Even with parallelization, no approach is known for the blossom algorithm to achieve a polylog runtime while preserving the same output as the sequential one, as detailed later.

\begin{figure*}
    \centering
    \includegraphics[width=7.0in]{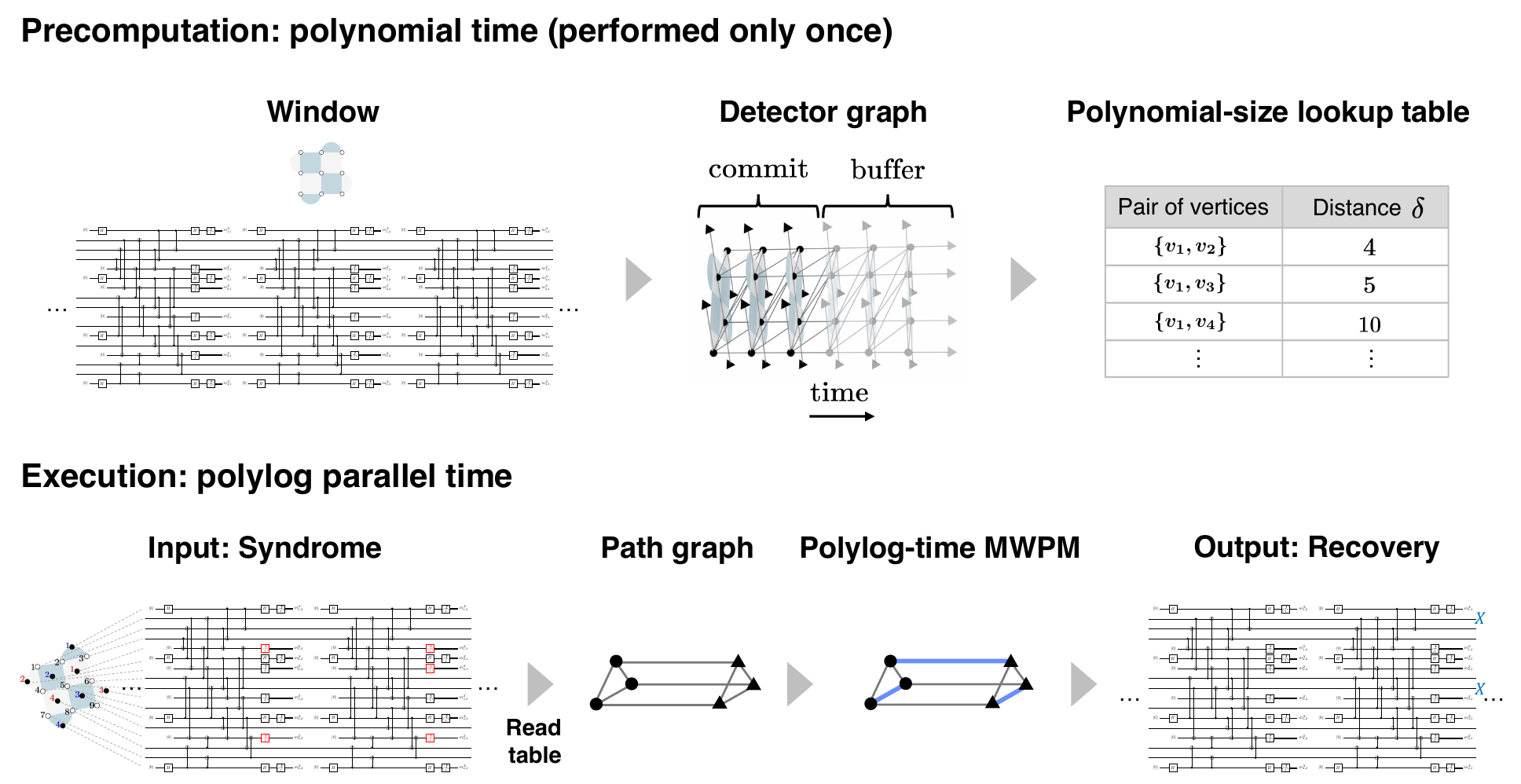}
    \caption{Our strategy for executing MWPM decoding within polylog parallel runtime in terms of the code size. The figure illustrates the case of 2D surface codes for simplicity. In this case, our decoding strategy can employ sliding window decoding. As a precomputation, we compute the distances $\delta$ (i.e., the shortest path lengths) between all pairs of vertices in detector graphs within polynomial time and store them in a polynomial-size lookup table. Given syndrome measurement outcomes as input, our parallel decoder constructs a path graph by reading from this lookup table and then applies an isolation-based polylog-time parallel algorithm to find an MWPM in the path graph. From the MWPM, the decoder outputs an estimate of a recovery operation on the code block, which will coincide with the output of conventional blossom-algorithm-based polynomial-time MWPM decoders. When translating the MWPM into the recovery operation, we also use a precomputed, polynomial-size lookup table that maps each vertex pair in the detector graphs to its corresponding recovery operation (not shown in the figure). By leveraging the polylog-time parallel algorithm for the MWPM finding and polynomial-size lookup tables, this decoding strategy achieves polylog parallel runtime.}
    \label{fig:main}
\end{figure*}

Our main innovation is the development of a fundamentally different strategy for the MWPM decoding, which allows for substantially higher parallelism while preserving the same output as the blossom-algorithm-based MWPM decoders to ensure the overhead bounds.
In our approach, finding the MWPM is based on the (derandomized) isolation lemma and matrix determinant computations---techniques originally introduced in the context of theoretical computer science to study computational complexities of graph problems for parallelized and randomized algorithms~\cite{vazirani1987matching,DAHLHAUS199879,4568269,10.1007/978-3-540-70918-3_42,datta2010deterministically,TEWARI20121,10.1145/2897518.2897564,goldwasser_et_al:LIPIcs.ICALP.2017.87,8104102}.
Progressing beyond the original complexity-theoretical motivation behind this technique, our key contribution is the total algorithm design for solving the entire decoding problem (see also Fig.~\ref{fig:main}), finding a useful application of this technique in addressing the unique challenge of FTQC\@.
As we will show, the original randomized parallel algorithm may not be straightforwardly used for the decoder to achieve a polylog runtime with a reasonable number of classical processors.
By developing solutions to this obstacle, we show that our overall MWPM decoding strategy achieves a polylog parallel runtime in terms of the problem size with a reasonable number of classical processors, substantially outperforming the known variants of the blossom algorithms~\cite{doi:10.1063/1.1499754,DBLP:journals/qic/Fowler15a,higgott2023sparse,wu2023fusion} when parallelized.

In the following, we first argue that an improved runtime of the decoder will lead to a doubly-polylog-time-overhead fault-tolerant protocol, and then describe our decoding strategy, its upper bound of runtime, and that of the number of parallel classical processes.
In addition to the theoretically guaranteed upper bounds, we also provide a numerical result to estimate the required number of parallel classical processes more precisely, which implies that a sublinear scaling of parallelism may indeed suffice in a practical regime.
These results open a new frontier of low-overhead FTQC, substantially surpassing the conventional polylog time overhead.

\section*{Results}

\paragraph*{Fault-tolerant protocol with doubly polylog time overhead}
The goal in this work is to shorten the time overhead $\frac {D_\mathrm{FT}(m)}{D(m)}$ below the conventional polylog scaling, in particular, to achieve~\eqref{eq:doubly_polylog_time_overhead}, while explicitly accounting for the runtime of classical decoding.
A quantum code with $n$ physical qubits (also called the code size), $k$ logical qubits, and distance $d$ is denoted by an $[[n,k,d]]$ code.
For a minimum-weight decoder, it has been proven that fault-tolerant protocols for $[[n,k,d]]$ topological codes with $d=\Theta(\poly(n))$ can exponentially suppress the logical error rate $\exp[-\Theta(d)]$ as we increase the code distance $d$~\cite{gottesman2014faulttolerant,PhysRevLett.109.180502,tamiya2024polylogtimeconstantspaceoverheadfaulttolerantquantum}.
Since the code size $n$ grows as $d$ increases, suppressing the logical error rate below $\lessapprox \frac{\epsilon(m)}{W(m)D(m)}$ requires a polylog code size~\cite{tamiya2024polylogtimeconstantspaceoverheadfaulttolerantquantum,PhysRevLett.109.180502}, i.e.,
\begin{align}
\label{eq:n_m}
n(m)=\Theta\qty(\polylog\qty(\frac{W(m)D(m)}{\epsilon(m)})),
\end{align}
where we may omit the argument to write $n$ if it is obvious from the context.
When $k=O(1)$, the polylog code size typically results in a polylog space overhead as $\frac{W_\mathrm{FT}(m)}{W(m)}=\Theta\qty(\frac{n(m)}{k(m)})=\Theta\qty(\polylog\qty(\frac{W(m)D(m)}{\epsilon(m)}))$, which will be the case for our protocol.

In our protocol, we employ the $[[n,k=1,d=\Theta(n^{1/3})]]$ three-dimensional (3D) subsystem surface code\footnote{This family of codes is also called 3D subsystem toric codes in Ref.~\cite{kubica2022single}, yet we call it 3D subsystem surface codes since the codes are defined with open boundary conditions.} originally introduced in Ref.~\cite{kubica2022single} as a single-shot-decodable topological code; whereas Ref.~\cite{kubica2022single} does not explicitly provide a protocol to implement logical operations, we show a complete protocol to implement all required logical operations for this code, i.e., state preparation, measurements, and gates (see Methods for details).
The protocol's time overhead is determined by three factors---the longest quantum-circuit depth $T_\mathrm{gate}(m)$ among implementing logical gates, the longest depth $T_\mathrm{SE}(m)$ of syndrome extraction (SE) rounds per logical gate, and the longest depth $T_\mathrm{dec}(m)$ for waiting for the classical decoding per logical gate---given by
\begin{align}
\label{eq:time_overhead}
    \frac{D_\mathrm{FT}(m)}{D(m)}=O\qty(T_\mathrm{gate}(m)+T_\mathrm{SE}(m)+T_\mathrm{dec}(m)),
\end{align}
where the argument $m$ may be omitted if there is no dependence on $m$.
Here, the reason $T_\mathrm{dec}(m)$ is included in~\eqref{eq:time_overhead} is that decoding must finish before proceeding to the next logical gate, especially when applying non-Clifford gates, as discussed in Supplementary Materials~\ref{supple:backlog}.
For the 3D subsystem surface codes, our logical gate implementations have $T_\mathrm{gate}=O(1)$ quantum depths through gate teleportation and transversal gates (up to qubit swaps), and each syndrome extraction requires only a single round, i.e., $T_\mathrm{SE}=O(1)$, due to the single-shot property shown in Ref.~\cite{kubica2022single}.
This implies that the dominant contribution to the time overhead~\eqref{eq:time_overhead} is indeed the runtime $T_\mathrm{dec}(m)$ of the classical decoder, which cannot be ignored in minimizing the time overhead.
Following the single-shot decoding procedure in Ref.~\cite{kubica2022single}, the decoder for the 3D subsystem surface codes in our protocol solves the problem of finding MWPMs in graphs of size at most $O(\poly(n))$, which is performed at most constant times per logical gate.
For our purpose, it is insufficient to use conventional MWPM decoding algorithms based on the blossom algorithm due to the polynomial decoding time $T_\mathrm{dec}(m)=O(\poly(n(m)))$ in $n$, which results in a polylog time overhead due to~\eqref{eq:n_m} and~\eqref{eq:time_overhead}.
By contrast, we will develop below an alternative approach to achieve the MWPM decoding within polylog parallel time
\begin{align}
\label{eq:decoder_runtime}
    T_\mathrm{dec}(m)=O(\polylog(n(m))).
\end{align}

\begin{figure}
\includegraphics[width=0.7\linewidth]{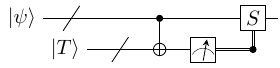}
\caption{\label{fig:teleportation} A circuit for $T$-gate teleportation.
}
\end{figure}

Even if~\eqref{eq:time_overhead} exhibits a doubly polylog time overhead, one cannot immediately conclude that the runtime of the entire computation achieves a doubly polylog time overhead; one must also address the backlog problem~\cite{T10}.
The backlog problem refers to the phenomenon where, if a decoder cannot keep up with syndrome generation, the computation will experience an exponential slowdown especially when applying non-Clifford gates (see Supplementary Materials~\ref{supple:backlog} for more details).
In the case of the 2D surface codes, fortunately, the parallel window decoding (also known as sandwich decoding and modular decoding)~\cite{https://doi.org/10.48550/arxiv.2209.08552,https://doi.org/10.48550/arxiv.2209.09219,bombin2023modular} enables arbitrary high decoding throughput regardless of the speed of a decoder, avoiding the backlog problem.
However, in the case of single-shot decodable codes such as the 3D subsystem surface codes, we cannot employ the parallel window decoding to avoid the backlog problem, because the analyses of single-shot error correction rely on the assumption that the correction of the previous round has been performed before the next round of decoding is performed~\cite{https://doi.org/10.48550/arxiv.2209.08552}.
Thus, if the decoding time per round is longer than the time required to generate a single round of syndrome, the backlog problem occurs~\cite{https://doi.org/10.48550/arxiv.2209.08552}.
This means that avoiding the backlog problem in an asymptotic regime naively requires a constant-time decoder with a theoretical threshold guarantee, which does not currently exist for the 3D subsystem surface codes.

In this work, we propose an alternative approach to avoid the backlog problem, using code concatenation. 
Imagine applying $T$ gates by gate teleportation shown in Fig.~\ref{fig:teleportation}.
Conventionally, while decoders that process the syndromes necessary to determine the auxiliary block's logical measurement outcome are running, syndrome extraction is repeatedly performed on the data block, and these syndromes must be processed when applying the next $T$ gate, causing the backlog problem.
Here, we instead propose that while the decoders are running, syndrome measurements on the data block are not repeated, and our protocol concatenates this topological-code protocol with a conventional fault-tolerant protocol using concatenated Steane codes~\cite{G} to ensure that the rate of error accumulation remains below the constant threshold of the protocol for the 3D subsystem surface code.
Since the error accumulation while waiting for the polylog decoding runtime is small, the overhead factor from this concatenated-code protocol is not dominant.
Overall, due to~\eqref{eq:n_m},~\eqref{eq:time_overhead} and~\eqref{eq:decoder_runtime}, our protocol achieves the doubly polylog time overhead~\eqref{eq:doubly_polylog_time_overhead} while avoiding the backlog problem.
See Methods for details of our protocol and the overhead analysis, as well as upper and lower bounds of time overheads of other protocols.

\paragraph*{Polylog-time parallel minimum-weight perfect matching decoder}
We describe our MWPM decoding strategy illustrated in Fig.~\ref{fig:main} (see Methods for details).
For $[[n,k,d]]$ surface codes with $d=\Theta\qty(\poly(n))$, the decoder typically handles $O\qty(\poly(n))$ syndrome measurement outcomes obtained from a certain part of the circuit consisting of at most a constant number of gate gadgets and SE gadgets, which we call a window for the decoding.
Our decoder is designed to achieve the polylog parallel runtime in~\eqref{eq:decoder_runtime} for this situation using the surface codes.
As discussed above, by applying this decoder to the 3D subsystem surface codes, we obtain a fault-tolerant protocol achieving the doubly polylog time overhead in~\eqref{eq:doubly_polylog_time_overhead}.
At the same time, this MWPM decoding is also applicable to the conventional $[[n,k=1,d=\Theta(\sqrt{n})]]$ 2D surface codes; in this case, notably, our MWPM decoding can be feasibly executed during the $d=\Theta(\poly(n))$ rounds of syndrome extraction for arbitrarily large $n$ since the decoder's runtime scales faster than $d$ rounds of syndrome extraction.
This makes it possible to apply the conventional MWPM decoding procedure to the 2D surface codes online in chronological order (also known as sliding window decoding)~\cite{doi:10.1063/1.1499754}, thereby making it possible to ensure the threshold existence through the conventional proof techniques in Refs.~\cite{gottesman2014faulttolerant,PhysRevLett.109.180502,tamiya2024polylogtimeconstantspaceoverheadfaulttolerantquantum}.

Conventionally, MWPM decoders rely on variants of the blossom algorithm~\cite{Edmonds_1965,edmonds1965maximum} to find an MWPM in a certain graph within polynomial time in $n$; however, a fundamental limitation is that the blossom algorithm itself is inherently difficult to parallelize to achieve polylogarithmic runtime, despite extensive efforts to accelerate it~\cite{4567800,kolmogorov2009blossom,DEZSO201123}.
For parallelization, Ref.~\cite{DBLP:journals/qic/Fowler15a} argued that the errors of the surface codes may, on average, cluster into $O(1)$-size subregions, but the MWPM decoder in Ref.~\cite{DBLP:journals/qic/Fowler15a} still requires $O(\poly(n))$ parallel runtime in the worst case; also problematically, existing proofs of error suppression $\exp(-\Theta(d))$ for topological codes essentially rely on finding an MWPM as a globally optimal solution over the entire $O\qty(\poly(n))$-size window~\cite{gottesman2014faulttolerant,PhysRevLett.109.180502}, rather than only within its local $O(1)$-size subregions.
State-of-the-art MWPM decoders, such as sparse blossom~\cite{higgott2023sparse} and fusion blossom~\cite{wu2023fusion}, can partially parallelize the blossom-algorithm-based decoding strategy but still have $O\qty(\poly(n))$ runtime for decoding the $O\qty(\poly(n))$-size window.
In contrast, we present an MWPM decoding strategy that finds a globally optimal solution within a polylog parallel runtime not on average but with a worst-case guarantee.

We describe our decoding strategy applicable to surface codes defined both in two and three spatial dimensions.
For feasibility, it is essential to recall that the fault-tolerant protocol supports a finite universal gate set; hence, the decoder is invoked for finite different types of windows that may include combinations of, at most, a constant number of logical operations.
Given the syndrome measurement outcomes in each window type, the task of decoding is to output a recovery operation on the code block by estimating the faulty locations in the window.

For each window type, we define a detector graph as in Ref.~\cite{higgott2023sparse}, which is precomputed prior to executing FTQC\@.
When performing multiple rounds of syndrome extraction, as in the 2D surface codes, the parity of each pair of successive syndrome measurement outcomes along time direction is called a detector, which is said to be active if the two outcomes differ.
An active detector is also called a detection event.
In the case of single-shot decoding, as in the 3D subsystem surface codes, each measurement outcome from the single-round syndrome extraction is used as a detector, which is considered active if the syndrome value is flipped.
For simplicity, we will present the decoding for $X$ errors below, but the decoding for $Z$ errors can be performed independently of that for $X$ errors to guarantee the error suppression.\footnote{Taking into account the correlation between $X$ and $Z$ errors may improve decoding performance~\cite{fowler2013optimal}, but the presented strategy suffices for our theoretical results.}
A detector graph is defined as a weighted simple graph $\mathcal{G}=(\mathcal{V},\mathcal{E})$, where the set $\mathcal{V}$ of vertices includes the vertices representing the detectors and boundary vertices.
If we have a faulty location in the window, $X$ errors occurring at the faulty location may flip (i.e., activate) one or two detectors~\cite{10.5555/2011383.2011385,PhysRevA.83.020302}.
In the conventional definition, an edge between two detectors in $\mathcal{E}$ has a non-negative real-valued weight given by the negative logarithm of an upper bound of the probability of errors activating the two detectors at the end of the edge, while the weight of an edge between a detector and a boundary vertex is given by that activating the single detector at one of the ends of the edge.
However, for the feasibility of the MWPM algorithm, we here represent the weights of edges of the detector graph with finite-digit integers, scaling them by a constant factor and rounding them up if necessary.
Given the finite logical gate set of the fault-tolerant protocol, we identify, in advance, a finite set of detector graphs defined for all possible window types.

Given syndrome measurement outcomes, using the detector graph, the decoder computes a path graph $\overline{\mathcal{G}}=\qty(\overline{\mathcal{V}},\overline{\mathcal{E}})$~\cite{higgott2023sparse,10.5555/2011362.2011368,PhysRevA.89.022326}, an $O(\poly(n))$-size weighted simple graph in which the decoder will find an MWPM\@; importantly, the polylog-time decoder requires that the path graph must be constructed within polylog time in parallel.\footnote{While Ref.~\cite{higgott2023sparse} presents a path graph as a complete graph, we define a path graph as a graph to find an MWPM therein. The path graphs in our work are not necessarily complete graphs since our construction of path graphs follows the convention of Refs.~\cite{10.5555/2011362.2011368,PhysRevA.89.022326}.}
In particular, from syndrome measurement outcomes, the decoder forms a subset $\mathcal{A}\subseteq\mathcal{V}$ of vertices in the detector graph representing detection events, which we call active-detector vertices.
The decoder then lists the closest boundary vertex in the detector graph for each active-detector vertex to form the set $\mathcal{A}'$ of these corresponding boundary vertices, where $|\mathcal{A}'|=|\mathcal{A}|$.
The set $\overline{\mathcal{V}}\coloneqq\mathcal{A}\cup\mathcal{A}'$ of vertices of the path graph $\overline{\mathcal{G}}$ consists of these active-detector and corresponding boundary vertices.
The set $\overline{\mathcal{E}}=\mathcal{E}_\mathcal{A}\cup\mathcal{E}_{\mathcal{A}\mathcal{A}'}\cup\mathcal{E}_\mathcal{A'}$ of weighted edges is composed of edges with weights representing the shortest path length in the detector graph between all pairs of active-detector vertices in $\mathcal{A}$, those between each active-detector vertex in $\mathcal{A}$ and the corresponding closest boundary vertex in $\mathcal{A}'$, and weight-zero edges between all pairs of boundary vertices in $\mathcal{A}'$.

One could use Dijkstra's algorithm~\cite{dijkstra1959note,10.1145/28869.28874,10756107} to compute the shortest path length between each pair of vertices in the $O(\poly(n))$-size detector graph within $O(\poly(n))$ time, thereby determining the weight of each edge in $\overline{\mathcal{E}}$ up to scaling and rounding; however, problematically, it is unknown how to parallelize Dijkstra's algorithm to achieve $O(\polylog(n))$ parallel runtime, making it difficult to perform while executing FTQC\@.
To resolve this issue, we propose to use an $O(\poly(n))$-size lookup table: for each pair of vertices in the detector graph, we store the length of the shortest path between the pair.
Since the detector graph has only $O(\poly(n))$ vertices, we can prepare this lookup table using Dijkstra's algorithm within $O(\poly(n))$ time.\footnote{An upper bound of the size of the lookup table is quadratic in terms of the detector graph size since it stores the path lengths between all pairs of its vertices. However, for efficiency in practice, its compression may be feasible under reasonable assumptions, such as symmetry in the syndrome-extraction circuit and physical error rates.}
The use of lookup tables to accelerate surface-code decoders was also proposed in Ref.~\cite{10.1145/3503222.3507707}, but the proposal in Ref.~\cite{10.1145/3503222.3507707} was to store all possible inputs and outputs of entire decoding process in an exponential-size lookup table, which may be infeasible on large scales; by contrast, our proposal is fundamentally different: we only store a polynomial-size lookup table to assist the MWPM decoders, which suffices to achieve the overall polylog runtime of our MWPM decoding.
Note that Ref.~\cite{liao2023wit} also employs a polynomial-size lookup table of the shortest-path lengths to accelerate a greedy decoder; however, their goal is practical speed-up, whereas we use such a table to obtain a theoretical upper bound on the decoder's runtime complexity while guaranteeing the threshold existence.
With this lookup table prepared in advance, our decoder constructs the path graph from the given syndrome measurement outcomes by reading weights in the lookup table in parallel.
Additionally, we precompute another polynomial-size lookup table in the same way, which maps each pair of vertices in the detector graph to the corresponding recovery operation on the surface code block, so that the decoder can output the recovery operation from the MWPM of the path graph by reading the lookup table in parallel.
See Methods for details.

At this point, the decoding task reduces to finding an MWPM in the path graph, which we require to be achieved within $O(\polylog(n))$ parallel runtime.
One could find the MWPM using the blossom algorithm~\cite{Edmonds_1965,edmonds1965maximum}, but as discussed above, it is unknown how to parallelize the blossom algorithm to achieve $O(\polylog(n))$ time.
By contrast, we employ another approach to find an MWPM by parallelizable algorithms.
Our approach is based on the isolation-based MWPM algorithm---a celebrated result in theoretical computer science~\cite{vazirani1987matching}, which shows that if the graph $\overline{\mathcal{G}}=\qty(\overline{\mathcal{V}},\overline{\mathcal{E}})$ has a unique MWPM (i.e., the MWPM is isolated), then the MWPM can be found by computing the matrix determinants in parallel for a certain $O\qty(\poly\qty(\left|\overline{\mathcal{V}}\right|))$-size set of $O\qty(\left|\overline{\mathcal{V}}\right|)\times O\qty(\left|\overline{\mathcal{V}}\right|)$ matrices (see Methods for details).
In our case of $\left|\overline{\mathcal{V}}\right|=O\qty(\poly(n))$, the computation of these determinants can be performed in $O(\polylog(n))$ runtime using $O(\poly(n))$ parallel processes via Samuelson-Berkowitz algorithm~\cite{BERKOWITZ1984147} (see also Methods for other parallel algorithms for this), where the arithmetics, i.e., addition and multiplication, in computing the determinants can also be parallelized~\cite{patterson2020computer,10.1007/978-3-540-45209-6_127}.

In this approach, an MWPM should be isolated for the feasibility of finding the MWPM, but the isolation may not always hold for the path graph in general; the remaining issue is how to achieve this.
To resolve this issue, Ref.~\cite{vazirani1987matching} proposed to add a small random perturbation to the weight of each edge to ensure, with a high probability, that the MWPM in the resulting weight-perturbed graph becomes isolated while remaining the same as one of the MWPMs in the original graph.
This probabilistic argument suggests that, by trying almost all possibilities of these perturbations in parallel, we can find the MWPM deterministically---without affecting the logical error rate of decoding at all---since at least one of the weight-perturbed path graphs will contain an isolated MWPM\@.
More recent breakthrough results~\cite{8104102} prove an even stronger guarantee: for any given graph $\overline{\mathcal{G}}=\qty(\overline{\mathcal{V}},\overline{\mathcal{E}})$ with arbitrary topology, it indeed suffices to use a quasi-polynomial-size $O\qty(\text{quasi-poly}\qty(\left|\overline{\mathcal{V}}\right|))$ set of weight-perturbed graphs to ensure that at least one of the graphs in the set contains the isolated MWPM as desired, where quasi-polynomial $O(\text{quasi-poly}(x))\coloneqq\exp[O(\polylog(x))]$ is slightly larger than polynomial but substantially smaller than exponential.
Therefore, given the path graph of size $\left|\overline{\mathcal{V}}\right|=O(\poly(n))$, by trying all such weight-perturbed path graphs in parallel using $O(\text{quasi-poly}(n))$ subprocesses, the decoder can deterministically find its MWPM within $O(\polylog(n))$ parallel runtime.
Once the MWPM is found, the decoder outputs the corresponding recovery operations by reading the precomputed lookup table, as discussed above.
See also Methods for details.

In summary, the overall decoding strategy shown above achieves the polylog-time parallel MWPM decoding due to the use of lookup tables and the isolation-based algorithm for finding an MWPM\@.
It may require precomputation upon designing the fault-tolerant protocol, but the precomputation is feasible within polynomial computational resources; it uses $O(\poly(n))$-time classical computation to prepare the $O(\poly(n))$-size lookup table.
Then, due to~\eqref{eq:n_m}, given syndrome measurement outcomes in a window, our MWPM decoder outputs the recovery operations within
$T_\mathrm{dec}(m)=O(\polylog(n(m)))=O\qty(\polylog\qty(\polylog\qty(\frac{W(m)D(m)}{\epsilon(m)})))$ 
parallel runtime per decoding an $O(\poly(n))$-size window, using $O(\text{quasi-poly}(n))=O\qty(\text{quasi-polylog}\qty(\frac{W(m)D(m)}{\epsilon(m)}))$ parallel classical processors.
It should be noted that the required number of parallel processors is quasi-polylog in terms of the size of the original circuit, substantially smaller than the size of the original circuit itself.
As discussed above, by applying this polylog-time MWPM decoder to decoding the 3D subsystem surface codes in our protocol, we achieve doubly-polylog-time-overhead FTQC, as summarized in the following theorem.
\begin{theorem}[Doubly-polylog-time-overhead FTQC]
For fault-tolerant protocols with $[[n,k,d]]$ surface codes discussed above, we construct an MWPM decoder achieving $O(\polylog(n))$ parallel runtime per decoding an $O(\poly(n))$-size window.
Using this decoder, we have the fault-tolerant protocol achieving doubly polylog time overhead $\frac{D_\mathrm{FT}(m)}{D(m)}=O\qty(\polylog\qty(\polylog\qty(\frac{W(m)D(m)}{\epsilon(m)})))$.
\end{theorem}

\paragraph*{Sublinear-scaling conjecture on the required size of the set of weight-perturbed path graphs}

\begin{figure}
    \centering
    \includegraphics[width=\linewidth]{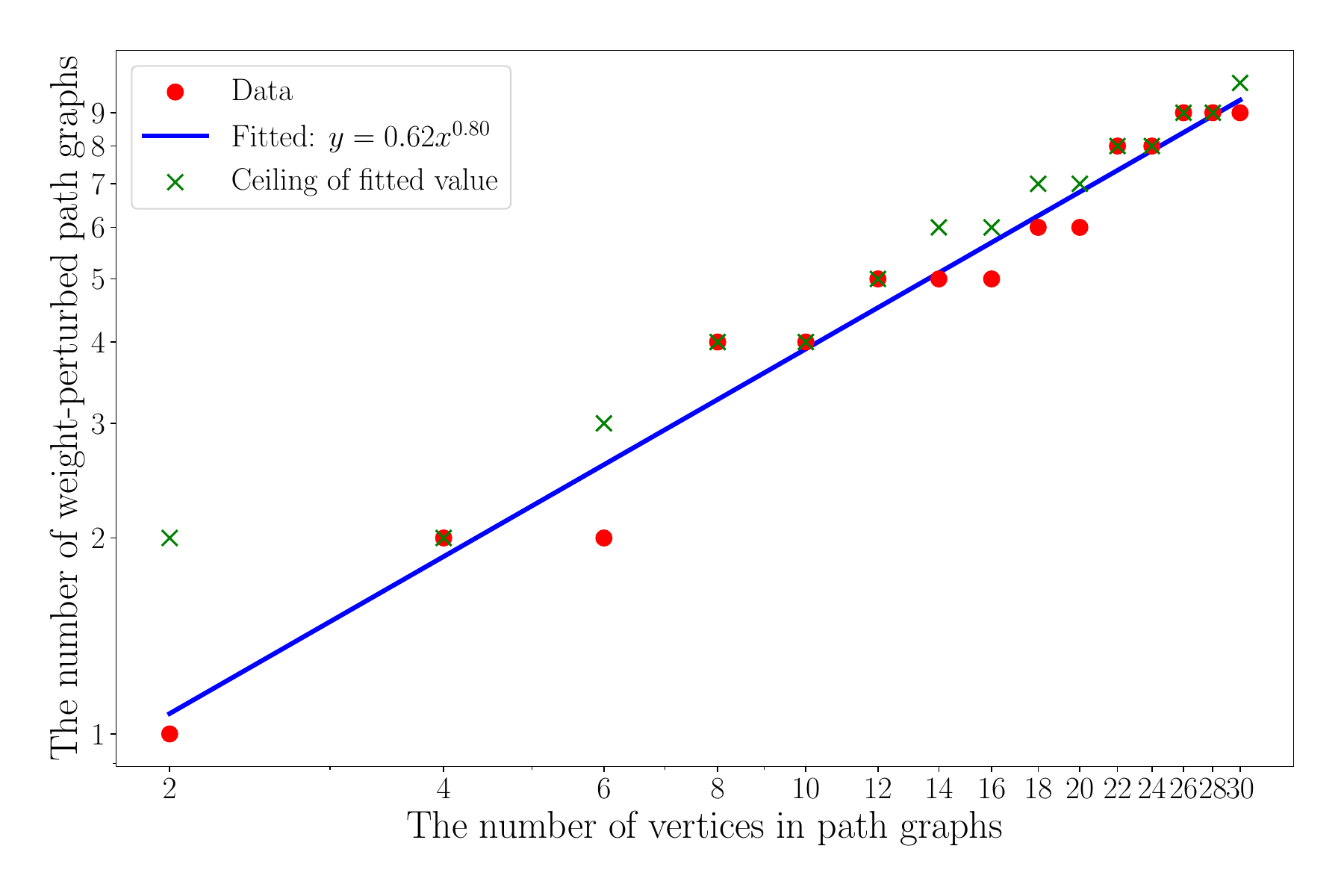}
    \caption{Upper bounds of the required size $y$ of the set of weight-perturbed path graphs to find the MWPM feasibly with our decoding strategy for each number $x=\left|\overline{\mathcal{V}}\right|\in\{2,4,\ldots,30\}$ of vertices in path graphs $\overline{\mathcal{G}}=\qty(\overline{\mathcal{V}},\overline{\mathcal{E}})$ (red circles), where $x$ is even by construction of the path graphs. The blue line, obtained by fitting, shows that a sublinear scaling $y\leq\left\lceil0.62x^{0.80}\right\rceil$ holds in this practical regime. The green cross markers indicate upper bounds of the smallest required size of the set of weight-perturbed path graphs for our decoding algorithm, obtained by rounding up the values of the fitting line.
    }
    \label{fig:numerics_main}
\end{figure}

An apparent bottleneck in implementing the above decoding strategy is the potentially large number of parallel processes required to handle the quasi-polynomial-size set of weight-perturbed path graphs in terms of the number $\left|\overline{\mathcal{V}}\right|$ of vertices of the original path graph $\overline{\mathcal{G}}=\qty(\overline{\mathcal{V}},\overline{\mathcal{E}})$; however, this requirement appears to stem merely from the current proof techniques in Ref.~\cite{8104102}, which theoretically guarantees the existence of the isolated MWPM but likely overestimate the upper bound compared to the optimal number of required parallel processes.
To resolve this issue, instead of using the weight perturbation methods proposed in Ref.~\cite{8104102}, we propose an alternative method using a pseudo-random number generator with an appropriately chosen seed to deterministically construct the set of weight-perturbed path graphs.
Through numerical simulation, we demonstrate that with this method, a sublinear-size set, i.e., an $o\qty(\left|\overline{\mathcal{V}}\right|)$-size set, of weight-perturbed path graphs indeed suffices for our decoding strategy to feasibly find the MWPM in a practically relevant regime.

To demonstrate this, we performed a numerical simulation of a conventional memory experiment for the fault-tolerant protocol with the $[[n,1,d=\sqrt{n}]]$ 2D rotated surface codes under the circuit-level depolarizing error model at a physical error rate $p=0.1\%$ (see Methods for details).
We repeated this simulation $10^5$ times for each $d=3, 4,\ldots,11$, constructing the original path graphs $\overline{\mathcal{G}}=\qty(\overline{\mathcal{V}},\overline{\mathcal{E}})$ from the syndrome measurement outcomes.
For all original path graphs with $\left|\overline{\mathcal{V}}\right|\leq 30$, we constructed sets of weight-perturbed path graphs of various sizes using a Mersenne twister~\cite{10.1145/272991.272995} as a pseudo-random number generator, instead of the construction in Ref.~\cite{8104102} (see Methods for details).
In Fig.~\ref{fig:numerics_main}, for each $\left|\overline{\mathcal{V}}\right|$, we plot an upper bound of the minimum required size of this set for our isolation-based decoding strategy to successfully find the MWPMs in all the path graphs encountered in the simulation.
The plot suggests that while the required size may increase as $\left|\overline{\mathcal{V}}\right|$ grows, its growth is at most only sublinear in $\left|\overline{\mathcal{V}}\right|$ in this regime.

Based on this numerical evidence, we conjecture that for any path graph $\overline{\mathcal{G}}=\qty(\overline{\mathcal{V}},\overline{\mathcal{E}})$ appearing in the decoding problem, there exists an $o\qty(\left|\overline{\mathcal{V}}\right|)$-size set of weight-perturbed path graphs such that at least one weight-perturbed path graph in the set contains an MWPM that is uniquely isolated and remains the same as one of the MWPMs in $\overline{\mathcal{G}}$.
We remark that a rigorous proof of this conjecture for arbitrary-topology graphs would imply the derandomization of polylog-time parallel randomized algorithms for finding MWPMs--- representing a big breakthrough in theoretical computer science~\cite{vazirani1987matching,DAHLHAUS199879,4568269,10.1007/978-3-540-70918-3_42,datta2010deterministically,TEWARI20121,10.1145/2897518.2897564,goldwasser_et_al:LIPIcs.ICALP.2017.87,8104102}; the contribution here is to provide numerical evidence supporting this conjecture in a regime relevant to its applications.

\section*{Discussion}

We have shown that FTQC is achievable within a doubly polylog time overhead, including the nonzero runtime of decoding, while maintaining the conventional polylog space overhead.
Conventionally, FTQC had polylog space and time overheads, and determining the ultimate bounds on time overheads has been hard due to the lack of techniques for surpassing polylog scaling.
We have argued that even with protocols for topological codes with single-shot features, the mere use of MWPM decoders based on polynomial-time blossom algorithms would still result in polylog time overheads; in contrast, our key contribution has been a polylog-time parallel MWPM decoder to eliminate the bottleneck and successfully surpass the polylog time overhead.

Our results point to a new frontier of low-overhead FTQC; on one hand, recent theoretical progress has proven that the constant space overhead is achievable with the (quasi-)polylog time overhead~\cite{yamasaki2022timeefficient,tamiya2024polylogtimeconstantspaceoverheadfaulttolerantquantum}, and on the other hand, this work shows that an even shorter, doubly polylog time overhead is achievable when allowing for polylog space overheads.
Apart from this, Ref.~\cite{wills2024constantoverheadmagicstatedistillation} has recently introduced and analyzed a protocol for constant-overhead magic state distillation, and based on these theoretical advances, it would be interesting to compare the cost of universality in FTQC, as in Ref.~\cite{PRXQuantum.2.020341}, in future research.
Furthermore, while we have focused on MWPM decoders, it would also be worthwhile to investigate the overheads of FTQC when using other decoders, such as cellular automaton (CA) decoders~\cite{Harrington_phd,PhysRevLett.123.020501,vasmer2021cellular} and self-correcting quantum computers~\cite{Bombin_2013_self_correcting}.
With these open directions, an exciting time has arrived to further explore new frontiers in the ultimate space and time overheads, as well as the fundamental space-time tradeoff in FTQC\@.

Finally, beyond the theoretical scope of this work, we also remark on the implementability of our polylog-time parallel MWPM decoder, which is an important avenue for future investigation.
To implement our decoder, it is essential to fully parallelize all necessary computations, including matrix determinant calculations, additions, and multiplications.
Thus, meaningful implementations of our decoding strategy may require hardware programming to ensure parallelization; such hardware-efficient realizations of MWPM decoders would themselves be a major achievement for practical FTQC~\cite{battistel2023real}.
Our contribution lies in providing the theoretical foundation for opening up the possibility of such polylog-time MPWM decoding, and it is exciting to investigate this new option from the experimental and computer-architectural perspectives as well.

\section*{Materials and Methods}

In Methods, we first discuss the requirements for achieving doubly-polylog-time-overhead FTQC, followed by a description of our fault-tolerant protocol that meets this criterion.
We then present our polylog-time parallel algorithm for MWPM decoding and provide details of our numerical simulation.
Finally, we analyze the upper and lower bounds of time overheads for various other fault-tolerant protocols.

\section*{Requirements for doubly-polylog-time-overhead FTQC}

The problem to be addressed in this work is to shorten the time overhead $\frac {D_\mathrm{FT}(m)}{D(m)}$ below the conventional polylog scaling, even if we take into account the cost of classical computation during executing FTQC such as that of decoding.
Topological codes, such as surface codes and color codes, are canonical examples of quantum low-density parity-check (LDPC) codes, where each stabilizer generator has a constant weight, and each physical qubit is involved in a constant number of stabilizer generators.
Using a minimum-weight decoder, fault-tolerant protocols for $[[n,k,d]]$ quantum LDPC codes with $d=\Theta(\poly(n))$ can exponentially suppress the logical error rate per logical operation, i.e., $\exp[-\Theta(d)]$, as we increase the code distance $d$~\cite{gottesman2014faulttolerant,PhysRevLett.109.180502,tamiya2024polylogtimeconstantspaceoverheadfaulttolerantquantum},\footnote{The proof of the threshold theorem needs to handle the local stochastic Pauli model defined for the entire fault-tolerant circuit rather than a single code block. As pointed out in Ref.~\cite{tamiya2024polylogtimeconstantspaceoverheadfaulttolerantquantum}, the argument in Ref.~\cite{gottesman2014faulttolerant} overlooks the correlations between a code block in the fault-tolerant circuit and the rest of the fault-tolerant circuit surrounding the code block.
The argument in Ref.~\cite{PhysRevLett.109.180502} also considers a quantum memory of a single code block without addressing the correlations in the entire fault-tolerant circuit.
However, incorporating the analysis in Ref.~\cite{gottesman2014faulttolerant} with the technique of partial circuit reduction introduced in Ref.~\cite{tamiya2024polylogtimeconstantspaceoverheadfaulttolerantquantum}, the proof can be completed, as discussed in Ref.~\cite{tamiya2024polylogtimeconstantspaceoverheadfaulttolerantquantum}.} but increasing $d$ requires growing the code size $n$; in this case, to suppress it below $\lessapprox \frac{\epsilon(m)}{W(m)D(m)}$, we need a polylog code size~\cite{tamiya2024polylogtimeconstantspaceoverheadfaulttolerantquantum,PhysRevLett.109.180502}
\begin{align}
\label{eq:n_m_methods}
n(m)=\Theta\qty(\polylog\qty(\frac{W(m)D(m)}{\epsilon(m)})).
\end{align}

The protocols with the quantum LDPC codes compile the original circuit into the fault-tolerant circuit by replacing each operation with the corresponding gadget to implement the operation at the logical level and then inserting an error correction (EC) gadget between each adjacent pair of the gadgets implementing the logical operations to extract syndromes of all stabilizer generators.
Using the syndrome measurement outcomes, the decoder estimates the locations of physical errors and outputs recovery operations on the physical qubits of the code block for quantum error correction.
A part of the circuit composed of a gate gadget followed by an EC gadget is called a gate rectangle~\cite{10.5555/2011665.2011666}, which includes syndrome extraction and the wait operations while running the decoder.
Note that in practice, particularly for Clifford circuits, it is possible to perform the next gadget without waiting for the decoding to complete. However, as described in the main text, when applying non-Clifford gates, we wait until the decoding is finished to avoid the backlog problem.
For this reason, the definition of the EC gadget here includes wait operations until decoding is completed.
The maximal depth of the gate rectangles in the circuit is called the logical gate time.\footnote{Some protocols, such as lattice surgery~\cite{horsman2012surface,Litinski2019gameofsurfacecodes},
may not have a clear separation between a gate gadget and syndrome extraction, but the logical gate time can be defined as the required depth of the fault-tolerant circuit per gate of the original circuit.}
Since preparations and measurements appear only once at the beginning and end of the original circuit, respectively, their time overheads do not affect the asymptotic scaling of the overall time overhead.
Therefore, the time overhead of the protocols with quantum LDPC codes is determined by the logical gate time, i.e.,
\begin{align}
\label{eq:time_overhead_methods}
\frac {D_\mathrm{FT}(m)}{D(m)}=O\qty(T_\mathrm{gate}(m)+T_\mathrm{SE}(m)+T_\mathrm{dec}(m)),
\end{align}
where $T_\mathrm{gate}(m)$ is the longest depth among the gate gadgets for the universal gate set in use, $T_\mathrm{SE}(m)$ is that of syndrome extraction (SE) per logical gate, and $T_\mathrm{dec}(m)$ is that of waiting for the decoder per logical gate.

To shorten the time overhead, it is vital to shorten all $T_\mathrm{gate}(m)$, $T_\mathrm{SE}(m)$, and $T_\mathrm{dec}(m)$ in~\eqref{eq:time_overhead} simultaneously, which causes a significant challenge.
For some of the gates in a universal gate set, transversal gate implementation may achieve $T_\mathrm{gate}=O(1)$, where $T_\mathrm{SE}(m)$ and $T_\mathrm{dec}(m)$ dominate the time overhead.
For universal quantum computation, we need to combine it with additional techniques, such as gate teleportation~\cite{gottesmanchuang1999,PhysRevA.62.052316} and gauge fixing~\cite{PhysRevLett.111.090505}, which also achieve
\begin{align}
\label{eq:T_gate_methods}
     T_\mathrm{gate}=O(1),
\end{align}
and $T_\mathrm{SE}(m)$ and $T_\mathrm{dec}(m)$ become dominant as well.
In the case of the conventional protocols with the $[[n,k=1,d=\Theta(\sqrt{n})]]$ 2D surface codes, the logical gate time includes $O(d)$ rounds of syndrome extraction to ensure the correctability of syndrome measurement errors~\cite{doi:10.1063/1.1499754}, thus incurring at least polylog time overhead due to $T_\mathrm{SE}(m)=O(d(m))=O\qty(\polylog\qty(\frac{W(m)D(m)}{\epsilon(m)}))$, where we use~\eqref{eq:n_m_methods}.
To avoid the issue of growing $T_\mathrm{SE}(m)$, Ref.~\cite{PhysRevX.5.031043} proposed to use topological codes equipped with the capability of single-shot decoding, with which we can correct measurement errors only within a single round of syndrome extraction, i.e.,
\begin{align}
\label{eq:single-shot}
    T_\mathrm{SE}=O\qty(1).
\end{align}
By combining it with the gauge fixing to implement all gates in a universal gate set within constant time $T_\mathrm{gate}=O(1)$~\cite{bombin2015gauge}, it was shown in Ref.~\cite{PhysRevX.5.031043} that a protocol with 3D subsystem color codes\footnote{This family of codes is also called 3D gauge color codes in Refs.~\cite{bombin2015gauge,PhysRevX.5.031043} while we call it 3D subsystem color codes for consistency with its surface-code version.} can achieve
\begin{align}
\label{eq:3d_subsystem_color_code}
    T_\mathrm{gate}+T_\mathrm{SE}=O(1).
\end{align}
More recently, it was shown in Ref.~\cite{kubica2022single} that a protocol with the 3D subsystem surface codes can also achieve~\eqref{eq:single-shot} by single-shot decoding, serving as an alternative promising candidate for achieving~\eqref{eq:3d_subsystem_color_code} while protocols for logical gate implementation on the 3D subsystem surface codes have not been shown explicitly in Ref.~\cite{kubica2022single}.
Despite these advances, the last essential element in~\eqref{eq:time_overhead}, i.e., the runtime $T_\mathrm{dec}(m)$ of decoding, was assumed to be instantaneous in these works and thus was not explicitly upper bounded prior to our work.

We here analyze the runtime of classical decoders in these protocols to clarify the requirement for surpassing the polylog time overhead.
For representative protocols with topological codes, the only known, feasible way to bound the time overheads with the theoretical guarantee is to use the property of the minimum-weight decoder~\cite{gottesman2014faulttolerant,PhysRevLett.109.180502}, which estimates the locations of the minimum-weight (i.e., highest-probability) physical errors in the fault-tolerant circuit.
Alternatively, a maximal likelihood decoder would find the highest-probability logical errors to achieve an even better error suppression but may require an exponential runtime for typical topological codes~\cite{doi:10.1063/1.1499754,PhysRevA.90.032326,PhysRevResearch.6.023055}, which remains infeasible in general.
For special types of quantum LDPC codes such as quantum expander codes~\cite{7354429,10.1145/3188745.3188886} and asymptotically good quantum LDPC codes~\cite{10.1145/3519935.3520017,9996782,10.1145/3564246.3585101}, one can use efficient single-shot decoders~\cite{8555154,Grospellier,gu2024single}, but these decoders do not apply to topological codes.\footnote{Later in Methods, we will show that a protocol with quantum expander codes can also achieve a doubly polylog time overhead while maintaining a polylog space overhead, offering an alternative approach for surpassing the conventional polylog time overhead. At the same time, we will argue that the protocol with topological codes presented in the main text is a stronger candidate for realizing doubly-polylog-time-overhead FTQC, since it offers better implementability due to the structured nature of topological codes defined on a regular lattice, compared to the less structured expander graphs used in quantum expander codes.}
Heuristic decoders, such as union-find decoders~\cite{Delfosse2021almostlineartime}, belief propagation with ordered statistics post-processing (BP-OSD)~\cite{Panteleev2021degeneratequantum}, and renormalization-group (RG) decoders~\cite{PhysRevLett.104.050504}, may also work in practice but are insufficient for theoretically guaranteeing the existence of thresholds and overhead bounds.
Cellular-automaton (CA) decoders yield provable thresholds for certain families of topological codes~\cite{Harrington_phd,PhysRevLett.123.020501,vasmer2021cellular}, but it is unknown whether time-efficient methods for gate implementation and syndrome extraction are available for these codes.
By contrast, the minimum-weight decoder provides a theoretical guarantee of the required overheads for error suppression~\cite{gottesman2014faulttolerant,PhysRevLett.109.180502} and can be feasibly implemented for representative topological codes by variants of the blossom algorithm~\cite{Edmonds_1965,edmonds1965maximum}, which finds an MWPM in a given graph within polynomial time in terms of the graph size.
Here, the size of this graph in the decoding problem is upper bounded by a polynomial of the code block size.

Among the topological codes with single-shot features achieving~\eqref{eq:single-shot},  progress following the original analysis of the single-shot decodability of the 3D subsystem color codes in Ref.~\cite{PhysRevX.5.031043} suggests color-code decoders with a polynomial runtime $O(\poly(n))$ in terms of the code size $n$, by mapping the decoding problem into subproblems where MWPM algorithms are feasible~\cite{PhysRevA.89.012317,Kubica2023efficientcolorcode,PRXQuantum.3.010310}.
However, although these decoders use MWPM algorithms as a subroutine, they do not necessarily find the minimum-weight errors in the original decoding problem defined for the color code itself.
Thus, these decoders cannot be used to achieve theoretically guaranteed single-shot error correction under the existing proof in Ref.~\cite{PhysRevX.5.031043}, which relies on the assumption that the decoder finds minimum-weight errors.
On the other hand, for the 3D subsystem surface codes employed in this work, MWPM decoders can be used to achieve single-shot error correction because it has been shown in Ref.~\cite{kubica2022single} that applying MWPM decoding twice is enough for single-shot error correction in this case.

However, the existing polynomial-time MWPM decoders are insufficient to surpass the polylog time overhead since we have $T_\mathrm{dec}(m)=O(\poly(n(m)))=O\qty(\polylog\qty(\frac{W(m)D(m)}{\epsilon(m)}))$ due to~\eqref{eq:n_m_methods}.
Achieving the doubly polylog time overhead requires a polylog-time parallel MWPM decoder
\begin{align}
\label{eq:decoder_runtime_methods}
    T_\mathrm{dec}(m)&=O(\polylog(n(m)))\\
    &=O\qty(\polylog\qty(\polylog\qty(\frac{W(m)D(m)}{\epsilon(m)}))).
\end{align}
If one establishes a way to combine an MWPM decoder achieving~\eqref{eq:decoder_runtime_methods} with a protocol achieving~\eqref{eq:3d_subsystem_color_code} via single-shot decoding, the overall protocol will achieve doubly polylog time overhead, i.e.,
\begin{align}
\label{eq:doubly_polylog_time_overhead_methods}
    &T_\mathrm{gate}(m)+T_\mathrm{SE}(m)+T_\mathrm{dec}(m)\nonumber\\
    &=O(1)+O\qty(\polylog\qty(\polylog\qty(\frac{W(m)D(m)}{\epsilon(m)}))).
\end{align}
In the rest, we will address the problem of fulfilling this requirement by constructing a fault-tolerant protocol and a polylog-time parallel MWPM decoder as desired.

\subsection*{Description of the fault-tolerant protocol achieving a doubly polylog time overhead}
We describe our fault-tolerant protocol that achieves a doubly polylog time overhead while maintaining a polylog space overhead.
Our protocol exploits the $[[n,k=1,d=\Theta(n^{1/3})]]$ 3D subsystem surface codes introduced in Ref.~\cite{kubica2022single} to convert the original circuit into an intermediate circuit, which is composed of gadgets, i.e., circuits to implement logical operations and quantum error correction, for the 3D subsystem surface codes;
then, the protocol further uses the $[[7^L,1,3^L]]$ concatenated Steane codes~\cite{G,S3} to convert the intermediate circuit into the fault-tolerant circuit, where $L$ represents the concatenation level.

Our protocol uses the set of elementary operations composed of preparation of $\ket{0}$, measurement in the $Z$ basis $\{\ket{0},\ket{1}\}$, and a universal gate set composed of the Hadamard ($H$) gate, the controlled-$\textsc{NOT}$ ($\textsc{CNOT}$) gate, and the $T$ gate, along with the wait operation (i.e., the identity gate).
We assume that the original circuit is given in this gate set $\{H,T,\textsc{CNOT}\}$, starting from preparation of $\ket{0}$ and ending with measurement in the $Z$ basis.

To compile the original circuit into the intermediate circuit using the 3D subsystem surface codes, our fault-tolerant protocol replaces each elementary operation in the original circuit with the corresponding gadget and then inserts the EC gadget between each pair of adjacent gadgets.
In our protocol, the intermediate circuit is written in terms of $\{H,T,\textsc{CNOT}\}$.
While we refer to Ref.~\cite{kubica2022single} for the definition and single-shot property of the 3D subsystem surface codes, the gadgets for the 3D subsystem surface codes have not been presented explicitly in Ref.~\cite{kubica2022single}; to address this point, we will specify constructions of all the gadgets below.
For the code block size $n(m)$, the worst-case depth for preparation of the logical $\ket{0}$ state and the logical measurement in the $Z$ basis are denoted by, respectively,
\begin{align}
\label{eq:t_prep_methods}
    T_\mathrm{prep}(m)>0
\end{align}
and
\begin{align}
\label{eq:t_meas_methods}
    T_\mathrm{meas}(m)>0.
\end{align}
All gate gadgets presented below are implementable within a constant depth:
\begin{align}
\label{eq:t_gate_methods}
    T_\mathrm{gate}=O(1).
\end{align}
Also, the EC gadgets presented below are implementable by performing a single-round syndrome extraction, i.e.,
\begin{align}
\label{eq:t_se_methods}
    T_\mathrm{SE}=O(1)
\end{align}
and running the MWPM decoder twice.
We define, for our protocol,
\begin{align}
\label{eq:t_dec_methods}
    T_\mathrm{dec}(m)>0
\end{align}
as the worst-case depth for these two executions of the MWPM decoding.
Then, the worst-case depth of the EC gadget is given by
\begin{align}
\label{eq:t_ec_methods}
   T_\mathrm{EC}(m)\coloneqq T_\mathrm{SE}+T_\mathrm{dec}(m).
\end{align}

The detailed constructions of the gadgets for the 3D subsystem surface codes in our protocol are given as follows.
\begin{itemize}
\item 
The preparation gadget prepares the logical $\ket{0}$ state of a code block by initializing each qubit of the code block in $\ket{0}$ and performing a single round of syndrome extraction, followed by the single-shot decoding shown in Ref.~\cite{kubica2022single}.
In this single-shot decoding, the MWPM decoding is performed twice: first, to correct the measurement errors, and second, to estimate the errors on qubits in the code block.
Reference~\cite{kubica2022single} analyzes the single-shot decoding under the assumption that the input state is already in the code space; thus, its argument does not directly apply to state preparation. 
A similar single-shot error correction protocol for the 3D subsystem color codes is presented in Ref.~\cite{PhysRevX.5.031043}, but its proof also assumes code-state input and does not detail why the state preparation can be done in a single-shot manner.
Here, we explain why the single-shot decoding is applicable for state preparation in detail.
Let us first consider the case where there are no measurement errors. After a single round of syndrome extraction on $\ket{0}^{\otimes n}$, each $X$ gauge measurement outcome is individually random because these gauge operators anti-commute with the single-qubit $Z$ stabilizers of $\ket{0}$. However, due to the redundancy of the gauge operators, they are not completely random; they correlate in such a way that there exists a corresponding physical error configuration on physical qubits in the code block. In other words, the obtained measurement outcomes of the gauge operators satisfy the consistency conditions~\cite{kubica2022single} that always hold in the absence of measurement errors.
Consequently, measurement errors can still be detected just as they would be for an input that is already in the code space. Because $\ket{\bar{0}}$ is stabilized by a logical $Z$ operator, a recovery operation for the initially random stabilizer values does not lead to a logical error; residual errors are determined by an incorrect recovery operation caused by measurement errors. Therefore, the single-shot decoding arguments in Ref.~\cite{kubica2022single} that assume code-state input remain applicable for the state preparation by only considering measurement-error parameters, guaranteeing that the small number of measurement errors does not lead to large residual errors.
While the MWPM decoding is carried out classically, the gadget performs wait operations to account for the classical processing time.
\item The $Z$-basis measurement gadget performs the logical $Z$-basis measurement for a code block by measuring all the qubits of the code block in the $Z$ basis, followed by performing the MWPM decoder to estimate the logical measurement outcome, as in the conventional measurement procedure for the 2D surface codes~\cite{doi:10.1063/1.1499754}.
The gadget also includes the wait operations to wait for the MWPM decoder's runtime.
Since we directly measure the qubits in the code block, it is unnecessary to use multiround syndrome extraction or single-shot decoding to overcome the effect of measurement errors.
Instead, the syndrome values are directly computed from the measurement outcomes, allowing the MWPM decoding to be performed on the code block.
\item  The $H$-gate gadget performs a logical $H$ gate on a code block by transversally applying the $H$ gate to each qubit of the code block, followed by swapping the qubits of the code block, similar to the fold-transversal implementation of the logical $H$ gate of the 2D surface codes~\cite{PhysRevA.94.042316}.
To specify this qubit swap, we refer to Fig.~1 of Ref.~\cite{kubica2022single} as an illustration of the 3D subsystem surface code defined on a cubic lattice with an open boundary condition.
In this figure, the cubic volumes are colored in red and blue in a checkerboard pattern, where each $X$ (and $Z$) stabilizer generator is placed on a red (and blue, respectively) volume.
A bare logical $Z$ (and $X$) operator is supported on the front or rear (and right or left, respectively) boundary of the cubic lattice, where the logical qubit is encoded.
The transversal $H$ gates exchange the $X$ and $Z$ stabilizer generators placed in the checkerboard pattern and also exchange the bare logical $Z$ and $X$ operators appearing alternately on these four boundaries.
Therefore, by applying the swap gates to rotate the lattice $\pi/2$ around the vertical axis after the transversal $H$ gates, we obtain the 3D subsystem surface code defined on the same lattice with the logical $H$ gate applied as desired.
\item The $\textsc{CNOT}$-gate gadget performs a logical $\textsc{CNOT}$ gate on two code blocks by transversally applying the $\textsc{CNOT}$ gate to each pair of qubits between the two blocks.
The transversal implementation works because the 3D subsystem surface code is a Calderbank-Shor-Steane (CSS) code~\cite{G}.
\item The $T$-gate gadget performs a logical $T$ gate on a code block by the standard teleportation-based implementation, where a high-fidelity magic state is prepared as an auxiliary state, followed by gate teleportation. 
The circuit of gate teleportation is shown in Fig.~\ref{fig:teleportation}.
The high-fidelity magic state for quantum LDPC codes is prepared in a fault-tolerant way using the concatenated Steane codes, in the same way as the constant-space-overhead protocols with the quantum LDPC codes in Refs.~\cite{gottesman2014faulttolerant,8555154,Grospellier,tamiya2024polylogtimeconstantspaceoverheadfaulttolerantquantum} (see Appendix~A of Ref.~\cite{tamiya2024polylogtimeconstantspaceoverheadfaulttolerantquantum} for details of the state-preparation protocol).
Whereas the fault-tolerant protocols in Refs.~\cite{gottesman2014faulttolerant,8555154,Grospellier,tamiya2024polylogtimeconstantspaceoverheadfaulttolerantquantum} allow only limited parallelization in preparing the auxiliary states to maintain constant space overhead, our protocol fully parallelizes high-fidelity magic state preparation by permitting polylog space overhead, leading to~\eqref{eq:t_gate_methods}. 
\item The wait gadget performs the logical identity gate on a code block, implemented by applying identity gates.
\item  Finally, the EC gadget performs quantum error correction for a code block by conducting a single round of syndrome extraction, followed by performing the single-shot decoding~\cite{kubica2022single}.
While the MWPM decoding is running, the wait operations are performed.
\end{itemize}

We note that for 3D stabilizer surface codes defined on certain combinations of lattices, a logical $CCZ$ gate can be implemented by transversally applying $CCZ$ gates between the three code blocks; for example, such combinations are known for one 3D stabilizer surface code defined on a cubic lattice and two 3D stabilizer surface codes defined on a tetrahedral-octahedral lattice (in the Kitaev picture)~\cite{PhysRevA.100.012312}.
The 3D subsystem surface code defined on a cubic lattice, which we employ in this work, can be converted to a 3D stabilizer surface code on the tetrahedral-octahedral lattice through gauge fixing~\cite{kubica2022single}.
However, it remains unknown in the literature whether a logical $CCZ$ gate can be transversally implemented for three 3D stabilizer surface codes all defined on the tetrahedral-octahedral lattices.
Additionally, it also remains unknown whether one of the required 3D stabilizer surface codes for the transversal $CCZ$ gates in Ref.~\cite{PhysRevA.100.012312}, specifically the one defined on a cubic lattice (in the Kitaev picture), can be decoded within polynomial time with a theoretical guarantee, e.g., via MWPM decoding, making the analysis of overhead bounds challenging.
An alternative protocol in Ref.~\cite{iverson2020aspects} for 3D surface codes defined on different 3D lattices may achieve universality by gauge fixing, allowing for transversal implementation of the $T$ gate as a non-Clifford gate.
However, it remains unclear how to perform single-shot decoding with MWPM decoders for this code.
Furthermore, the corresponding 3D subsystem surface code introduced in Ref.~\cite{iverson2020aspects} is a non-CSS code, which is incompatible with our protocol.
For these reasons, in this work, we do not adopt protocols achieving universality via gauge fixing; instead, as shown above, we implement the $T$ gate via gate teleportation, which suffices to achieve our desired space and time overhead.
That being said, it would also be interesting to find a way to achieve single-shot error correction with MWPM decoders and universality by gauge fixing, for instance, by inventing novel combinations of subsystem and stabilizer surface codes defined on more favorable lattices.
We leave such developments for future work.

Given the intermediate circuit, our protocol further uses the fault-tolerant protocol with concatenated Steane codes to compile the intermediate circuit into the fault-tolerant circuit.
For the description of the protocol with concatenated Steane codes, we refer to standard references such as Ref.~\cite{G}, where we use the gate set $\{H,T,\textsc{CNOT}\}$ at any concatenation level.
For detailed construction and analysis of each gadget, see also Ref.~\cite{tamiya2024polylogtimeconstantspaceoverheadfaulttolerantquantum}.
In the intermediate circuit, the EC gadget includes not only the syndrome extraction but also the wait operations to account for the classical decoder's runtime.
In our setting, where classical computation is non-instantaneous classical computation---as is often the case in practical scenarios---the decoder for the 3D subsystem surface codes has a growing runtime as the size of the code block increases for scaling the original circuit.
In this case, errors may accumulate while waiting for the runtime of the decoder.
To guarantee the existence of a threshold, it is essential to further reduce the error rate in the intermediate circuit using the concatenated codes, as done in our protocol.
In particular, the rate of error accumulation while waiting for the decoding in the intermediate circuit must not exceed the threshold of the protocol for the 3D subsystem surface codes.
As we will analyze below, the use of the concatenated-code protocol does not impact the overall scaling of the time overhead.
Thus, the overall fault-tolerant circuit will maintain a doubly polylog time overhead as given in~\eqref{eq:doubly_polylog_time_overhead_methods} compared to the original circuit.

In the following, we analyze the time overhead of our protocol.
Let $W(m)$ and $D(m)$ denote the width and depth, respectively, of the original circuit, and let $\epsilon(m)$ be the target error.
As presented in the main text, we assume, as $m\to\infty$,
\begin{align}
\label{eq:W_assumption_methods}
    W(m)&\to\infty,\\
\label{eq:D_assumption_methods}
    D(m)&=O\qty(\poly\qty(W(m))),\\
\label{eq:epsilon_assumption_methods}
    \epsilon(m)&=\frac{1}{O\qty(\poly\qty(W(m)))},
\end{align}
which includes the conventional setting with fixed target error $\epsilon(m)=O(1)$ as a special case.
Let $W_\mathrm{int}(m)$ and $D_\mathrm{int}(m)$ denote the width and depth, respectively, of the intermediate circuit.
To achieve the overall target error $\epsilon(m)$ in implementing the $W(m)$-width $D(m)$-depth original circuit, the required size of each code block of the 3D subsystem surface code is $n(m)=\Theta\qty(\polylog\qty(\frac{W(m)D(m)}{\epsilon(m)}))$ as discussed in~\eqref{eq:n_m}.

Since each gadget for the 3D subsystem surface code is implemented with $\Theta(n(m))$ qubits for the $n(m)$-size code block,
the width of the intermediate circuit is
\begin{align}
    W_\mathrm{int}(m)=\Theta\qty(n(m)W(m)).
\end{align}
Hence, the space overhead of the intermediate circuit is
\begin{align}
\label{eq:space_overhead_intermediate}
    \frac{W_\mathrm{int}(m)}{W(m)}&=\Theta(n(m))\\
    &=\Theta\qty(\polylog\qty(\frac{W(m)D(m)}{\epsilon(m)})).
\end{align}

The depth of the intermediate circuit is given by
\begin{align}
\label{eq:D_int_methods}
    D_\mathrm{int}(m)&=O(T_\mathrm{prep}(m)+T_\mathrm{meas}(m)+\notag\\
    &\quad\qty(D(m)-2)T_\mathrm{gate}+\qty(D(m)-1)T_\mathrm{EC}(m)),
\end{align}
where the depth $D(m)$ of the original circuit includes a single depth of preparations, a parallel sequence of gates, and a single depth of measurements, so the depth of gates is $D(m)-2$, and the EC gadgets are inserted at $D(m)-1$ time steps in between.
Our MWPM decoder constructed below will have a polylog parallel runtime in terms of the code size $n(m)$ in the worst case, i.e., $T_\mathrm{dec}(m)=O(\polylog(n(m))$.

In this case, the worst-case depths of the preparation gadget, $T_\mathrm{prep}(m)$, and of the measurement gadget, $T_\mathrm{meas}(m)$, are both dominated by the decoder's runtime, i.e.,
\begin{align}
    T_\mathrm{prep}(m)&=O(1)+O\qty(T_\mathrm{dec}(m))\\
    &=O\qty(\polylog(n(m)))\\
    &=O\qty(\polylog\qty(\polylog\qty(\frac{W(m)D(m)}{\epsilon(m)}))),
\end{align}
\begin{align}
    T_\mathrm{meas}(m)&=O(1)+O\qty(T_\mathrm{dec}(m))\\
    &=O\qty(\polylog(n(m)))\\
    &=O\qty(\polylog\qty(\polylog\qty(\frac{W(m)D(m)}{\epsilon(m)}))).
\end{align}
The worst-case depth of each gate gadget, $T_\mathrm{gate}$, is given by~\eqref{eq:t_gate_methods}, and that of syndrome extraction, $T_\mathrm{SE}$, is given by~\eqref{eq:t_se_methods}, which are in a constant order $O(1)$.
Then, that  of the EC gadget, $T_\mathrm{EC}(m)$, is also dominated by the runtime of decoding, i.e.,
\begin{align}
    T_\mathrm{EC}(m)&=O(1)+O\qty(T_\mathrm{dec}(m))\\
    &=O\qty(\polylog(n(m)))\\
    &=O\qty(\polylog\qty(\polylog\qty(\frac{W(m)D(m)}{\epsilon(m)}))).
\end{align}
Therefore, due to~\eqref{eq:D_int_methods}, the time overhead of the intermediate circuit compared to the original circuit is determined by
\begin{align}
    \frac{D_\mathrm{int}(m)}{D(m)}&=O\qty(T_\mathrm{dec}(m)+\frac{T_\mathrm{prep}(m)+T_\mathrm{meas}(m)}{D(m)})\\
\label{eq:time_overhead_intermediate}
    &=O\left(\polylog\qty(\polylog\qty(\frac{W(m)D(m)}{\epsilon(m)}))\right).
\end{align}

For the feasibility of error suppression with the 3D subsystem surface code, the rate of errors that accumulate while waiting for the decoder, which takes $O\qty(T_\mathrm{dec}(m))$, is required to surpass the constant threshold of the fault-tolerant protocol with the 3D subsystem surface code, which we write as $p_{\mathrm{th},\mathrm{int}}$.
That is, the error rate $p_\mathrm{int}(m)$ of each operation of the intermediate circuit is required to be suppressed slightly to overcome the $O\qty(T_\mathrm{dec}(m))$ accumulation, i.e.,
\begin{align}
    p_\mathrm{int}(m)&=\Theta\qty(\frac{p_{\mathrm{th},\mathrm{int}}}{T_\mathrm{dec}(m)})\\
\label{eq:p_int}
    &=\Theta\qty(\frac{1}{\polylog\qty(\polylog\qty(\frac{W(m)D(m)}{\epsilon(m)}))}),
\end{align}
where the constant $p_{\mathrm{th},\mathrm{int}}$ is omitted in the second line and henceforth.
Let $p_\mathrm{ph}$ be the physical error rate; then, we will reduce this physical error rate $p_\mathrm{ph}$ to $p_\mathrm{int}(m)$ using the protocol with concatenated Steane codes.
For the concatenation level $L(m)$, the logical error rate $p_L(m)$ of the protocol with concatenated Steane codes decreases doubly exponentially in $L(m)$, i.e., $p_L(m)=\exp\qty(-\Theta\qty(2^{L(m)}))$~\cite{G}.
Thus, to achieve $p_L(m)\leq p_\mathrm{int}(m)$ for $p_\mathrm{int}(m)$ in~\eqref{eq:p_int} as required, we take the concatenation level on the order of
\begin{widetext}
\begin{align}
    L(m)
    &=\Theta\qty(\log\qty(\polylog\qty(\frac{1}{p_\mathrm{int}(m)})))\\
\label{eq:L}
    &=\Theta\qty(\log\qty(\polylog\qty(\polylog\qty(\polylog\qty(\frac{W(m)D(m)}{\epsilon(m)}))))).
\end{align}
\end{widetext}
On the other hand, the space and time overhead of the protocol with concatenated Steane codes grows exponentially in $L(m)$~\cite{G}, but since $L(m)$ in our case is as small as~\eqref{eq:L}, these overheads are given by
\begin{widetext}
\begin{align}
    \frac{W_\mathrm{FT}(m)}{W_\mathrm{int}(m)}&=\exp(\Theta(L(m)))\\
\label{eq:space_overhead_physical}
    &=\Theta\qty(\polylog\qty(\polylog\qty(\polylog\qty(\frac{W(m)D(m)}{\epsilon(m)})))),\\
    \frac{D_\mathrm{FT}(m)}{D_\mathrm{int}(m)}&=
    \exp(\Theta(L(m)))\\
\label{eq:time_overhead_physical}
    &=\Theta\qty(\polylog\qty(\polylog\qty(\polylog\qty(\frac{W(m)D(m)}{\epsilon(m)})))).
\end{align}
\end{widetext}

Consequently, due to~\eqref{eq:space_overhead_intermediate},~\eqref{eq:time_overhead_intermediate},~\eqref{eq:space_overhead_physical}, and~\eqref{eq:time_overhead_physical}, our fault-tolerant protocol achieves the doubly polylog time overhead while maintaining the polylog space overhead, i.e.,
\begin{align}
    \frac{D_\mathrm{FT}(m)}{D(m)}&=O\qty(\polylog\qty(\polylog\qty(\frac{W(m)D(m)}{\epsilon(m)}))),\\
    \frac{W_\mathrm{FT}(m)}{W(m)}&=\Theta\qty(\polylog\qty(\frac{W(m)D(m)}{\epsilon(m)})),
\end{align}
where the triply polylogarithmic factors $\Theta\qty(\polylog\qty(\polylog\qty(\polylog\qty(\frac{W(m)D(m)}{\epsilon(m)}))))$ of the space and time overheads of the concatenated-code protocol in~\eqref{eq:space_overhead_physical} and~\eqref{eq:time_overhead_physical} do not affect the leading factors of the overheads of the overall protocol.
Note that to determine the required concatenation level given by~\eqref{eq:L}, we must know the worst-case decoding time $T_\mathrm{dec}(m)$ before executing FTQC.
This is feasible because given the original circuit and target error $\epsilon(m)$, both the maximum number of vertices and the maximum of weight values in the path graphs are predetermined, allowing us to evaluate $T_\mathrm{dec}(m)$ in advance.
In practice, the number of vertices in the path graphs that appear while executing FTQC rarely reaches its maximum, so the gadgets do not necessarily have to wait for the worst-case runtime of the MWPM decoder.
Nevertheless, choosing the concatenation level as in~\eqref{eq:L} ensures the existence of the threshold, even if we account for the worst-case runtime.

Consequently, as we have argued in~\eqref{eq:doubly_polylog_time_overhead_methods}, the polylog-time MWPM decoder is a general requirement for achieving the doubly polylog time overhead by topological-code protocols with single-shot features; here, by precisely analyzing our specific protocol, we indeed see that the time overhead of our protocol is dominated by the polylog runtime $T_\mathrm{dec}(m)=O\qty(\polylog(n(m)))$ of the MWPM decoder even with accounting for all details of the protocol, such as the depth of preparation and measurement gadgets and the concatenated-code protocol.

\subsection*{Description of the polylog-time parallel algorithm for MWPM decoding}
\label{sec:polylog_MWPM}
We describe the details of our polylog-time parallel algorithm for the MWPM decoding.
Since our decoder works for both 2D and 3D surface codes with the code size $n$, the following explanation is presented in a way that applies to both.

\subsubsection*{Task of decoding}
We here provide a precise definition of the task of decoding relevant to this work; in particular, we consider the minimum-weight decoding.
Following the convention of Refs.~\cite{gottesman2014faulttolerant,tamiya2024polylogtimeconstantspaceoverheadfaulttolerantquantum}, we consider a local stochastic Pauli error model, which is a general error model including an independent Pauli error model as a special case.
To define this error model more precisely, let $C$ denote the set of all the locations of the fault-tolerant circuit, where a location refers to one of the operations in the circuit, i.e., preparations, gates, measurements, and wait operations.
A set of faulty locations is given by a random variable $F \subseteq C$ of locations of the circuit, where any Pauli errors may occur.
In particular, we say that the fault-tolerant circuit experiences a local stochastic Pauli error model with error parameters $\{p_j\}_{j\in C}$ if the following conditions hold~\cite{gottesman2014faulttolerant,tamiya2024polylogtimeconstantspaceoverheadfaulttolerantquantum}.
\begin{enumerate}
    \item Each physical location $j\in C$ has an error parameter $p_j\in(0,1]$ such that for any set $A\subseteq C$, the probability that $F$ contains $A$ satisfies
\begin{equation}
\label{eq: local stochastic error model on locations}
    \Pr[F\supseteq A]\leq \prod_{j\in A} p_j.
\end{equation}
\item The faulty operations in $F$ may suffer from errors represented by the assignment of arbitrary Pauli operators, which are applied to the state at each time step (between the locations) in the following way.
\begin{itemize}
    \item If a location in $F$ is a state preparation, any Pauli operator may be applied to the qubit after the $\ket{0}$-state preparation is performed.
    \item If a location in $F$ is a gate (or a wait, i.e., the identity gate), then after the gate operation is performed, any Pauli operators may be applied to the qubit(s) on which the gate operation has acted.
    \item If a location in $F$ is a measurement, a bit-flip (or identity) operation may be applied to the measurement outcome.
\end{itemize}
The locations in $C\backslash F$ behave in the same way as the case without faults. 
\end{enumerate}

As presented in the main text, we assume that the fault-tolerant circuit experiences the local stochastic Pauli error model at the physical level.
Then, after applying the fault-tolerant protocol of the concatenated Steane code, the intermediate circuit implemented at the logical level of the concatenated Steane code also experiences the local stochastic Pauli error model~\cite{G}.
In our protocol, measurement outcomes in the EC gadgets of the intermediate circuit for the 3D subsystem surface codes are to be used for the task of MWPM decoding.
Note that even if one imposes a stronger assumption that identical and ideally distributed (IID) errors occur in the physical circuit, the logical errors will be correlated~\cite{G}; still, the analysis of single-shot decoding for the 3D subsystem surface codes in Ref.~\cite{kubica2022single} assumes local stochastic Pauli errors, allowing for such correlations.

Given the syndrome measurement outcomes, the task of the minimum-weight decoding for $X$ (and $Z$) errors is to estimate the locations of $X$ (and $Z$, respectively) errors in the circuit that are consistent with the syndrome values and achieve the maximum among the upper bounds of their probability on the right-hand side of~\eqref{eq: local stochastic error model on locations}, so that the decoder should output a recovery operation, i.e., Pauli gates on the code block to correct the estimated Pauli errors~\cite{gottesman2014faulttolerant}.
If there is more than one maximum, the decoder may return one of them.
As formulated below, the maximum upper bound of the error probability will be equivalent to the minimum of its negative logarithm.
With the minimum-weight decoding, we have the threshold theorem for the protocols with quantum LDPC codes~\cite{gottesman2014faulttolerant,PhysRevLett.109.180502,tamiya2024polylogtimeconstantspaceoverheadfaulttolerantquantum}, which provides a theoretical guarantee on the threshold existence and overhead bounds as discussed in the main text.

\subsubsection*{Definition of detector graphs}
Here we provide a precise definition of detector graphs, following the notation in Ref.~\cite{higgott2023sparse}.
A detector is a parity of measurement outcomes that is deterministically equal to $0$ when no errors are present. 
In the single-shot decoding for the 3D subsystem surface codes, we take a single syndrome measurement outcome as a detector, as there is only one round of syndrome measurements in each syndrome extraction.
The single-shot decoding of the 3D subsystem surface codes in Ref.~\cite{kubica2022single} constructs two graphs: a qubit graph, which corresponds to a detector graph, and a measurement graph, which includes detectors and measurement outcomes of gauge operators. 
The presentation in our work focuses on a detector graph since the measurement graph can be treated analogously to the detector graph, as shown in Ref.~\cite{kubica2022single}.
In the case of the 2D surface codes, there are multiple rounds of syndrome measurements within an EC gadget.
In such a case, the parity of the measurement outcomes from two consecutive rounds corresponding to the same stabilizer generator is taken as a detector. 
In particular, let $R>1$ be the number of measurement rounds in a syndrome extraction, and consider a sequence of measurement outcomes in the EC gadget
\begin{align}
    m_{j,0},m_{j,1},\ldots,m_{j,R},
\end{align}
where $j$ and $r$ in $m_{j,r} \in \{0,1\}$ represent an index of the stabilizer generator and the syndrome measurement round, respectively.
Then, the detectors are given by
\begin{align}
    \hat{m}_{j,r-1}&\coloneqq m_{j,r-1}+m_{j,r}\mod 2
\end{align}
for $r\in\{1,\ldots,R\}$.
A detection event refers to a detector with the outcome $1$, i.e., an active detector.

A detector graph is defined as a weighted simple graph
\begin{align}
    \mathcal{G}\coloneqq(\mathcal{V},\mathcal{E})
\end{align}
with a weight denoted by $w$.
A vertex $v \in \mathcal{V}$ represents either a detector, a space-like boundary vertex, or a time-like boundary vertex.
In our protocol, for the code size $n$, we form an $O(\poly(n))$-size window, from which we obtain the detector graph of size
\begin{align}
\label{eq:V}
    \qty|\mathcal{V}|=O(\poly(n)).
\end{align}
An edge $e \in \mathcal{E}$ connects either two detectors, or one detector and one (space- or time-like) boundary vertex.
An edge connecting two detectors represents errors occurring at certain faulty locations that activate the two detectors.
An edge connecting one detector and one boundary vertex represents errors at certain faulty locations that activate the single detector.
In particular, a time-like boundary vertex is incident with an edge representing measurement errors, and a space-like boundary vertex is incident with an edge representing errors at locations that do not lead to measurement errors.
Since the degree of each vertex of the detector graph is bounded,
for the code size $n$, the number of edges of the detector graph is
\begin{align}
\label{eq:E}
    \qty|\mathcal{E}|=O\qty(\qty|\mathcal{V}|).
\end{align}
Note that there may be a set of multiple locations such that errors occurring at any of these locations activate the same set of detectors, which are represented by the same edge.

As for the edge weights, conventionally, the weight of an edge between two detectors is defined by the negative logarithm of an upper bound of the probability of errors activating the two detectors, while the weight of an edge between a detector and a boundary vertex represents that activating the single detector.
That is, the conventional definition of each edge weight $w(e)$ is
\begin{align}
    \label{edge_weights}
    w(e)\coloneqq -\log\qty(\sum_{j}p_j),
\end{align}
where $p_j$ is the error parameter at each location $j$ for the error-probability upper bounds on the right-hand side of~\eqref{eq: local stochastic error model on locations}, and the sum is taken over the set $\{j\}$ of locations such that errors occurring at the location $j$ activate the set of detectors incident with the edge $e$.
Note that under an IID error model, the probability of an edge being triggered would be calculated by taking into account all odd numbers of detection events that activate the same set of detectors; however, under the local stochastic Pauli error model, we cannot calculate in this way due to the correlations of errors.
To address this, we provide an upper bound of this probability in terms of the error parameters in~\eqref{edge_weights}, applying the union bound to account for the correlated errors.
In contrast to the conventional definition of the edge weights in~\eqref{edge_weights}, for the feasibility of our decoding algorithm, we here define the edge weight of the detector graph as an integer upper bounding this real-valued weight, up to an appropriate rescaling.
In particular, the integer weight is obtained by multiplying the weight in~\eqref{edge_weights} by a chosen constant $C$ and then rounding it up if necessary, i.e., 
\begin{align}
    \label{edge_weights_integer}
    w(e) = \left\lceil -C\log\qty(\sum_{j}p_j) \right\rceil.
\end{align}

\subsubsection*{Definition of path graphs}

\begin{figure}
\includegraphics[width=0.6\linewidth]{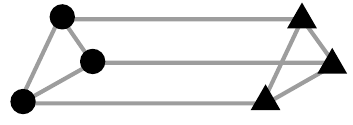}
\caption{\label{fig:comp_graph2}An example of a path graph when the number of detection events is $3$. The circles correspond to detection events and the triangles correspond to boundary vertices.}
\end{figure}

Given a detector graph $\cG$ and a set of detection events, we define a path graph, which is a graph to find an MWPM therein.
The path graph is denoted by
\begin{equation}
    \overline{\cG}=\qty(\cD,\overline{\cE}),
\end{equation}
which is a simple graph with a weight assigned to each edge.
Note that the size of the path graph $\overline{\cG}$ may vary probabilistically depending on the detection events, while its worst-case size will be deterministically upper bounded by that of the detector graph.
In Fig.~\ref{fig:comp_graph2}, we show a path graph for the case where the number of detection events is $3$.
As mentioned in the main text, the path graph in our definition is not necessarily a complete graph, unlike the presentation in Ref.~\cite{higgott2023sparse}.

To define $\overline{\cG}$, let $\mathcal{A}\subset\mathcal{V}$ be the set of vertices representing detection events in $\cG$, which we call active-detector vertices.
By definition, we have
\begin{align}
\label{eq:A}
    \qty|\mathcal{A}|<\qty|\mathcal{V}|.
\end{align}
Note that $\qty|\mathcal{A}|\neq\qty|\mathcal{V}|$ always holds, since $\mathcal{V}$ includes not only detectors but also boundary vertices, whereas $\mathcal{A}$ is a subset of detectors.
For each $v \in \mathcal{A}$, we pick the nearest boundary vertex in $\cG$ from the detector represented by $v$, and form a (multi)set $\mathcal{A}'$ of these corresponding boundary vertices, where $|\mathcal{A}'|=|\mathcal{A}|$.
This way of incorporating boundary vertices in the graph follows the conventional construction in Refs.~\cite{10.5555/2011362.2011368,PhysRevA.89.022326}.
Then, the set of vertices $\overline{\mathcal{V}}$ in $\overline{\mathcal{G}}$ is given by a union 
\begin{align}
\overline{\mathcal{V}}\coloneqq\mathcal{A}\cup\mathcal{A}'.  
\end{align}
Due to~\eqref{eq:A}, the size of the path graph is bounded by that of the detector graph as
\begin{align}
\label{eq:V_bound}
    \qty|\overline{\mathcal{V}}|=2\qty|\mathcal{A}|< 2\qty|\mathcal{V}|.
\end{align}

As for the edges of the path graph, let $\mathcal{E}_\mathcal{A}$ be the set of weighted edges between any pair of active-detector vertices in $\mathcal{A}$, and $\mathcal{E}_\mathcal{A'}$ be the set of weight-$0$ edges between any pair of boundary vertices in $\mathcal{A}'$, where
\begin{align}
\label{eq:E_A}
    |\mathcal{E}_{\mathcal{A}}|=|\mathcal{E}_{\mathcal{A}}'|=\frac{|\mathcal{A}|(|\mathcal{A}|-1)}{2}.
\end{align}
Additionally, for each detection event in $\mathcal{A}$ and the corresponding boundary vertex in $\mathcal{A}'$, we introduce a weighted edge connecting them, to form the set $\mathcal{E}_{\mathcal{A}\mathcal{A}'}$ of these weighted edges, where
\begin{align}
\label{eq:E_AA}
    \qty|\mathcal{E}_{\mathcal{A}\mathcal{A}'}|=|\mathcal{A}|=|\mathcal{A}'|. 
\end{align}
We then give the set $\overline{\mathcal{E}}$ as a union
\begin{align}
\overline{\mathcal{E}}=\mathcal{E}_\mathcal{A} \cup \mathcal{E}_{\mathcal{A}\mathcal{A'}} \cup \mathcal{E}_\mathcal{A'}.
\end{align}
Due to~\eqref{eq:A},~\eqref{eq:E_A}, and~\eqref{eq:E_AA}, the number of edges in the path graph is given by
\begin{align}
\label{eq:num_E}
\qty|\overline{\mathcal{E}}|=|\mathcal{E}_{\mathcal{A}}|+|\mathcal{E}_{\mathcal{A}\mathcal{A'}}|+|\mathcal{E}_{\mathcal{A}'}|=|\mathcal{A}|^2<|\mathcal{V}|^2.
\end{align}

The weight of an edge $\{u,v\}\in\mathcal{E}_\mathcal{A} \cup \mathcal{E}_{\mathcal{A}\mathcal{A'}}$ is defined by the distance $\delta (u,v)$ between the pair of vertices on the detector graph $\cG$ represented by $u$ and $v$, i.e, the sum of the weights along the shortest path between them on $\cG$:
\begin{align}
    \label{edge_weights_scale}
    w(\{u,v\})=\delta(u,v)=O\qty(\qty|\mathcal{V}|),
\end{align}
where we note that the distance $\delta (u,v)$ between any pair of vertices in a graph scales at most the number of vertices of the graph.

\subsubsection*{A polynomial-size lookup table for constructing path graphs}
We discuss how to construct path graphs in our protocol.
As presented in Ref.~\cite{higgott2022pymatching}, a conventional way of constructing a path graph $\overline{\cG}$ from a detector graph $\mathcal{G}$ and a set of detection events is to execute Dijkstra's algorithm~\cite{dijkstra1959note} to identify the edge weights~\eqref{edge_weights_scale} of $\overline{\cG}$ by finding the shortest paths in $\mathcal{G}$ between any two detection events and from each detection event to a boundary node after we obtain the detection events. 
In a graph that contains $V$ nodes and $E$ edges, the worst-case complexity of Dijkstra's algorithm is $O(V\log V+E)$~\cite{siek2001boost}.
Thus, the complexity of constructing a path graph from $\mathcal{G}=(\mathcal{V},\mathcal{E})$ and a set of detection events of size $|\mathcal{A}|$ is
\begin{align}
    O(|\mathcal{A}|(|\mathcal{V}|\log |\mathcal{V}|+|\mathcal{E}|))&=O(|\mathcal{V}|^2\log |\mathcal{V}|)\\
    &=O\qty(\poly(n)),
\end{align}
where we use~\eqref{eq:V},~\eqref{eq:E}, and~\eqref{eq:A}.
This is a polynomial time, yet it grows faster than the desired polylog time. 
Therefore, if we execute this process during executing FTQC, we would not be able to achieve the MWPM decoding within polylog time in terms of the code size $n$ with a worst-case guarantee.

To address this issue, we here prepare a polynomial-size lookup table in advance and construct the path graph within polylog time in $n$ at the execution time of FTQC assisted by the lookup table.
In this lookup table, we store the weights~\eqref{edge_weights_scale} for all possible pairs of detectors in the detector graph and for all possible pairs of a detector and its nearest boundary vertex.
This can be constructed as a preprocessing, i.e., prior to executing FTQC\@, by exhaustively executing Dijkstra's algorithm only once at the time of the protocol design.
To analyze the size of the lookup table, we see that the total number of the shortest paths stored in the lookup table is dominated by the number of all possible pairs of detectors in the detector graph, alongside the number of detectors themselves to account for their corresponding boundary vertices; hence, the space complexity for classical computers to store all the weights in the lookup table is
\begin{align}
\label{eq:table_size}
     O(|\mathcal{V}|^2)=O(\poly(n)).
\end{align}
The classical runtime to compute all the polynomial number of distances $\delta$ to construct the lookup table also scales polynomially due to the above runtime argument on Dijkstra's algorithm, i.e.,\footnote{One could also use more efficient shortest-path algorithms than conventional Dijkstra's algorithms, e.g., the one in Ref.~\cite{10353179}. Investigation of the advantages of using such alternative algorithms is left for future work.}
\begin{align}
\label{eq:table_runtime}
     O(|\mathcal{V}|^2\log\qty|\mathcal{V}|)=O(\poly(n)).
\end{align}
Note that as discussed in~\eqref{eq:n_m_methods}, the fault-tolerant protocols require only a polylog code size in terms of the size of the original circuit, and hence, the required runtime for preparing the lookup table is exponentially shorter than the polynomial time of quantum computation represented by the original circuit.
Then, upon obtaining detection events, our decoder constructs a path graph in constant parallel time by reading this lookup table.

\subsubsection*{A parallel algorithm for finding an MWPM in a path graph} 

Given a path graph $\overline{\cG}=\qty(\cD,\overline{\cE})$ with edge weights $w: \overline{\cE}\to\mathbb{Z}$, we describe an algorithm for finding an MWPM in $\overline{\cG}$ within polylog parallel time in $\qty|\cD|$, i.e., in the code size $n$ due to~\eqref{eq:V} and~\eqref{eq:V_bound}.
A perfect matching $M\subseteq\overline{\cE}$ refers to a subset of edges such that every vertex $v\in\cD$ is incident to exactly one edge in $M$.
An MWPM $M^\ast$ is a perfect matching that has the minimum weight $\sum_{e\in M^\ast}w(e)$ among all perfect matchings.
To achieve the parallelization in finding an MWPM, we use a linear algebraic approach based on Refs.~\cite{8104102,vazirani1987matching}, rather than Edmonds' blossom algorithm. 
This approach uses a Tutte matrix $A$ of $\overline{\cG}$, which is a $\qty|\cD|\times \qty|\cD|$ matrix with its $(i,j)$ element $A_{i,j}$ given by
\begin{equation}
\label{eq:tutte}
    A_{i,j}=\begin{cases}
    x_{i,j}&i<j,\,\{i,j\}\in \overline{\cE},\\
    -x_{j,i}&i>j,\,\{i,j\}\in \overline{\cE},\\
    0&\text{otherwise},
    \end{cases}
\end{equation}
where $x_{i,j}$ for each $i,j$ is a (formal) variable, and, here and henceforth, the labels of the vertices of the path graph are relabelled from~\eqref{edge_weights_scale} so that they start at $1$.
The matrix $A$ has the property that the graph $\overline{\cG}$ has a perfect matching if and only if $\det(A)\neq 0$, where the left-hand side is a polynomial of the variables $\{x_{i,j}\}$, although this property itself is not directly useful here because the path graph $\overline{\cG}$ always has a perfect matching in our construction.

In the approach of Refs.~\cite{8104102,vazirani1987matching}, a sufficient condition for the algorithm to successfully find an MWPM $M^\ast$ is that the MWPM $M^\ast$ is isolated; in other words, $M^\ast$ is a unique MWPM\@.
To achieve the isolation, it suffices to perturb the weights slightly so that one of the MWPMs in $\overline{\cG}$ becomes the unique MWPM after the weight perturbation.
For this purpose,
we use a weight-perturbation function $W:\overline{\cE}\to\mathbb{N}$ satisfying
\begin{align}
\label{eq:W}
    1\leq W(e)\leq W_{\max}\coloneqq\max_{e \in \overline{\cE}}\{W(e)\}, 
\end{align}
which represents a perturbation to each edge $e\in \overline{\cE}$ for achieving the isolation.
We then define a modified weight $\widetilde{w}:\overline{\cE} \to\mathbb{Z}$ as
\begin{align}
\label{eq:tilde_w}
\widetilde{w}(e)&\coloneqq \tilde{C} w(e)+W(e)\\
\label{eq:tilde_w_bound}
&=O\qty(\poly\qty(\qty|\mathcal{V}|,W_{\max})),
\end{align}
where we take the factor $\tilde{C}$ as\footnote{Originally, Ref.~\cite{vazirani1987matching} proposed to use $W_{\max}\qty|\cD|/2$ as $\tilde{C}$, but in~\eqref{eq:tilde_C}, we improve this up to subleading terms, while still guaranteeing the isolation.}
\begin{align}
\tilde{C}&\coloneqq\frac{\qty|\cD|}{2} \qty(W_{\max}-1)+1,
\label{eq:tilde_C}
\end{align}
and the bound~\eqref{eq:tilde_w_bound} follows from~\eqref{eq:V_bound} and~\eqref{edge_weights_scale}.
The modified weight in~\eqref{eq:tilde_w} ensures that a perfect matching that was not originally one of the MWPMs in $\overline{\mathcal{G}}$ with the weight $w$ never becomes an MWPM with $\tilde{w}$ after the perturbation.
To see that this is ensured, let $M'$ be any perfect matching other than the MWPMs in $\overline{\mathcal{G}}$\@.
The weights of $M^\ast$ and $M'$ after the perturbation are, respectively:
\begin{align}
    &\sum_{e\in M^\ast}\tilde{w}(e)\leq\tilde{C}\sum_{e\in M^\ast}w(e)+\frac{\qty|\cD|}{2}W_{\max},\\
    &\sum_{e\in M'}\tilde{w}(e)\geq\tilde{C}\sum_{e\in M'}w(e)+\frac{\qty|\cD|}{2},
\end{align}
where $|M^\ast|=|M'|=\qty|\cD|/2$.
Then, the difference in weight between $M'$ and $M^\ast$ after the perturbation is
\begin{align}
&\sum_{e\in M'}\tilde{w}(e)-\sum_{e\in M^\ast}\tilde{w}(e)\nonumber\\
&\geq\tilde{C}\qty( \sum_{e\in M'}w(e)-\sum_{e\in M^\ast}w(e))-\frac{\qty|\cD|}{2}(W_{\max}-1)\\
&\geq\tilde{C}-\frac{\qty|\cD|}{2}(W_{\max}-1)\\
&\geq 1,
\end{align}
which shows that any perfect matching $M'$ that was not originally an MWPM does not become an MWPM after the perturbation.

The perturbation function $W$ should be chosen in such a way that an MWPM becomes isolated, that is, the problem of minimizing
\begin{equation}
    \sum_{e\in M}\widetilde{w}(e)
\end{equation}
for all perfect matchings $M$ in $\overline{\cG}$ must have a unique solution
\begin{equation}
\label{eq:M_ast}
    M^\ast\subseteq \overline{\cE}.
\end{equation}
We will provide such a choice of $W$ later, after describing the algorithm to find an MWPM assuming that the MWPM is isolated by the perturbation.

Under this assumption of the isolated MWPM, using the modified weight $\widetilde{w}$ in~\eqref{eq:tilde_w}, for each edge $\{i,j\}\in \overline{\cE}$, we assign
\begin{align}
x_{i,j}=2^{\widetilde{w}\qty(\{i,j\})},
\end{align}
where $x_{i,j}$ is a variable of the Tutte matrix in~\eqref{eq:tutte}.
With this assignment, from $A$, we obtain a matrix $B$ defined as
\begin{align}
    B_{i,j}=\begin{cases}
    2^{\widetilde{w}\qty(\{i,j\})}&i<j,\,\{i,j\}\in \overline{\cE},\\
    -2^{\widetilde{w}\qty(\{i,j\})}&i>j,\,\{i,j\}\in \overline{\cE},\\
    0&\text{otherwise},
    \end{cases}
\end{align}
As shown in Lemma~2 of Ref.~\cite{vazirani1987matching}, 
if $M^\ast$ in~\eqref{eq:M_ast} is the isolated MWPM in $\overline{\cG}$, then it holds that
\begin{equation}
\label{eq:isolation_condition}
    \det(B)\neq 0.
\end{equation}
In this case, we obtain the weight of the MWPM by computing\footnote{We can efficiently compute $w^\ast$ in~\eqref{eq:w_ast} by looking at the lower digits of the binary representation of $\det(B)$.}
\begin{equation}
\label{eq:w_ast}
    w^\ast=\max\qty{w\in\mathbb{Z}:\frac{\det(B)}{2^{2w}}\in\mathbb{N}}.
\end{equation}
Furthermore, for each edge $\{i,j\}\in \overline{\cE}$, let $B_{\mathrm{sub}}^{(i,j)}$ denote a $\qty(\qty|\cD|-1)\times\qty(\qty|\cD|-1)$ submatrix of $B$ obtained by removing the $i$th row and the $j$th column from $B$.
As shown in Lemma~3 of Ref.~\cite{vazirani1987matching}, if it holds for the the minor $\det(B_{\mathrm{sub}}^{(i,j)})$ that\footnote{We can efficiently check the condition~\eqref{eq:B_minor} by comparing the binary representations of $\det(B_{\mathrm{sub}}^{(i,j)})$ and $2^{\widetilde{w}(\{i,j\})}/2^{2w^\ast}$.}
\begin{equation}
\label{eq:B_minor}
    \frac{2^{\widetilde{w}(\{i,j\})}\det(B_{\mathrm{sub}}^{(i,j)})}{2^{2w^\ast}}~\text{is odd},
\end{equation}
then the edge $\{i,j\}$ is in the MWPM, i.e., $\{i,j\}\in M^\ast$; otherwise, $\{i,j\}\notin M^\ast$.

For a path graph of size $\qty|\overline{\mathcal{V}}|$ and its weight perturbation with $W_{\max}$, we bound the runtime of this algorithm for finding the isolated MWPM as follows.
The determinant of $O\qty(\qty|\cD|)\times O\qty(\qty|\cD|)$ matrices $B$ in~\eqref{eq:w_ast} and $B_\mathrm{sub}^{(i,j)}$ in~\eqref{eq:B_minor} can be computed in parallel within $O\qty(\log^2\qty(\qty|\cD|))$ runtime using $O\qty(\poly\qty(\qty|\cD|))$ classical processors; in particular, several different algorithms achieving such parallelization are known, such as the Samuelson-Berkowitz algorithm~\cite{BERKOWITZ1984147}, the Faddeev-LeVerrier algorithm~\cite{doi:10.1137/0205040,PREPARATA1978148}, and combinatorial algorithms~\cite{10.5555/866057}.\footnote{Gaussian elimination is widely used as a method for computing the determinant, but it remains unknown whether the Gaussian elimination can be parallelized to achieve polylog runtime in general.}
Among these algorithms, the Samuelson-Berkowitz algorithm~\cite{BERKOWITZ1984147} is numerically stable since it does not use division at all, using $O\qty(\qty|\cD|^4)$ processors with a reasonably small constant factor~\cite{johansson2020fast}.
In Supplementary Materials~\ref{sec:determinant}, we summarize the Samuelson-Berkowitz algorithm for convenience.
Apart from this, an improved version of the Faddeev-LeVerrier algorithm constructed in Ref.~\cite{PREPARATA1978148} uses $O\qty(\qty|\cD|^{3.5})$ processors while the Faddeev-LeVerrier algorithm includes a division and thus may be numerically unstable in the presence of rounding error in representing real numbers~\cite{johansson2020fast}.
The combinatorial algorithms in Ref.~\cite{10.5555/866057} require $O\qty(\qty|\cD|^6)$ processors.
In these determinant computations, addition and multiplication of two $N$-bit integers, which are in the order of $O(2^N)$, can be performed within $O(\log(N))$ parallel runtime, using $O(\poly(N))$ parallel processors~\cite{patterson2020computer,10.1007/978-3-540-45209-6_127}; in our case, we have $N=O\qty(\poly\qty(\qty|\overline{\mathcal{V}}|,W_{\max}))$ as shown in~\eqref{eq:tilde_w_bound}.
As a whole, given a path graph of size $\qty|\overline{\mathcal{V}}|$ and its weight perturbation with $W_{\max}$, the runtime of the above algorithm for finding the isolated MWPM is
\begin{align}
\label{eq:decoder_runtime_bound}
    O\qty(\polylog\qty(\qty|\overline{\mathcal{V}}|,W_{\max})),
\end{align}
with the required number of parallel processors bounded by
\begin{align}
\label{eq:decoder_processors_bound}
O\qty(\poly\qty(\qty|\overline{\mathcal{V}}|,W_{\max})).
\end{align}

\subsubsection*{Choice of weight-perturbation functions}
\label{sec:weight}

We now discuss the ways to choose the weight-perturbation function $W:E\to\mathbb{N}$ for the modified weight $\widetilde{w}$ in~\eqref{eq:tilde_w} to achieve the isolation for the parallel algorithm to successfully find the MWPM in a given graph.

We first show that a straightforward application of randomized algorithms for finding the MWPM is insufficient for achieving doubly-polylog-time-overhead FTQC\@.
As shown in Ref.~\cite{vazirani1987matching},
one way to achieve isolation is to randomly sample
\begin{equation}
    W(e)\in\{1,2,\ldots,W_{\max}\}
\end{equation}
for each edge $e\in \overline{\cE}$ from a uniform distribution over $\{1,2,\ldots,W_{\max}\}$ for some fixed $W_{\max}>0$.
Lemma~1 of Ref.~\cite{vazirani1987matching} analyzes the case of $W_{\max}=2\qty|\overline{\cE}|$ and shows that
\begin{align}
    \Pr\qty[\text{Isolation fails to hold}]\leq\frac{1}{2}.
\end{align}
Applying the same analysis to a general choice of $W_{\max}\in\{1,2,3,\ldots\}$, we obtain
\begin{align}
\label{eq:isolation_inequality}
    \Pr\qty[\text{Isolation fails to hold}]\leq\frac{\qty|\overline{\cE}|}{W_{\max}}.
\end{align}
If isolation fails to hold, the decoder may not function correctly, potentially increasing the logical error rate of the fault-tolerant protocol.
To ensure that the logical error rate is bounded by $\lessapprox \frac{\epsilon(m)}{W(m)D(m)}$ as required in~\eqref{eq:n_m_methods}, due to~\eqref{eq:V} and~\eqref{eq:V_bound}, at least we need to take
\begin{equation}
\label{eq:W_max_large}
    W_{\max}=\widetilde{\Theta}\qty(\frac{W(m)D(m)}{\epsilon(m)}),
\end{equation}
where $\widetilde{\Theta}$ may omit a polylog factor.
However, under this choice, the MWPM decoder's runtime in~\eqref{eq:decoder_runtime_bound} would become $O\qty(\polylog\qty(\frac{W(m)D(m)}{\epsilon(m)}))$; comparing this with the requirement in~\eqref{eq:decoder_runtime_methods}, we see that the choice of $W_{\max}$ in~\eqref{eq:W_max_large} renders this straightforward randomized approach insufficient for our purpose.

One alternative way to avoid the runtime $O\qty(\polylog\qty(\frac{W(m)D(m)}{\epsilon(m)}))$ in the randomized approach is to use a smaller $W_{\max}$ than~\eqref{eq:W_max_large} while trying multiple sets of random perturbations simultaneously in parallel and selecting the one that successfully finds an MWPM\@.
Although reducing $W_{\max}$ increases the probability of isolation failure, this can be compensated for by parallelizing many independent attempts.
Suppose that for each $m$, the number of edges in the path graph $\qty|\overline{\cE}|$, as in~\eqref{eq:n_m_methods},~\eqref{eq:V},~\eqref{eq:V_bound}, and~\eqref{eq:num_E}, scales as
\begin{equation}
\label{eq:V_large_parallel}
    \qty|\overline{\cE}|=O\qty(\log^a\qty(\frac{W(m)D(m)}{\epsilon(m)}))
\end{equation}
for some constant $a \in \mathbb{R}$. If we take 
\begin{equation}
\label{eq:W_max_large_parallel}
    W_{\max}=\Theta\qty(\log^b\qty(\frac{W(m)D(m)}{\epsilon(m)}))
\end{equation}
for some constant $b \in \mathbb{R}$, then the runtime of finding an isolated MWPM in~\eqref{eq:decoder_runtime_bound} becomes
\begin{align}
\label{eq:runtime_large_parallel}
   &O\qty(\polylog\qty(\qty|\cD|,W_{\max}))\nonumber\\
   &=O\qty(\polylog\qty(\qty|\overline{\cE}|,W_{\max}))\nonumber\\
   &=O\qty(\polylog\qty(\polylog\qty(\frac{W(m)D(m)}{\epsilon(m)}))), 
\end{align}
as desired.
At the same time, the right-hand side of \eqref{eq:isolation_inequality} becomes
\begin{equation}
\label{eq:VW_max_large_parallel}
    \frac{\qty|\overline{\cE}|}{W_{\max}}=O\qty(\frac{1}{\log^{b-a}\qty(\frac{W(m)D(m)}{\epsilon(m)})}),
\end{equation}
which is larger than the required logical error rate $\lessapprox \frac{\epsilon(m)}{W(m)D(m)}$ in the asymptotic regime and thus does not directly meet our requirements.
However, for $l\in \mathbb{N}$, by attempting $l$ random perturbations simultaneously via parallel repetition, the upper bound of the  probability that all these attempts fail to isolate decreases exponentially as
\begin{equation}
\label{eq:simultaneous_perturb}
    \frac{\qty|\overline{\cE}|}{W_{\max}}=O\qty(\frac{1}{\log^{l(b-a)}\qty(\frac{W(m)D(m)}{\epsilon(m)})}).
\end{equation}
By choosing $l$ in~\eqref{eq:simultaneous_perturb} as
\begin{equation}
\label{eq:VW_max_large_parallel_k}
    l=\Theta\qty(\frac{\log\qty(\frac{W(m)D(m)}{\epsilon(m)})}{\log\qty(\log\qty(\frac{W(m)D(m)}{\epsilon(m)}))}),
\end{equation}
we can ensure that the upper bound~\eqref{eq:simultaneous_perturb} of the probability of all attempts falling to isolate is suppressed below $\lessapprox \frac{\epsilon(m)}{W(m)D(m)}$ as desired.

The required number of classical processors for this parallel randomized perturbation is given by~\eqref{eq:decoder_processors_bound} multiplied by $l$ and hence is only polylog in the size of the original circuit, i.e.,
\begin{align}
O\qty(\poly\qty(\qty|\overline{\mathcal{V}}|,W_{\max}))\cdot l=O\qty(\polylog\qty(\frac{W(m)D(m)}{\epsilon(m)})),
\end{align}
due to~\eqref{eq:V_large_parallel},~\eqref{eq:W_max_large_parallel}, and~\eqref{eq:VW_max_large_parallel_k}.
To find the MWPM, our algorithm checks whether each $w^\ast$ in~\eqref{eq:w_ast} is equal to the weight of $M^\ast$, where $M^\ast$ is obtained by calculating~\eqref{eq:B_minor} for all edges $\{i,j\}$ in parallel. 
Note that this condition directly confirms whether the algorithm's output is an MWPM, rather than checking if the MWPM is isolated. Indeed, we do not have to check whether the MWPM is isolated as long as the algorithm outputs an MWPM that satisfies our checking condition.
Trying all the randomly chosen weight-perturbation functions simultaneously using the classical processors in parallel, we find an MWPM in the original path graph $\overline{\cG}$ within polylog time in $\qty|\cD|$, as shown in~\eqref{eq:runtime_large_parallel}.

Although this improved randomized approach may allow us to bound the logical error rate below $\lessapprox \frac{\epsilon(m)}{W(m)D(m)}$ in many cases, a deterministic (derandomized) approach is even more preferable, as it guarantees that the logical error rate is not affected at all.
This could save space overhead, since achieving the target logical error rate no longer requires a further increase in code distance to compensate for the higher logical error rate introduced by isolation failures.
To achieve a deterministic approach, we use a more recent technique in Ref.~\cite{8104102}. This protocol chooses $W$ from a carefully designed set of functions, whose size is at most quasi-polynomial in $\qty|\cD|$, ensuring that at least one of them will successfully isolate an MWPM\@.
In the same way as the parallel randomized method above, by trying all these deterministically chosen weight-perturbation functions in parallel, we can deterministically find an MWPM\@.

To explicitly obtain the quasi-polynomial-size set of weight-perturbation functions for isolation, following Ref.~\cite{8104102}, we define a function $w_k:\overline{\cE}\to\mathbb{Z}$ of each edge $e_j\in \overline{\cE}=\{e_1,\ldots,e_{\qty|\overline{\cE}|}\}$ as
\begin{align}
    w_k\qty(e_j)\coloneqq{\qty(4\qty|\overline{\mathcal{V}}|^2+1)}^j\bmod k,
\end{align}
which satisfies $w_k\qty(e_j)<k$.
For functions $w,w^\prime:\overline{\cE}\to\mathbb{Z}$ and $t>\max_{e_j\in \overline{\cE}}\qty{w^\prime(e_j)}$,
we let $w\circ w^\prime:\overline{\cE}\to\mathbb{Z}$ denote
\begin{equation}
    \qty(w\circ w^\prime)(e_j)\coloneqq \qty|\overline{\mathcal{V}}|t\cdot w(e_j)+w^\prime(e_j).
\end{equation}
For $t\in\{7,8,\ldots\}$, we define a set of functions
\begin{align}
    \mathcal{W}_t\coloneqq\qty{w_k:k=2,\ldots,t},
\end{align}
with size $|\mathcal{W}_t|=t-1$.
Using functions $w_k\in\mathcal{W}_t$ satisfying $\max_{e_j\in \overline{\cE}}\{w_k(e_j)\}<t$, we define a set of functions
\begin{align}
\label{eq:W_t_s}
    \mathcal{W}_t^{s}\coloneqq\qty{\qty(\qty(w_{1}\circ w_{2})\circ \cdots)\circ w_{s}:w_{1},\ldots,w_{s}\in\mathcal{W}_t},
\end{align}
with size
\begin{align}
    |\mathcal{W}_t^s|=|\mathcal{W}_t|^s\leq t^s.
\end{align}
We use each function in this set $\mathcal{W}_t^{s}$ as $W$ for each weight perturbation in~\eqref{eq:tilde_w}, where any function $W\in\mathcal{W}_t^{s}$ satisfies
\begin{align}
\label{eq:W_max_bound}
    0\leq W(e_j)\leq W_{\max}\coloneqq \qty(\qty|\overline{\mathcal{V}}|t)^s;
\end{align}
in this case, the number of weight-perturbed graphs is bounded by
\begin{align}
\label{eq:W_bound}
    |\mathcal{W}_t^s|\leq W_{\max}.
\end{align}

Therefore, to achieve the isolation deterministically, we attempt, in parallel, all the functions
\begin{equation}
    W\in\mathcal{W}_t^{s}
\end{equation}
in the set $\mathcal{W}_t^{s}$ of~\eqref{eq:W_t_s} for an appropriate choice of $s$ and $t$, so that at least one of the modified weights in~\eqref{eq:tilde_w} should achieve the isolation with a theoretical guarantee.
One possible choice of such $s$ and $t$ is given in Ref.~\cite{8104102} by
\begin{align}
\label{eq:alpha}
    s&=\left\lceil\log_2\qty(\qty|\cD|)+1\right\rceil\cdot\left\lceil\log_2\qty(\qty|\cD|)\right\rceil,\\
    t&=\qty|\cD|^{20},
\end{align}
where $\lceil{}\cdots{}\rceil$ is the ceiling function; in this case, due to~\eqref{eq:W_bound}, the required number of weight-perturbed graphs is upper bounded by
\begin{align}
\label{eq:parallelization_upper_bound}
    W_{\max}=\qty|\cD|^{21\left\lceil\log_2\qty(\qty|\cD|)+1\right\rceil\cdot\left\lceil\log_2\qty(\qty|\cD|)\right\rceil}.
\end{align}
Remarkably, this is substantially smaller than~\eqref{eq:W_max_large} in the case of randomized algorithms.

We analyze the overall computational resources required for finding the MWPM in this deterministic algorithm, in the worst case with a theoretical guarantee.
Since all the weight-perturbed path graphs can be tried in parallel, due to~\eqref{eq:n_m_methods},~\eqref{eq:V},~\eqref{eq:V_bound},~\eqref{eq:decoder_runtime_bound} and~\eqref{eq:parallelization_upper_bound}, the parallel runtime is upper bounded by 
\begin{align}
&O\qty(\polylog\qty(\qty|\overline{\mathcal{V}}|,W_{\max}))\nonumber\\
&=O\qty(\polylog\qty(\qty|\overline{\mathcal{V}}|))\\
&=O\qty(\polylog\qty(\qty|\mathcal{V}|))\\
&=O\qty(\polylog\qty(n))\\
&=O\qty(\polylog\qty(\polylog\qty(\frac{W(m)D(m)}{\epsilon(m)}))),
\end{align}
achieving the requirement for our fault-tolerant protocol.
Similarly, with~\eqref{eq:decoder_processors_bound},
the required number of parallel processors is bounded by
\begin{align}
\label{eq:number_processor}
&O\qty(\poly\qty(\qty|\overline{\mathcal{V}}|,W_{\max}))\nonumber\\
&=O\qty(\text{quasi-poly}\qty(\qty|\overline{\mathcal{V}}|))\\
&=O\qty(\text{quasi-poly}\qty(\qty|\mathcal{V}|))\\
&=O\qty(\text{quasi-poly}\qty(n))\\
&=O\qty(\text{quasi-polylog}\qty(\frac{W(m)D(m)}{\epsilon(m)})).
\end{align}
Notably, whereas the quasi-polynomial number of processors in $\qty|\cD|$ may be required to attain the theoretical guarantee, $\qty|\cD|$ is indeed only polylog in the size of the original circuit; thus, only a quasi-polylog number of classical parallel processors per code block are needed for our protocol, which is still substantially smaller compared to the original circuit itself.

Finally, we remark that~\eqref{eq:parallelization_upper_bound} is merely an analytical upper bound to provide a theoretical guarantee, which is potentially largely overestimated since the analysis in Ref.~\cite{8104102} does not aim to optimize it; as discussed later in Methods, we indeed see through a numerical simulation that the optimal number of weight-perturbed graphs can be much smaller than~\eqref{eq:parallelization_upper_bound}, scaling sublinearly as $o\qty(\qty|\cD|)$ in a regime relevant in practice.
For such an optimization, we propose designing the set of weight-perturbation functions using pseudo-random functions, such as a Mersenne twister~\cite{10.1145/272991.272995}, which remains heuristic in the sense that we will demonstrate the achievability of deterministic isolation only numerically, unlike the above method from Ref.~\cite{8104102}, which allows for a theoretical guarantee.
In particular, for each edge $e_j\in \overline{\cE}$, we choose a seed $s_{\qty|\cD|,e_j,k}$, where $k$ is the index of the sets of modified weights for each edge $e_j$ of a graph with $\qty|\cD|$ vertices.
We then employ a pseudo-random function $f\qty(x,s_{\qty|\cD|,e_j,k})$ with this seed $s_{\qty|\cD|,e_j,k}$ to define
\begin{equation}
\label{eq:pseudo_random_perturbation}
    W(e_j)=f\qty(j,s_{\qty|\cD|,e_j,k}).
\end{equation}
The proposal here is to fix several choices of seeds for each edge in advance (in some heuristic way, e.g., just randomly) so that the isolation can be achieved practically without failure.
Further details of this optimization will be presented later in Methods, along with specifics of our numerical simulation.

\subsubsection*{Polynomial-size lookup tables for obtaining recovery operations}
Once an MWPM in a path graph is obtained, the decoder in our protocol has to output the corresponding recovery operation.
In particular, the edges in the MWPM are mapped into their counterparts in the detector graph, which are then further mapped into specific error locations and ultimately into a recovery operation given by Pauli gates on the code block.
During the execution of FTQC, we need to perform all these steps without slowing down the polylog-time MWPM decoding.

To achieve this, we propose to prepare a polynomial-size lookup table for this mapping in advance and use it at the execution time to return the mapping within constant time, similar to the procedure for constructing a path graph using a lookup table as described above.
Identifying the relevant edges in the detector graph relies on knowledge of the shortest paths between any two detectors and between any detector and its nearest boundary vertex.
This information is acquired through the aforementioned exhaustive Dijkstra search, and the runtime for this is $O(\poly(n))$ as in~\eqref{eq:table_runtime}.
The errors associated with the obtained edges in the detector graph are propagated in a stabilizer circuit in the window to derive the corresponding recovery operation. 
Storing the recovery operations corresponding to each detector graph edge in a lookup table requires memory proportional to the number of edges in the detector graph, i.e., $O(|\mathcal{E}|)=O(\poly(n))$ due to~\eqref{eq:V} and~\eqref{eq:E}.
The computation of propagating the Pauli errors is performed within polynomial time due to the Gottesman-Knill theorem~\cite{PhysRevA.70.052328}, leading to an $O(\poly(n))$ overall runtime for preparing the lookup table.
Assisted by this polynomial-size lookup table, which is precomputed within polynomial time, our decoder outputs the recovery operation from the MWPM found within polylog parallel time.

\subsection*{Numerical simulation}
Here, we numerically estimate the optimal number of classical parallel processors required to execute the most computationally intensive part of our MWPM decoding algorithm, i.e., the parallelization over a set of weight-perturbed path graphs.
Our numerical simulation provides evidence that the entire decoding algorithm can be executed using only $O(\poly(n))$ classical processors with modest weight perturbation, even though the current proof technique based on Ref.~\cite{8104102} provides a quasi-polynomial upper bound of the size of the weight perturbation and the required number of classical processors, as discussed above in~\eqref{eq:parallelization_upper_bound} and~\eqref{eq:number_processor}.

\begin{figure*}
\includegraphics[width=7.0in]{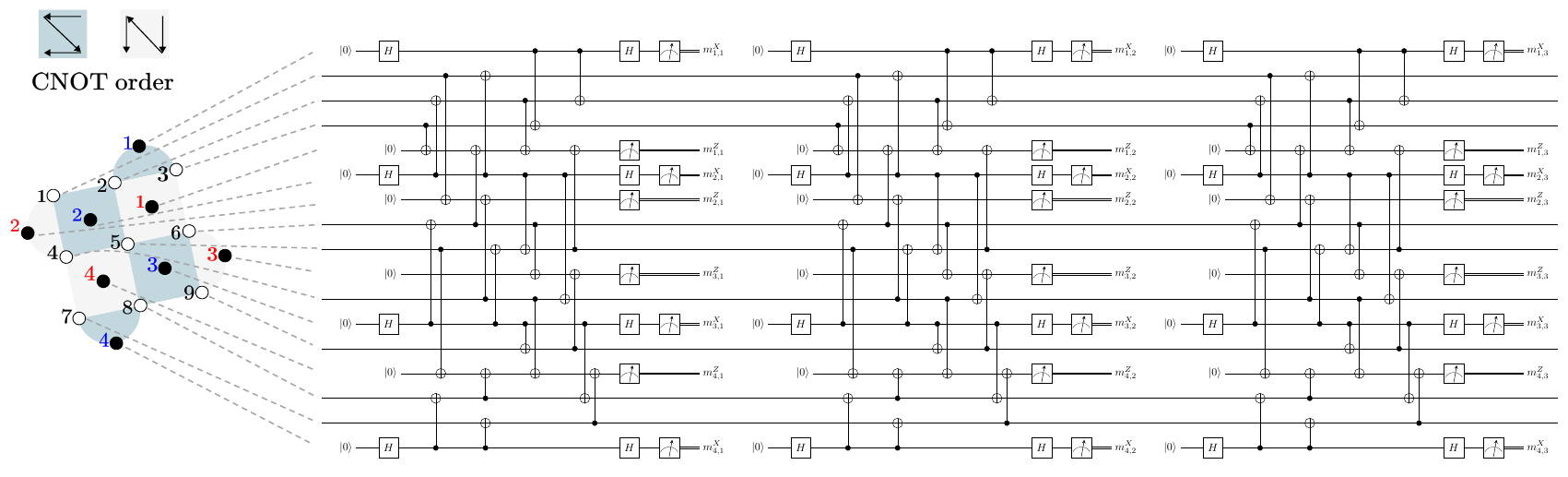}
\caption{\label{fig:gadget}An EC gadget for the $d=3$ 2D rotated surface code. White and black circles represent data and syndrome qubits, respectively. Dark and light faces correspond to $X$-type and $Z$-type stabilizer generators, respectively. The \textsc{CNOT} order employed here does not decrease the effective distance of the 2D rotated surface codes~\cite{PhysRevA.90.062320}.}
\end{figure*}

Specifically, we performed a Monte Carlo simulation to estimate, for each number of vertices $\qty|\cD|$ of path graphs encountered in the simulation, the smallest number of weight-perturbed path graphs required such that our algorithm successfully outputs an MWPM in at least one of these weight-perturbed path graphs.
The simulation was performed for the $[[n,1,d=\sqrt{n}]]$ 2D rotated surface codes under the IID circuit-level depolarizing error model with the physical error rate $p=10^{-3}$.
We simulated memory experiments with $d$ rounds of syndrome extraction, starting from a codeword and ending with measurements, without employing the sliding window decoding.
The syndrome extraction circuit used for $d=3$ is shown in Fig.~\ref{fig:gadget}.
Since the primary goal of this numerical simulation is not the demonstration of low-overhead FTQC but the estimation of the optimal number of weight-perturbed path graphs for our algorithm, we did not fully parallelize the execution of our algorithm in this numerical simulation.

To construct the lookup table for the $[[n,1,d=\sqrt{n}]]$ surface codes for each $d\in\{3, 4,\ldots,11\}$, the syndrome extraction circuits are generated using Stim~\cite{gidney2021stim}.
Stim's detector error model is converted into a graph in PyMatching~\cite{higgott2023sparse}. 
It is then converted into a graph in NetworkX~\cite{hagberg2008exploring}. 
Using NetworkX, we performed the exhaustive Dijkstra search to construct the lookup table.

Once we construct the lookup table, detection events are sampled by Stim.
Then, a path graph is constructed by reading the lookup table.
For the value of the normalization constant $C$ in~\eqref{edge_weights_integer}, we take $C=10$ for efficiency of the numerical simulation.
To output the same solution as that obtained by solving the original matching problem defined with real-valued weights using a blossom-based algorithm, it is not necessary to convert every digit of the given real-valued weights into integers; we can take $C$ as a modest finite value such that an original MWPM always remains an MWPM after scaling the weights to integers. 
Thus, in our numerical simulation, we assume that the true real-valued weights can be represented with significant digits such that the MWPM remains unchanged when the weights are scaled to integers with $C = 10$. 
This choice of $C$ provides an upper bound on the required number of weight-perturbed path graphs in the case where real-valued weights with an arbitrarily larger number of digits are given, because there would be more isolated MWPMs as the number of digits of the given weights increases.
For the determinant calculation, we implemented an algorithm in C++ manually. 

In the Monte Carlo simulation,
for each path graph encountered, we applied the perturbations proposed in~\eqref{eq:pseudo_random_perturbation} to its edge weights and computed the MWPM solution by our polylog-time MWPM algorithm.
Here, the Mersenne twister~\cite{10.1145/272991.272995} was used as the pseudo-random function satisfying~\eqref{eq:W} with the initial value of $W_{\max}\in\{2,3,\ldots\}$ set as $2$, and each seed $s_{\qty|\cD|,e_j,k}$ of the perturbation was randomly chosen from an unsigned $32$-bit integer.
We then checked whether $w^\ast$ in~\eqref{eq:w_ast} agreed with the weight of $M^\ast$, where $M^\ast$ is obtained by calculating~\eqref{eq:B_minor} for all $\{i,j\}$. 
If they do not agree, the output we obtained was not an MWPM\@.
In such cases, we applied perturbations using a different seed sequence, up to a maximum of $W_{\max}$ attempts; since it is theoretically guaranteed by~\eqref{eq:W_bound} that a suitable set of $W_{\max}$ weight-perturbed path graphs exists for each $W_{\max}$, our goal here is to heuristically construct such a set using the pseudo-random function.
If all $W_{\max}$ perturbations with different seed sequences failed to output an MWPM, $W_{\max}$ was incremented by $1$, and the procedure was repeated to determine the minimum $W_{\max}$ required to find the MWPM for the given path graph. 
If the matching solution was an MWPM of the path graph, we record the values of such $W_{\max}$ and $\qty|\cD|$.
This process was repeated $10^5$ times for each code distance $d\in\{3,4,\ldots,11\}$ to determine the minimum $W_{\max}$ required to output the MWPM for all the path graphs encountered in the simulation at each $\qty|\cD|$.
Finally, for each $\qty|\cD|$, we selected the largest $W_{\max}$ among those obtained for each $d$, thereby determining the smallest $W_{\max}$ required to guarantee an MWPM solution for each $\qty|\cD|$, independent of the code distance.
It is important to note that we computed the smallest $W_{\max}$ for the algorithm to output an MWPM, rather than the smallest $W_{\max}$ for the MWPM to be isolated; the reason is that, for the purpose of the decoding, we do not care whether it is indeed isolated as long as the output is the MWPM as desired.

In Fig.~\ref{fig:numerics_main}, we plot the smallest number of weight-perturbed path graphs required to deterministically output an MWPM for each $\qty|\cD|$, where the values of the vertical axis correspond to the smallest $W_{\max}$.
We see that the scaling of the smallest $W_{\max}$ is
\begin{align}
\label{eq:W_max_numerics}
    \min W_{\max}\leq\left\lceil0.62\qty|\cD|^{0.80}\right\rceil,
\end{align}
which is sublinear in $\qty|\cD|$.
Under the assumption that the algorithm can find the MWPM with $W_{\max}$ at least as given by the right-hand side of~\eqref{eq:W_max_numerics},
it suffices for the MWPM decoding to perform weight perturbations using
\begin{align}
    \label{eq:W_max_numerics_2}
W_{\max}=\left\lceil0.62\qty|\cD|^{0.80}\right\rceil
\end{align}
illustrated by the green cross markers in Fig.~\ref{fig:numerics_main}.
In our setting, the smallest $W_{\max}$ depends only on the number of vertices, minimized over all the obtained path graphs.
When the normalization constant $C$ is chosen as a larger value, the required number of weight-perturbed path graphs is upper bounded by the value given by~\eqref{eq:W_max_numerics_2}.
The path graphs appearing in the MWPM decoding have a particular topology as illustrated in Fig.~\ref{fig:comp_graph2}.
Given this structure, in other scenarios such as employing sliding window decoding or using a 3D subsystem surface code, one can reasonably expect a similar scaling of $W_{\max}$ to the one numerically reported in our setting, provided that the structure of the detector graphs resembles. 
The numerical results demonstrate that the part of the decoding identified in our analysis as a potential computational bottleneck---which theoretically requires a quasi-polynomial number of classical processors to provide the worst-case guarantee---can, in practice, be handled by at most only a sublinear number of processors, even within the deterministic approach. 
Consequently, under this assumption, the entire decoding algorithm can be executed in polylog time using at most only a polynomial number of classical parallel processors in the practically relevant regime.

\subsection*{Upper and lower bounds of time overheads of fault-tolerant protocols}

Finally, we discuss the upper and lower bounds of time overheads of various fault-tolerant protocols beyond those argued in the main text.
Two approaches exist to achieve the task of FTQC; one is with the concatenated codes, and the other with the quantum LDPC codes.
Here, we analyze achievable upper bounds of time overhead of FTQC in these two approaches, apart from the case of topological codes discussed in the main text.
Then, we also provide a lower bound of time overheads of FTQC\@.

Conventionally, the time overhead of fault-tolerant protocols with concatenated codes grows polylogarithmically~\cite{G}.
In the fault-tolerant protocols with concatenated codes, the fault-tolerant circuit is obtained from the original circuit by replacing each operation with the corresponding gadget and inserting EC gadgets recursively $L$ times~\cite{G,yamasaki2022timeefficient,10.1145/258533.258579,10.1137/S0097539799359385}, where $L$ is the concatenation level.
With this procedure, the logical error rate can be suppressed doubly exponentially in $L$, and then, to suppress it below $\lessapprox \frac{\epsilon(m)}{W(m)D(m)}$ as in~\eqref{eq:n_m_methods}, the required concatenation level is $L(m)=\Theta\qty(\log\qty(\polylog\qty(\frac{W(m)D(m)}{\epsilon(m)})))$.
In this case, even if each gadget has a constant depth $c$, the time overhead per operation becomes at least $c^{L(m)}=\Theta\qty(\polylog\qty(\frac{W(m)D(m)}{\epsilon(m)}))$, growing polylogarithmically.\footnote{
Section~8 of Ref.~\cite{gottesman2014faulttolerant} claims that constant time overhead is achievable by the protocol with the concatenated Steane codes, but for the reasons presented here, this concatenated-code protocol incurs a polylog time overhead.
}
It is unknown how to avoid this polylog growth of time overhead in protocols with concatenated codes.

By contrast, protocols with quantum LDPC codes do not suffer from this inherent time-overhead issue of code concatenation and thus have the potential for shortening time overheads, as shown in the main text using topological codes.
We here show another protocol to achieve doubly polylog time overheads using quantum expander codes, which are $[[n,k=\Theta(n),d=\Theta(\sqrt{n})]]$ quantum LDPC codes~\cite{7354429,10.1145/3188745.3188886} defined with an expander graph.
In Ref.~\cite{tamiya2024polylogtimeconstantspaceoverheadfaulttolerantquantum}, it is proven that we have a fault-tolerant protocol achieving a constant space overhead $O(1)$ and a polylog time overhead $\Theta\qty(\polylog\qty(\frac{W(m)D(m)}{\epsilon(m)}))$ using a hybrid approach that combines quantum expander codes and concatenated Steane codes.
This approach uses quantum expander codes as a quantum memory, with logical gates for quantum expander codes implemented by gate teleportation; for this purpose, the protocol exploits concatenated Steane codes to fault-tolerantly prepare auxiliary quantum states required for the gate teleportation, similar to our protocol.
To achieve the constant space overhead $O(1)$, the protocol in Ref.~\cite{tamiya2024polylogtimeconstantspaceoverheadfaulttolerantquantum} limited gate parallelism.
By contrast, the idea here is that by allowing for a conventional polylog space overhead in this hybrid approach, we can design an alternative protocol that achieves a doubly polylog time overhead.\footnote{
In our setting, which explicitly accounts for the runtime of classical computation for decoders, it is not straightforward to design protocols based on those developed for settings that do not fully incorporate the runtime of decoders, such as those in Refs.~\cite{gottesman2014faulttolerant,8555154,Grospellier,PhysRevX.5.031043,nguyen2024quantumfaulttoleranceconstantspace}. However, Ref.~\cite{tamiya2024polylogtimeconstantspaceoverheadfaulttolerantquantum} provides a method for modifying the protocols in Refs.~\cite{gottesman2014faulttolerant,8555154,Grospellier} so that the resulting protocols properly account for the decoder’s runtime. Using this method, one could potentially identify a suitable modification of the protocol in Ref.~\cite{nguyen2024quantumfaulttoleranceconstantspace}. Assuming such a modification is possible, the resulting protocol could then also be used to design an alternative protocol to achieve doubly polylog time overhead, by following the procedure outlined---currently without proof---in the Introduction of Ref.~\cite{nguyen2024quantumfaulttoleranceconstantspace} according to Ref.~\cite{ChrisPersonal}, in the same spirit as the space-time trade-off protocol design presented here based on Ref.~\cite{tamiya2024polylogtimeconstantspaceoverheadfaulttolerantquantum}.
}

In particular, for simplicity of implementation, consider using the quantum expander codes as subsystem codes that have a constant number $\Theta(1)$ of logical qubits per code block, leaving the other logical qubits as gauge qubits that are not used for storing logical information; that is, we have $[[n,k=\Theta(1),d=\Theta(\sqrt{n})]]$ codes, which we call subsystem quantum expander codes.
Then, using the same circuit compilation procedure as Ref.~\cite{tamiya2024polylogtimeconstantspaceoverheadfaulttolerantquantum} with complete gate parallelization (i.e., without limiting gate parallelism to achieve $O(1)$ space overheads), we obtain a fault-tolerant protocol using subsystem quantum expander codes as quantum memory and concatenated Steane codes for gate teleportation, which may require a polylog space overhead  $\Theta\qty(\polylog\qty(\frac{W(m)D(m)}{\epsilon(m)}))$~\cite{tamiya2024polylogtimeconstantspaceoverheadfaulttolerantquantum} but now achieves complete gate parallelism.
As discussed in~\eqref{eq:T_gate_methods}, by parallelizing preparations of the auxiliary states, gate teleportation implements each logical gate within a constant depth, i.e., $T_\mathrm{gate}=O(1)$, except for the runtime of decoding.
Due to the single-shot decodability of quantum expander codes, the syndrome extraction can be performed within a single round, i.e., $T_\mathrm{SE}=O\qty(1)$~\cite{8555154,Grospellier,tamiya2024polylogtimeconstantspaceoverheadfaulttolerantquantum}.
The runtime of each gate teleportation is then dominated by measuring code blocks followed by invoking the logarithmic-time small-set-flip decoder $T_\mathrm{dec}(m)=O(\log(n(m)))=O\qty(\log\qty(\polylog\qty(\frac{W(m)D(m)}{\epsilon(m)})))$~\cite{8555154,Grospellier,tamiya2024polylogtimeconstantspaceoverheadfaulttolerantquantum} to determine the logical measurement outcome required for the gate teleportation.
As a whole, similar to the analysis of the topological-code protocol in the main text, the fault-tolerant protocol with subsystem quantum expander codes also achieves the doubly polylog time overhead
\begin{align}
    &\frac{D_\mathrm{FT}(m)}{D(m)}\nonumber\\
    &=O\qty(T_\mathrm{gate}+T_\mathrm{SE}+T_\mathrm{dec}(m))\\
    &=O(1)+O\qty(\polylog\qty(\polylog\qty(\frac{W(m)D(m)}{\epsilon(m)}))).
\end{align}

Whereas both this protocol and the topological-code protocol presented in the main text achieve doubly polylog time overheads, we believe that the protocol presented in the main text holds a primary advantage in its better implementability, owing to the structured nature of the topological code defined on a regular lattice and the concatenated code organized in a hierarchical manner.
While fault-tolerant protocols based on quantum expander codes can, in principle, be implemented across various hardware architectures~\cite{sunami2024scalablenetworkingneutralatomqubits,PhysRevLett.129.050504,xu2024constant}, expander codes rely on expander graphs, which lack a naturally useful structure for practical implementation.
This structural complexity can pose additional challenges compared to topological and concatenated codes, which are more naturally suited for scalable architectures.
As discussed in the main text, a promising direction for future research is to explore experimental implementations of doubly-polylog-time-overhead fault-tolerant protocols, by examining their feasibility across different quantum computing platforms.

Finally, we discuss the lower bounds of time overheads of FTQC\@.
When we take into account the runtime of classical computation in executing FTQC,
the analysis of lower bounds of time overheads of FTQC seems to have a similar flavor to studying circuit lower bounds appearing in the P versus NP problem~\cite{RAZBOROV199724}, so it seems challenging to provide the time-overhead lower bounds unconditionally without additional assumptions.
One way to provide such a time-overhead lower bound is to use the fact that taking a parity of an $n$-bit string---a commonly appearing subroutine in fault-tolerant protocols---can be classically performed in $O(\log(n))$ parallel runtime but cannot be performed in $O(1)$ parallel runtime with a polynomial number of processors~\cite{furst1984parity,4568121,10.1145/12130.12132,razborov1987lower,10.1145/28395.28404}.
In particular, under the assumption that
\begin{enumerate}
    \item the code size $n$ grows on large scales,
    \item the fault-tolerant protocol includes classical computation of parity of bit strings of growing length in $n$ per implementing logical gates, such as gate teleportation and lattice surgery,
    \item and the number of classical processors is at most subexponential $\exp[o(n)]$ per code block,
\end{enumerate}
then, due to the lower bound of circuit complexity of computing parity~\cite{razborov1987lower}, we conclude that constant-time-overhead FTQC is unattainable with any such protocol; in other words, the time overhead is lower bounded by $\omega(1)$.
The protocols presented in this work satisfy these assumptions; therefore, surpassing the time-overhead lower bound established here would require a fundamentally different fault-tolerant protocol that deviates from the assumptions and techniques used in this work.

Based on these observations, we conjecture that a fundamental obstacle may exist that unconditionally prohibits constant-time-overhead FTQC, even without imposing assumptions about the specifics of the protocols.
An unconditional and rigorous proof of such time-overhead lower bounds for FTQC would represent a major theoretical breakthrough, significantly advancing our understanding of the ultimate overheads of FTQC\@.
Nevertheless, our contribution in this work is to substantially narrow the gap between the upper and lower bounds of the time overheads of FTQC\@.
By surpassing the conventional polylog time overhead, while explicitly accounting for the decoder's runtime, we bring FTQC significantly closer to its fundamental limits, paving the way for a deeper understanding of the ultimate space-time tradeoffs in FTQC\@.

\bibliography{main.bib}

\begin{thebibliography}{116}%
\makeatletter
\providecommand \@ifxundefined [1]{%
 \@ifx{#1\undefined}
}%
\providecommand \@ifnum [1]{%
 \ifnum #1\expandafter \@firstoftwo
 \else \expandafter \@secondoftwo
 \fi
}%
\providecommand \@ifx [1]{%
 \ifx #1\expandafter \@firstoftwo
 \else \expandafter \@secondoftwo
 \fi
}%
\providecommand \natexlab [1]{#1}%
\providecommand \enquote  [1]{``#1''}%
\providecommand \bibnamefont  [1]{#1}%
\providecommand \bibfnamefont [1]{#1}%
\providecommand \citenamefont [1]{#1}%
\providecommand \href@noop [0]{\@secondoftwo}%
\providecommand \href [0]{\begingroup \@sanitize@url \@href}%
\providecommand \@href[1]{\@@startlink{#1}\@@href}%
\providecommand \@@href[1]{\endgroup#1\@@endlink}%
\providecommand \@sanitize@url [0]{\catcode `\\12\catcode `\$12\catcode
  `\&12\catcode `\#12\catcode `\^12\catcode `\_12\catcode `\%12\relax}%
\providecommand \@@startlink[1]{}%
\providecommand \@@endlink[0]{}%
\providecommand \url  [0]{\begingroup\@sanitize@url \@url }%
\providecommand \@url [1]{\endgroup\@href {#1}{\urlprefix }}%
\providecommand \urlprefix  [0]{URL }%
\providecommand \Eprint [0]{\href }%
\providecommand \doibase [0]{https://doi.org/}%
\providecommand \selectlanguage [0]{\@gobble}%
\providecommand \bibinfo  [0]{\@secondoftwo}%
\providecommand \bibfield  [0]{\@secondoftwo}%
\providecommand \translation [1]{[#1]}%
\providecommand \BibitemOpen [0]{}%
\providecommand \bibitemStop [0]{}%
\providecommand \bibitemNoStop [0]{.\EOS\space}%
\providecommand \EOS [0]{\spacefactor3000\relax}%
\providecommand \BibitemShut  [1]{\csname bibitem#1\endcsname}%
\let\auto@bib@innerbib\@empty
\bibitem [{\citenamefont {Kitaev}(2003)}]{K2}%
  \BibitemOpen
  \bibfield  {author} {\bibinfo {author} {\bibfnamefont {A.}~\bibnamefont
  {Kitaev}},\ }\bibfield  {title} {\bibinfo {title} {Fault-tolerant quantum
  computation by anyons},\ }\href
  {https://doi.org/10.1016/S0003-4916(02)00018-0} {\bibfield  {journal}
  {\bibinfo  {journal} {Ann. Phys.}\ }\textbf {\bibinfo {volume} {303}},\
  \bibinfo {pages} {2 } (\bibinfo {year} {2003})}\BibitemShut {NoStop}%
\bibitem [{\citenamefont {Bravyi}\ and\ \citenamefont
  {Kitaev}(1998)}]{bravyi1998quantumcodeslatticeboundary}%
  \BibitemOpen
  \bibfield  {author} {\bibinfo {author} {\bibfnamefont {S.~B.}\ \bibnamefont
  {Bravyi}}\ and\ \bibinfo {author} {\bibfnamefont {A.~Y.}\ \bibnamefont
  {Kitaev}},\ }\href {https://arxiv.org/abs/quant-ph/9811052} {\bibinfo {title}
  {Quantum codes on a lattice with boundary}} (\bibinfo {year} {1998}),\
  \Eprint {https://arxiv.org/abs/quant-ph/9811052} {arXiv:quant-ph/9811052
  [quant-ph]} \BibitemShut {NoStop}%
\bibitem [{\citenamefont {Bombin}\ and\ \citenamefont
  {Martin-Delgado}(2006)}]{PhysRevLett.97.180501}%
  \BibitemOpen
  \bibfield  {author} {\bibinfo {author} {\bibfnamefont {H.}~\bibnamefont
  {Bombin}}\ and\ \bibinfo {author} {\bibfnamefont {M.~A.}\ \bibnamefont
  {Martin-Delgado}},\ }\bibfield  {title} {\bibinfo {title} {Topological
  quantum distillation},\ }\href
  {https://doi.org/10.1103/PhysRevLett.97.180501} {\bibfield  {journal}
  {\bibinfo  {journal} {Phys. Rev. Lett.}\ }\textbf {\bibinfo {volume} {97}},\
  \bibinfo {pages} {180501} (\bibinfo {year} {2006})}\BibitemShut {NoStop}%
\bibitem [{\citenamefont {Dennis}\ \emph {et~al.}(2002)\citenamefont {Dennis},
  \citenamefont {Kitaev}, \citenamefont {Landahl},\ and\ \citenamefont
  {Preskill}}]{doi:10.1063/1.1499754}%
  \BibitemOpen
  \bibfield  {author} {\bibinfo {author} {\bibfnamefont {E.}~\bibnamefont
  {Dennis}}, \bibinfo {author} {\bibfnamefont {A.}~\bibnamefont {Kitaev}},
  \bibinfo {author} {\bibfnamefont {A.}~\bibnamefont {Landahl}},\ and\ \bibinfo
  {author} {\bibfnamefont {J.}~\bibnamefont {Preskill}},\ }\bibfield  {title}
  {\bibinfo {title} {Topological quantum memory},\ }\href
  {https://doi.org/10.1063/1.1499754} {\bibfield  {journal} {\bibinfo
  {journal} {Journal of Mathematical Physics}\ }\textbf {\bibinfo {volume}
  {43}},\ \bibinfo {pages} {4452} (\bibinfo {year} {2002})}\BibitemShut
  {NoStop}%
\bibitem [{\citenamefont {Bombín}(2013)}]{Bombin_2013}%
  \BibitemOpen
  \bibfield  {author} {\bibinfo {author} {\bibfnamefont {H.}~\bibnamefont
  {Bombín}},\ }\bibinfo {title} {Topological codes},\ in\ \href
  {https://www.cambridge.org/core/books/abs/quantum-error-correction/topological-codes/7C4A21FD07F40BFF938611FAE98DAE9C}
  {\emph {\bibinfo {booktitle} {Quantum Error Correction}}},\ \bibinfo {editor}
  {edited by\ \bibinfo {editor} {\bibfnamefont {D.~A.}\ \bibnamefont {Lidar}}\
  and\ \bibinfo {editor} {\bibfnamefont {T.~A.}\ \bibnamefont {Brun}}}\
  (\bibinfo  {publisher} {Cambridge University Press},\ \bibinfo {year}
  {2013})\ p.\ \bibinfo {pages} {455–481}\BibitemShut {NoStop}%
\bibitem [{\citenamefont {Bomb\'{\i}n}(2015)}]{PhysRevX.5.031043}%
  \BibitemOpen
  \bibfield  {author} {\bibinfo {author} {\bibfnamefont {H.}~\bibnamefont
  {Bomb\'{\i}n}},\ }\bibfield  {title} {\bibinfo {title} {Single-shot
  fault-tolerant quantum error correction},\ }\href
  {https://doi.org/10.1103/PhysRevX.5.031043} {\bibfield  {journal} {\bibinfo
  {journal} {Phys. Rev. X}\ }\textbf {\bibinfo {volume} {5}},\ \bibinfo {pages}
  {031043} (\bibinfo {year} {2015})}\BibitemShut {NoStop}%
\bibitem [{\citenamefont {Bomb{\'\i}n}(2015)}]{bombin2015gauge}%
  \BibitemOpen
  \bibfield  {author} {\bibinfo {author} {\bibfnamefont {H.}~\bibnamefont
  {Bomb{\'\i}n}},\ }\bibfield  {title} {\bibinfo {title} {Gauge color codes:
  optimal transversal gates and gauge fixing in topological stabilizer codes},\
  }\href {https://iopscience.iop.org/article/10.1088/1367-2630/17/8/083002}
  {\bibfield  {journal} {\bibinfo  {journal} {New Journal of Physics}\ }\textbf
  {\bibinfo {volume} {17}},\ \bibinfo {pages} {083002} (\bibinfo {year}
  {2015})}\BibitemShut {NoStop}%
\bibitem [{\citenamefont {Kubica}\ and\ \citenamefont
  {Vasmer}(2022)}]{kubica2022single}%
  \BibitemOpen
  \bibfield  {author} {\bibinfo {author} {\bibfnamefont {A.}~\bibnamefont
  {Kubica}}\ and\ \bibinfo {author} {\bibfnamefont {M.}~\bibnamefont
  {Vasmer}},\ }\bibfield  {title} {\bibinfo {title} {Single-shot quantum error
  correction with the three-dimensional subsystem toric code},\ }\href
  {https://www.nature.com/articles/s41467-022-33923-4} {\bibfield  {journal}
  {\bibinfo  {journal} {Nature communications}\ }\textbf {\bibinfo {volume}
  {13}},\ \bibinfo {pages} {6272} (\bibinfo {year} {2022})}\BibitemShut
  {NoStop}%
\bibitem [{\citenamefont {Bridgeman}\ \emph {et~al.}(2024)\citenamefont
  {Bridgeman}, \citenamefont {Kubica},\ and\ \citenamefont
  {Vasmer}}]{PRXQuantum.5.020310}%
  \BibitemOpen
  \bibfield  {author} {\bibinfo {author} {\bibfnamefont {J.~C.}\ \bibnamefont
  {Bridgeman}}, \bibinfo {author} {\bibfnamefont {A.}~\bibnamefont {Kubica}},\
  and\ \bibinfo {author} {\bibfnamefont {M.}~\bibnamefont {Vasmer}},\
  }\bibfield  {title} {\bibinfo {title} {Lifting topological codes:
  Three-dimensional subsystem codes from two-dimensional anyon models},\ }\href
  {https://doi.org/10.1103/PRXQuantum.5.020310} {\bibfield  {journal} {\bibinfo
   {journal} {PRX Quantum}\ }\textbf {\bibinfo {volume} {5}},\ \bibinfo {pages}
  {020310} (\bibinfo {year} {2024})}\BibitemShut {NoStop}%
\bibitem [{\citenamefont {Campbell}(2019)}]{Campbell_2019}%
  \BibitemOpen
  \bibfield  {author} {\bibinfo {author} {\bibfnamefont {E.~T.}\ \bibnamefont
  {Campbell}},\ }\bibfield  {title} {\bibinfo {title} {A theory of single-shot
  error correction for adversarial noise},\ }\href
  {https://doi.org/10.1088/2058-9565/aafc8f} {\bibfield  {journal} {\bibinfo
  {journal} {Quantum Science and Technology}\ }\textbf {\bibinfo {volume}
  {4}},\ \bibinfo {pages} {025006} (\bibinfo {year} {2019})}\BibitemShut
  {NoStop}%
\bibitem [{\citenamefont {Brown}\ \emph {et~al.}(2016)\citenamefont {Brown},
  \citenamefont {Loss}, \citenamefont {Pachos}, \citenamefont {Self},\ and\
  \citenamefont {Wootton}}]{RevModPhys.88.045005}%
  \BibitemOpen
  \bibfield  {author} {\bibinfo {author} {\bibfnamefont {B.~J.}\ \bibnamefont
  {Brown}}, \bibinfo {author} {\bibfnamefont {D.}~\bibnamefont {Loss}},
  \bibinfo {author} {\bibfnamefont {J.~K.}\ \bibnamefont {Pachos}}, \bibinfo
  {author} {\bibfnamefont {C.~N.}\ \bibnamefont {Self}},\ and\ \bibinfo
  {author} {\bibfnamefont {J.~R.}\ \bibnamefont {Wootton}},\ }\bibfield
  {title} {\bibinfo {title} {Quantum memories at finite temperature},\ }\href
  {https://doi.org/10.1103/RevModPhys.88.045005} {\bibfield  {journal}
  {\bibinfo  {journal} {Rev. Mod. Phys.}\ }\textbf {\bibinfo {volume} {88}},\
  \bibinfo {pages} {045005} (\bibinfo {year} {2016})}\BibitemShut {NoStop}%
\bibitem [{\citenamefont {Quintavalle}\ \emph {et~al.}(2021)\citenamefont
  {Quintavalle}, \citenamefont {Vasmer}, \citenamefont {Roffe},\ and\
  \citenamefont {Campbell}}]{PRXQuantum.2.020340}%
  \BibitemOpen
  \bibfield  {author} {\bibinfo {author} {\bibfnamefont {A.~O.}\ \bibnamefont
  {Quintavalle}}, \bibinfo {author} {\bibfnamefont {M.}~\bibnamefont {Vasmer}},
  \bibinfo {author} {\bibfnamefont {J.}~\bibnamefont {Roffe}},\ and\ \bibinfo
  {author} {\bibfnamefont {E.~T.}\ \bibnamefont {Campbell}},\ }\bibfield
  {title} {\bibinfo {title} {Single-shot error correction of three-dimensional
  homological product codes},\ }\href
  {https://doi.org/10.1103/PRXQuantum.2.020340} {\bibfield  {journal} {\bibinfo
   {journal} {PRX Quantum}\ }\textbf {\bibinfo {volume} {2}},\ \bibinfo {pages}
  {020340} (\bibinfo {year} {2021})}\BibitemShut {NoStop}%
\bibitem [{\citenamefont {Aharonov}\ and\ \citenamefont
  {Ben-Or}(1997)}]{10.1145/258533.258579}%
  \BibitemOpen
  \bibfield  {author} {\bibinfo {author} {\bibfnamefont {D.}~\bibnamefont
  {Aharonov}}\ and\ \bibinfo {author} {\bibfnamefont {M.}~\bibnamefont
  {Ben-Or}},\ }\bibfield  {title} {\bibinfo {title} {Fault-tolerant quantum
  computation with constant error},\ }in\ \href
  {https://doi.org/10.1145/258533.258579} {\emph {\bibinfo {booktitle}
  {Proceedings of the Twenty-Ninth Annual ACM Symposium on Theory of
  Computing}}},\ \bibinfo {series and number} {STOC '97}\ (\bibinfo
  {publisher} {Association for Computing Machinery},\ \bibinfo {address} {New
  York, NY, USA},\ \bibinfo {year} {1997})\ p.\ \bibinfo {pages}
  {176–188}\BibitemShut {NoStop}%
\bibitem [{\citenamefont {Aharonov}\ and\ \citenamefont
  {Ben-Or}(2008)}]{10.1137/S0097539799359385}%
  \BibitemOpen
  \bibfield  {author} {\bibinfo {author} {\bibfnamefont {D.}~\bibnamefont
  {Aharonov}}\ and\ \bibinfo {author} {\bibfnamefont {M.}~\bibnamefont
  {Ben-Or}},\ }\bibfield  {title} {\bibinfo {title} {Fault-tolerant quantum
  computation with constant error rate},\ }\href
  {https://doi.org/10.1137/S0097539799359385} {\bibfield  {journal} {\bibinfo
  {journal} {SIAM J. Comput.}\ }\textbf {\bibinfo {volume} {38}},\ \bibinfo
  {pages} {1207–1282} (\bibinfo {year} {2008})}\BibitemShut {NoStop}%
\bibitem [{\citenamefont {Bluvstein}\ \emph {et~al.}(2024)\citenamefont
  {Bluvstein}, \citenamefont {Evered}, \citenamefont {Geim}, \citenamefont
  {Li}, \citenamefont {Zhou}, \citenamefont {Manovitz}, \citenamefont {Ebadi},
  \citenamefont {Cain}, \citenamefont {Kalinowski}, \citenamefont {Hangleiter}
  \emph {et~al.}}]{bluvstein2024logical}%
  \BibitemOpen
  \bibfield  {author} {\bibinfo {author} {\bibfnamefont {D.}~\bibnamefont
  {Bluvstein}}, \bibinfo {author} {\bibfnamefont {S.~J.}\ \bibnamefont
  {Evered}}, \bibinfo {author} {\bibfnamefont {A.~A.}\ \bibnamefont {Geim}},
  \bibinfo {author} {\bibfnamefont {S.~H.}\ \bibnamefont {Li}}, \bibinfo
  {author} {\bibfnamefont {H.}~\bibnamefont {Zhou}}, \bibinfo {author}
  {\bibfnamefont {T.}~\bibnamefont {Manovitz}}, \bibinfo {author}
  {\bibfnamefont {S.}~\bibnamefont {Ebadi}}, \bibinfo {author} {\bibfnamefont
  {M.}~\bibnamefont {Cain}}, \bibinfo {author} {\bibfnamefont {M.}~\bibnamefont
  {Kalinowski}}, \bibinfo {author} {\bibfnamefont {D.}~\bibnamefont
  {Hangleiter}}, \emph {et~al.},\ }\bibfield  {title} {\bibinfo {title}
  {Logical quantum processor based on reconfigurable atom arrays},\ }\href
  {https://www.nature.com/articles/s41586-023-06927-3} {\bibfield  {journal}
  {\bibinfo  {journal} {Nature}\ }\textbf {\bibinfo {volume} {626}},\ \bibinfo
  {pages} {58} (\bibinfo {year} {2024})}\BibitemShut {NoStop}%
\bibitem [{\citenamefont {Sunami}\ \emph {et~al.}(2024)\citenamefont {Sunami},
  \citenamefont {Tamiya}, \citenamefont {Inoue}, \citenamefont {Yamasaki},\
  and\ \citenamefont {Goban}}]{sunami2024scalablenetworkingneutralatomqubits}%
  \BibitemOpen
  \bibfield  {author} {\bibinfo {author} {\bibfnamefont {S.}~\bibnamefont
  {Sunami}}, \bibinfo {author} {\bibfnamefont {S.}~\bibnamefont {Tamiya}},
  \bibinfo {author} {\bibfnamefont {R.}~\bibnamefont {Inoue}}, \bibinfo
  {author} {\bibfnamefont {H.}~\bibnamefont {Yamasaki}},\ and\ \bibinfo
  {author} {\bibfnamefont {A.}~\bibnamefont {Goban}},\ }\href
  {https://arxiv.org/abs/2407.11111} {\bibinfo {title} {Scalable networking of
  neutral-atom qubits: Nanofiber-based approach for multiprocessor
  fault-tolerant quantum computer}} (\bibinfo {year} {2024}),\ \Eprint
  {https://arxiv.org/abs/2407.11111} {arXiv:2407.11111 [quant-ph]} \BibitemShut
  {NoStop}%
\bibitem [{\citenamefont {{Google Quantum AI}}(2023)}]{google2023suppressing}%
  \BibitemOpen
  \bibfield  {author} {\bibinfo {author} {\bibnamefont {{Google Quantum AI}}},\
  }\bibfield  {title} {\bibinfo {title} {Suppressing quantum errors by scaling
  a surface code logical qubit},\ }\href
  {https://www.nature.com/articles/s41586-022-05434-1} {\bibfield  {journal}
  {\bibinfo  {journal} {Nature}\ }\textbf {\bibinfo {volume} {614}},\ \bibinfo
  {pages} {676} (\bibinfo {year} {2023})}\BibitemShut {NoStop}%
\bibitem [{\citenamefont {Nielsen}\ and\ \citenamefont {Chuang}(2010)}]{N4}%
  \BibitemOpen
  \bibfield  {author} {\bibinfo {author} {\bibfnamefont {M.~A.}\ \bibnamefont
  {Nielsen}}\ and\ \bibinfo {author} {\bibfnamefont {I.~L.}\ \bibnamefont
  {Chuang}},\ }\href {https://doi.org/10.1017/CBO9780511976667} {\emph
  {\bibinfo {title} {Quantum Computation and Quantum Information: 10th
  Anniversary Edition}}}\ (\bibinfo  {publisher} {Cambridge University Press},\
  \bibinfo {address} {Cambridge},\ \bibinfo {year} {2010})\BibitemShut
  {NoStop}%
\bibitem [{\citenamefont {Gottesman}(2014)}]{gottesman2014faulttolerant}%
  \BibitemOpen
  \bibfield  {author} {\bibinfo {author} {\bibfnamefont {D.}~\bibnamefont
  {Gottesman}},\ }\bibfield  {title} {\bibinfo {title} {Fault-tolerant quantum
  computation with constant overhead},\ }\href
  {https://doi.org/10.26421/QIC14.15-16-5} {\bibfield  {journal} {\bibinfo
  {journal} {Quantum Info. Comput.}\ }\textbf {\bibinfo {volume} {14}},\
  \bibinfo {pages} {1338–1372} (\bibinfo {year} {2014})}\BibitemShut
  {NoStop}%
\bibitem [{\citenamefont {Gottesman}(2010)}]{G}%
  \BibitemOpen
  \bibfield  {author} {\bibinfo {author} {\bibfnamefont {D.}~\bibnamefont
  {Gottesman}},\ }\bibfield  {title} {\bibinfo {title} {An introduction to
  quantum error correction and fault-tolerant quantum computation},\ }in\ \href
  {https://doi.org/10.1090/psapm/068} {\emph {\bibinfo {booktitle} {Quantum
  information science and its contributions to mathematics}}},\ \bibinfo
  {series} {Proceedings of Symposia in Applied Mathematics}, Vol.~\bibinfo
  {volume} {68}\ (\bibinfo  {publisher} {American Mathematical Society},\
  \bibinfo {address} {Providence, Rhode Island},\ \bibinfo {year} {2010})\ pp.\
  \bibinfo {pages} {13--58}\BibitemShut {NoStop}%
\bibitem [{\citenamefont {Yamasaki}\ and\ \citenamefont
  {Koashi}(2024)}]{yamasaki2022timeefficient}%
  \BibitemOpen
  \bibfield  {author} {\bibinfo {author} {\bibfnamefont {H.}~\bibnamefont
  {Yamasaki}}\ and\ \bibinfo {author} {\bibfnamefont {M.}~\bibnamefont
  {Koashi}},\ }\bibfield  {title} {\bibinfo {title} {Time-efficient
  constant-space-overhead fault-tolerant quantum computation},\ }\href
  {https://link.springer.com/article/10.1038/s41567-023-02325-8} {\bibfield
  {journal} {\bibinfo  {journal} {Nature Physics}\ }\textbf {\bibinfo {volume}
  {20}},\ \bibinfo {pages} {247} (\bibinfo {year} {2024})}\BibitemShut
  {NoStop}%
\bibitem [{\citenamefont {Tamiya}\ \emph {et~al.}(2024)\citenamefont {Tamiya},
  \citenamefont {Koashi},\ and\ \citenamefont
  {Yamasaki}}]{tamiya2024polylogtimeconstantspaceoverheadfaulttolerantquantum}%
  \BibitemOpen
  \bibfield  {author} {\bibinfo {author} {\bibfnamefont {S.}~\bibnamefont
  {Tamiya}}, \bibinfo {author} {\bibfnamefont {M.}~\bibnamefont {Koashi}},\
  and\ \bibinfo {author} {\bibfnamefont {H.}~\bibnamefont {Yamasaki}},\ }\href
  {https://arxiv.org/abs/2411.03683} {\bibinfo {title} {Polylog-time- and
  constant-space-overhead fault-tolerant quantum computation with quantum
  low-density parity-check codes}} (\bibinfo {year} {2024}),\ \Eprint
  {https://arxiv.org/abs/2411.03683} {arXiv:2411.03683 [quant-ph]} \BibitemShut
  {NoStop}%
\bibitem [{\citenamefont {Kovalev}\ and\ \citenamefont
  {Pryadko}(2013)}]{PhysRevA.87.020304}%
  \BibitemOpen
  \bibfield  {author} {\bibinfo {author} {\bibfnamefont {A.~A.}\ \bibnamefont
  {Kovalev}}\ and\ \bibinfo {author} {\bibfnamefont {L.~P.}\ \bibnamefont
  {Pryadko}},\ }\bibfield  {title} {\bibinfo {title} {Fault tolerance of
  quantum low-density parity check codes with sublinear distance scaling},\
  }\href {https://doi.org/10.1103/PhysRevA.87.020304} {\bibfield  {journal}
  {\bibinfo  {journal} {Phys. Rev. A}\ }\textbf {\bibinfo {volume} {87}},\
  \bibinfo {pages} {020304} (\bibinfo {year} {2013})}\BibitemShut {NoStop}%
\bibitem [{\citenamefont {Fawzi}\ \emph
  {et~al.}(2018{\natexlab{a}})\citenamefont {Fawzi}, \citenamefont
  {Grospellier},\ and\ \citenamefont {Leverrier}}]{8555154}%
  \BibitemOpen
  \bibfield  {author} {\bibinfo {author} {\bibfnamefont {O.}~\bibnamefont
  {Fawzi}}, \bibinfo {author} {\bibfnamefont {A.}~\bibnamefont {Grospellier}},\
  and\ \bibinfo {author} {\bibfnamefont {A.}~\bibnamefont {Leverrier}},\
  }\bibfield  {title} {\bibinfo {title} {{ Constant Overhead Quantum
  Fault-Tolerance with Quantum Expander Codes }},\ }in\ \href
  {https://doi.org/10.1109/FOCS.2018.00076} {\emph {\bibinfo {booktitle} {2018
  IEEE 59th Annual Symposium on Foundations of Computer Science (FOCS)}}}\
  (\bibinfo  {publisher} {IEEE Computer Society},\ \bibinfo {address} {Los
  Alamitos, CA, USA},\ \bibinfo {year} {2018})\ pp.\ \bibinfo {pages}
  {743--754}\BibitemShut {NoStop}%
\bibitem [{\citenamefont {Grospellier}(2019)}]{Grospellier}%
  \BibitemOpen
  \bibfield  {author} {\bibinfo {author} {\bibfnamefont {A.}~\bibnamefont
  {Grospellier}},\ }\emph {\bibinfo {title} {Constant time decoding of quantum
  expander codes and application to fault-tolerant quantum computation}},\
  \href {https://tel.archives-ouvertes.fr/tel-03364419/document} {Ph.D.
  thesis},\ \bibinfo  {school} {Sorbonne Universi\'{e}} (\bibinfo {year}
  {2019})\BibitemShut {NoStop}%
\bibitem [{\citenamefont {Nguyen}\ and\ \citenamefont
  {Pattison}(2024)}]{nguyen2024quantumfaulttoleranceconstantspace}%
  \BibitemOpen
  \bibfield  {author} {\bibinfo {author} {\bibfnamefont {Q.~T.}\ \bibnamefont
  {Nguyen}}\ and\ \bibinfo {author} {\bibfnamefont {C.~A.}\ \bibnamefont
  {Pattison}},\ }\href {https://arxiv.org/abs/2411.03632v1} {\bibinfo {title}
  {Quantum fault tolerance with constant-space and logarithmic-time overheads}}
  (\bibinfo {year} {2024}),\ \Eprint {https://arxiv.org/abs/2411.03632v1}
  {arXiv:2411.03632v1 [quant-ph]} \BibitemShut {NoStop}%
\bibitem [{\citenamefont {Fowler}(2012)}]{PhysRevLett.109.180502}%
  \BibitemOpen
  \bibfield  {author} {\bibinfo {author} {\bibfnamefont {A.~G.}\ \bibnamefont
  {Fowler}},\ }\bibfield  {title} {\bibinfo {title} {Proof of finite surface
  code threshold for matching},\ }\href
  {https://doi.org/10.1103/PhysRevLett.109.180502} {\bibfield  {journal}
  {\bibinfo  {journal} {Phys. Rev. Lett.}\ }\textbf {\bibinfo {volume} {109}},\
  \bibinfo {pages} {180502} (\bibinfo {year} {2012})}\BibitemShut {NoStop}%
\bibitem [{\citenamefont {Wang}\ \emph {et~al.}(2010)\citenamefont {Wang},
  \citenamefont {Fowler}, \citenamefont {Stephens},\ and\ \citenamefont
  {Hollenberg}}]{10.5555/2011362.2011368}%
  \BibitemOpen
  \bibfield  {author} {\bibinfo {author} {\bibfnamefont {D.~S.}\ \bibnamefont
  {Wang}}, \bibinfo {author} {\bibfnamefont {A.~G.}\ \bibnamefont {Fowler}},
  \bibinfo {author} {\bibfnamefont {A.~M.}\ \bibnamefont {Stephens}},\ and\
  \bibinfo {author} {\bibfnamefont {L.~C.~L.}\ \bibnamefont {Hollenberg}},\
  }\bibfield  {title} {\bibinfo {title} {Threshold error rates for the toric
  and planar codes},\ }\href
  {https://www.rintonpress.com/journals/doi/QIC10.5-6-6.html} {\bibfield
  {journal} {\bibinfo  {journal} {Quantum Info. Comput.}\ }\textbf {\bibinfo
  {volume} {10}},\ \bibinfo {pages} {456–469} (\bibinfo {year}
  {2010})}\BibitemShut {NoStop}%
\bibitem [{\citenamefont {Fowler}\ \emph {et~al.}(2011)\citenamefont {Fowler},
  \citenamefont {Wang},\ and\ \citenamefont
  {Hollenberg}}]{10.5555/2011383.2011385}%
  \BibitemOpen
  \bibfield  {author} {\bibinfo {author} {\bibfnamefont {A.~G.}\ \bibnamefont
  {Fowler}}, \bibinfo {author} {\bibfnamefont {D.~S.}\ \bibnamefont {Wang}},\
  and\ \bibinfo {author} {\bibfnamefont {L.~C.~L.}\ \bibnamefont
  {Hollenberg}},\ }\bibfield  {title} {\bibinfo {title} {Surface code quantum
  error correction incorporating accurate error propagation},\ }\href
  {https://www.rintonpress.com/journals/doi/QIC11.1-2-2.html} {\bibfield
  {journal} {\bibinfo  {journal} {Quantum Info. Comput.}\ }\textbf {\bibinfo
  {volume} {11}},\ \bibinfo {pages} {8–18} (\bibinfo {year}
  {2011})}\BibitemShut {NoStop}%
\bibitem [{\citenamefont {Wang}\ \emph {et~al.}(2011)\citenamefont {Wang},
  \citenamefont {Fowler},\ and\ \citenamefont
  {Hollenberg}}]{PhysRevA.83.020302}%
  \BibitemOpen
  \bibfield  {author} {\bibinfo {author} {\bibfnamefont {D.~S.}\ \bibnamefont
  {Wang}}, \bibinfo {author} {\bibfnamefont {A.~G.}\ \bibnamefont {Fowler}},\
  and\ \bibinfo {author} {\bibfnamefont {L.~C.~L.}\ \bibnamefont
  {Hollenberg}},\ }\bibfield  {title} {\bibinfo {title} {Surface code quantum
  computing with error rates over 1\%},\ }\href
  {https://doi.org/10.1103/PhysRevA.83.020302} {\bibfield  {journal} {\bibinfo
  {journal} {Phys. Rev. A}\ }\textbf {\bibinfo {volume} {83}},\ \bibinfo
  {pages} {020302} (\bibinfo {year} {2011})}\BibitemShut {NoStop}%
\bibitem [{\citenamefont {Hutter}\ \emph {et~al.}(2014)\citenamefont {Hutter},
  \citenamefont {Wootton},\ and\ \citenamefont {Loss}}]{PhysRevA.89.022326}%
  \BibitemOpen
  \bibfield  {author} {\bibinfo {author} {\bibfnamefont {A.}~\bibnamefont
  {Hutter}}, \bibinfo {author} {\bibfnamefont {J.~R.}\ \bibnamefont
  {Wootton}},\ and\ \bibinfo {author} {\bibfnamefont {D.}~\bibnamefont
  {Loss}},\ }\bibfield  {title} {\bibinfo {title} {Efficient markov chain monte
  carlo algorithm for the surface code},\ }\href
  {https://doi.org/10.1103/PhysRevA.89.022326} {\bibfield  {journal} {\bibinfo
  {journal} {Phys. Rev. A}\ }\textbf {\bibinfo {volume} {89}},\ \bibinfo
  {pages} {022326} (\bibinfo {year} {2014})}\BibitemShut {NoStop}%
\bibitem [{\citenamefont {Fowler}(2015)}]{DBLP:journals/qic/Fowler15a}%
  \BibitemOpen
  \bibfield  {author} {\bibinfo {author} {\bibfnamefont {A.~G.}\ \bibnamefont
  {Fowler}},\ }\bibfield  {title} {\bibinfo {title} {Minimum weight perfect
  matching of fault-tolerant topological quantum error correction in average
  {O(1)} parallel time},\ }\href {https://doi.org/10.26421/QIC15.1-2-9}
  {\bibfield  {journal} {\bibinfo  {journal} {Quantum Inf. Comput.}\ }\textbf
  {\bibinfo {volume} {15}},\ \bibinfo {pages} {145} (\bibinfo {year}
  {2015})}\BibitemShut {NoStop}%
\bibitem [{\citenamefont {Higgott}\ and\ \citenamefont
  {Gidney}(2025)}]{higgott2023sparse}%
  \BibitemOpen
  \bibfield  {author} {\bibinfo {author} {\bibfnamefont {O.}~\bibnamefont
  {Higgott}}\ and\ \bibinfo {author} {\bibfnamefont {C.}~\bibnamefont
  {Gidney}},\ }\bibfield  {title} {\bibinfo {title} {Sparse {B}lossom:
  correcting a million errors per core second with minimum-weight matching},\
  }\href {https://doi.org/10.22331/q-2025-01-20-1600} {\bibfield  {journal}
  {\bibinfo  {journal} {{Quantum}}\ }\textbf {\bibinfo {volume} {9}},\ \bibinfo
  {pages} {1600} (\bibinfo {year} {2025})}\BibitemShut {NoStop}%
\bibitem [{\citenamefont {Wu}\ and\ \citenamefont
  {Zhong}(2023)}]{wu2023fusion}%
  \BibitemOpen
  \bibfield  {author} {\bibinfo {author} {\bibfnamefont {Y.}~\bibnamefont
  {Wu}}\ and\ \bibinfo {author} {\bibfnamefont {L.}~\bibnamefont {Zhong}},\
  }\href@noop {} {\bibinfo {title} {Fusion blossom: Fast mwpm decoders for
  qec}} (\bibinfo {year} {2023}),\ \Eprint {https://arxiv.org/abs/2305.08307}
  {arXiv:2305.08307 [quant-ph]} \BibitemShut {NoStop}%
\bibitem [{\citenamefont {Delfosse}(2014)}]{PhysRevA.89.012317}%
  \BibitemOpen
  \bibfield  {author} {\bibinfo {author} {\bibfnamefont {N.}~\bibnamefont
  {Delfosse}},\ }\bibfield  {title} {\bibinfo {title} {Decoding color codes by
  projection onto surface codes},\ }\href
  {https://doi.org/10.1103/PhysRevA.89.012317} {\bibfield  {journal} {\bibinfo
  {journal} {Phys. Rev. A}\ }\textbf {\bibinfo {volume} {89}},\ \bibinfo
  {pages} {012317} (\bibinfo {year} {2014})}\BibitemShut {NoStop}%
\bibitem [{\citenamefont {Sahay}\ and\ \citenamefont
  {Brown}(2022)}]{PRXQuantum.3.010310}%
  \BibitemOpen
  \bibfield  {author} {\bibinfo {author} {\bibfnamefont {K.}~\bibnamefont
  {Sahay}}\ and\ \bibinfo {author} {\bibfnamefont {B.~J.}\ \bibnamefont
  {Brown}},\ }\bibfield  {title} {\bibinfo {title} {Decoder for the triangular
  color code by matching on a m\"obius strip},\ }\href
  {https://doi.org/10.1103/PRXQuantum.3.010310} {\bibfield  {journal} {\bibinfo
   {journal} {PRX Quantum}\ }\textbf {\bibinfo {volume} {3}},\ \bibinfo {pages}
  {010310} (\bibinfo {year} {2022})}\BibitemShut {NoStop}%
\bibitem [{\citenamefont {Kubica}\ and\ \citenamefont
  {Delfosse}(2023)}]{Kubica2023efficientcolorcode}%
  \BibitemOpen
  \bibfield  {author} {\bibinfo {author} {\bibfnamefont {A.}~\bibnamefont
  {Kubica}}\ and\ \bibinfo {author} {\bibfnamefont {N.}~\bibnamefont
  {Delfosse}},\ }\bibfield  {title} {\bibinfo {title} {Efficient color code
  decoders in {$d\geq 2$} dimensions from toric code decoders},\ }\href
  {https://doi.org/10.22331/q-2023-02-21-929} {\bibfield  {journal} {\bibinfo
  {journal} {{Quantum}}\ }\textbf {\bibinfo {volume} {7}},\ \bibinfo {pages}
  {929} (\bibinfo {year} {2023})}\BibitemShut {NoStop}%
\bibitem [{\citenamefont {Edmonds}(1965{\natexlab{a}})}]{Edmonds_1965}%
  \BibitemOpen
  \bibfield  {author} {\bibinfo {author} {\bibfnamefont {J.}~\bibnamefont
  {Edmonds}},\ }\bibfield  {title} {\bibinfo {title} {Paths, trees, and
  flowers},\ }\href {https://doi.org/10.4153/CJM-1965-045-4} {\bibfield
  {journal} {\bibinfo  {journal} {Canadian Journal of Mathematics}\ }\textbf
  {\bibinfo {volume} {17}},\ \bibinfo {pages} {449–467} (\bibinfo {year}
  {1965}{\natexlab{a}})}\BibitemShut {NoStop}%
\bibitem [{\citenamefont {Edmonds}(1965{\natexlab{b}})}]{edmonds1965maximum}%
  \BibitemOpen
  \bibfield  {author} {\bibinfo {author} {\bibfnamefont {J.}~\bibnamefont
  {Edmonds}},\ }\bibfield  {title} {\bibinfo {title} {Maximum matching and a
  polyhedron with 0, 1-vertices},\ }\href
  {https://doi.org/10.6028/jres.069b.013} {\bibfield  {journal} {\bibinfo
  {journal} {Journal of research of the National Bureau of Standards B}\
  }\textbf {\bibinfo {volume} {69}},\ \bibinfo {pages} {55} (\bibinfo {year}
  {1965}{\natexlab{b}})}\BibitemShut {NoStop}%
\bibitem [{\citenamefont {Vazirani}\ and\ \citenamefont
  {Vazirani}(1987)}]{vazirani1987matching}%
  \BibitemOpen
  \bibfield  {author} {\bibinfo {author} {\bibfnamefont {K.~M.~U.}\
  \bibnamefont {Vazirani}}\ and\ \bibinfo {author} {\bibfnamefont
  {V.}~\bibnamefont {Vazirani}},\ }\bibfield  {title} {\bibinfo {title}
  {Matching is as easy as matrix inversion},\ }\href
  {https://link.springer.com/article/10.1007/BF02579206} {\bibfield  {journal}
  {\bibinfo  {journal} {Combinatorica}\ }\textbf {\bibinfo {volume} {7}},\
  \bibinfo {pages} {105} (\bibinfo {year} {1987})}\BibitemShut {NoStop}%
\bibitem [{\citenamefont {Dahlhaus}\ and\ \citenamefont
  {Karpinski}(1998)}]{DAHLHAUS199879}%
  \BibitemOpen
  \bibfield  {author} {\bibinfo {author} {\bibfnamefont {E.}~\bibnamefont
  {Dahlhaus}}\ and\ \bibinfo {author} {\bibfnamefont {M.}~\bibnamefont
  {Karpinski}},\ }\bibfield  {title} {\bibinfo {title} {Matching and
  multidimensional matching in chordal and strongly chordal graphs},\ }\href
  {https://doi.org/https://doi.org/10.1016/S0166-218X(98)00006-7} {\bibfield
  {journal} {\bibinfo  {journal} {Discrete Applied Mathematics}\ }\textbf
  {\bibinfo {volume} {84}},\ \bibinfo {pages} {79} (\bibinfo {year}
  {1998})}\BibitemShut {NoStop}%
\bibitem [{\citenamefont {Grigoriev}\ and\ \citenamefont
  {Karpinski}(1987)}]{4568269}%
  \BibitemOpen
  \bibfield  {author} {\bibinfo {author} {\bibfnamefont {D.~Y.}\ \bibnamefont
  {Grigoriev}}\ and\ \bibinfo {author} {\bibfnamefont {M.}~\bibnamefont
  {Karpinski}},\ }\bibfield  {title} {\bibinfo {title} {The matching problem
  for bipartite graphs with polynomially bounded permanents is in nc},\ }in\
  \href {https://doi.org/10.1109/SFCS.1987.56} {\emph {\bibinfo {booktitle}
  {28th Annual Symposium on Foundations of Computer Science (sfcs 1987)}}}\
  (\bibinfo {year} {1987})\ pp.\ \bibinfo {pages} {166--172}\BibitemShut
  {NoStop}%
\bibitem [{\citenamefont {Agrawal}\ \emph {et~al.}(2007)\citenamefont
  {Agrawal}, \citenamefont {Hoang},\ and\ \citenamefont
  {Thierauf}}]{10.1007/978-3-540-70918-3_42}%
  \BibitemOpen
  \bibfield  {author} {\bibinfo {author} {\bibfnamefont {M.}~\bibnamefont
  {Agrawal}}, \bibinfo {author} {\bibfnamefont {T.~M.}\ \bibnamefont {Hoang}},\
  and\ \bibinfo {author} {\bibfnamefont {T.}~\bibnamefont {Thierauf}},\
  }\bibfield  {title} {\bibinfo {title} {The polynomially bounded perfect
  matching problem is in nc2},\ }in\ \href
  {https://link.springer.com/chapter/10.1007/978-3-540-70918-3_42} {\emph
  {\bibinfo {booktitle} {STACS 2007}}},\ \bibinfo {editor} {edited by\ \bibinfo
  {editor} {\bibfnamefont {W.}~\bibnamefont {Thomas}}\ and\ \bibinfo {editor}
  {\bibfnamefont {P.}~\bibnamefont {Weil}}}\ (\bibinfo  {publisher} {Springer
  Berlin Heidelberg},\ \bibinfo {address} {Berlin, Heidelberg},\ \bibinfo
  {year} {2007})\ pp.\ \bibinfo {pages} {489--499}\BibitemShut {NoStop}%
\bibitem [{\citenamefont {Datta}\ \emph {et~al.}(2010)\citenamefont {Datta},
  \citenamefont {Kulkarni},\ and\ \citenamefont
  {Roy}}]{datta2010deterministically}%
  \BibitemOpen
  \bibfield  {author} {\bibinfo {author} {\bibfnamefont {S.}~\bibnamefont
  {Datta}}, \bibinfo {author} {\bibfnamefont {R.}~\bibnamefont {Kulkarni}},\
  and\ \bibinfo {author} {\bibfnamefont {S.}~\bibnamefont {Roy}},\ }\bibfield
  {title} {\bibinfo {title} {Deterministically isolating a perfect matching in
  bipartite planar graphs},\ }\href
  {https://link.springer.com/article/10.1007/s00224-009-9204-8} {\bibfield
  {journal} {\bibinfo  {journal} {Theory of Computing Systems}\ }\textbf
  {\bibinfo {volume} {47}},\ \bibinfo {pages} {737} (\bibinfo {year}
  {2010})}\BibitemShut {NoStop}%
\bibitem [{\citenamefont {Tewari}\ and\ \citenamefont
  {Vinodchandran}(2012)}]{TEWARI20121}%
  \BibitemOpen
  \bibfield  {author} {\bibinfo {author} {\bibfnamefont {R.}~\bibnamefont
  {Tewari}}\ and\ \bibinfo {author} {\bibfnamefont {N.}~\bibnamefont
  {Vinodchandran}},\ }\bibfield  {title} {\bibinfo {title} {Green's theorem and
  isolation in planar graphs},\ }\href
  {https://doi.org/https://doi.org/10.1016/j.ic.2012.03.002} {\bibfield
  {journal} {\bibinfo  {journal} {Information and Computation}\ }\textbf
  {\bibinfo {volume} {215}},\ \bibinfo {pages} {1} (\bibinfo {year}
  {2012})}\BibitemShut {NoStop}%
\bibitem [{\citenamefont {Fenner}\ \emph {et~al.}(2016)\citenamefont {Fenner},
  \citenamefont {Gurjar},\ and\ \citenamefont
  {Thierauf}}]{10.1145/2897518.2897564}%
  \BibitemOpen
  \bibfield  {author} {\bibinfo {author} {\bibfnamefont {S.}~\bibnamefont
  {Fenner}}, \bibinfo {author} {\bibfnamefont {R.}~\bibnamefont {Gurjar}},\
  and\ \bibinfo {author} {\bibfnamefont {T.}~\bibnamefont {Thierauf}},\
  }\bibfield  {title} {\bibinfo {title} {Bipartite perfect matching is in
  quasi-nc},\ }in\ \href {https://doi.org/10.1145/2897518.2897564} {\emph
  {\bibinfo {booktitle} {Proceedings of the Forty-Eighth Annual ACM Symposium
  on Theory of Computing}}},\ \bibinfo {series and number} {STOC '16}\
  (\bibinfo  {publisher} {Association for Computing Machinery},\ \bibinfo
  {address} {New York, NY, USA},\ \bibinfo {year} {2016})\ p.\ \bibinfo {pages}
  {754–763}\BibitemShut {NoStop}%
\bibitem [{\citenamefont {Goldwasser}\ and\ \citenamefont
  {Grossman}(2017)}]{goldwasser_et_al:LIPIcs.ICALP.2017.87}%
  \BibitemOpen
  \bibfield  {author} {\bibinfo {author} {\bibfnamefont {S.}~\bibnamefont
  {Goldwasser}}\ and\ \bibinfo {author} {\bibfnamefont {O.}~\bibnamefont
  {Grossman}},\ }\bibfield  {title} {\bibinfo {title} {{Bipartite Perfect
  Matching in Pseudo-Deterministic NC}},\ }in\ \href
  {https://doi.org/10.4230/LIPIcs.ICALP.2017.87} {\emph {\bibinfo {booktitle}
  {44th International Colloquium on Automata, Languages, and Programming (ICALP
  2017)}}},\ \bibinfo {series} {Leibniz International Proceedings in
  Informatics (LIPIcs)}, Vol.~\bibinfo {volume} {80},\ \bibinfo {editor}
  {edited by\ \bibinfo {editor} {\bibfnamefont {I.}~\bibnamefont
  {Chatzigiannakis}}, \bibinfo {editor} {\bibfnamefont {P.}~\bibnamefont
  {Indyk}}, \bibinfo {editor} {\bibfnamefont {F.}~\bibnamefont {Kuhn}},\ and\
  \bibinfo {editor} {\bibfnamefont {A.}~\bibnamefont {Muscholl}}}\ (\bibinfo
  {publisher} {Schloss Dagstuhl -- Leibniz-Zentrum f{\"u}r Informatik},\
  \bibinfo {address} {Dagstuhl, Germany},\ \bibinfo {year} {2017})\ pp.\
  \bibinfo {pages} {87:1--87:13}\BibitemShut {NoStop}%
\bibitem [{\citenamefont {Svensson}\ and\ \citenamefont
  {Tarnawski}(2017)}]{8104102}%
  \BibitemOpen
  \bibfield  {author} {\bibinfo {author} {\bibfnamefont {O.}~\bibnamefont
  {Svensson}}\ and\ \bibinfo {author} {\bibfnamefont {J.}~\bibnamefont
  {Tarnawski}},\ }\bibfield  {title} {\bibinfo {title} {The matching problem in
  general graphs is in quasi-nc},\ }in\ \href
  {https://doi.org/10.1109/FOCS.2017.70} {\emph {\bibinfo {booktitle} {2017
  IEEE 58th Annual Symposium on Foundations of Computer Science (FOCS)}}}\
  (\bibinfo {year} {2017})\ pp.\ \bibinfo {pages} {696--707}\BibitemShut
  {NoStop}%
\bibitem [{\citenamefont {Terhal}(2015)}]{T10}%
  \BibitemOpen
  \bibfield  {author} {\bibinfo {author} {\bibfnamefont {B.~M.}\ \bibnamefont
  {Terhal}},\ }\bibfield  {title} {\bibinfo {title} {Quantum error correction
  for quantum memories},\ }\href {https://doi.org/10.1103/RevModPhys.87.307}
  {\bibfield  {journal} {\bibinfo  {journal} {Rev. Mod. Phys.}\ }\textbf
  {\bibinfo {volume} {87}},\ \bibinfo {pages} {307} (\bibinfo {year}
  {2015})}\BibitemShut {NoStop}%
\bibitem [{\citenamefont {Skoric}\ \emph {et~al.}(2023)\citenamefont {Skoric},
  \citenamefont {Browne}, \citenamefont {Barnes}, \citenamefont {Gillespie},\
  and\ \citenamefont {Campbell}}]{https://doi.org/10.48550/arxiv.2209.08552}%
  \BibitemOpen
  \bibfield  {author} {\bibinfo {author} {\bibfnamefont {L.}~\bibnamefont
  {Skoric}}, \bibinfo {author} {\bibfnamefont {D.~E.}\ \bibnamefont {Browne}},
  \bibinfo {author} {\bibfnamefont {K.~M.}\ \bibnamefont {Barnes}}, \bibinfo
  {author} {\bibfnamefont {N.~I.}\ \bibnamefont {Gillespie}},\ and\ \bibinfo
  {author} {\bibfnamefont {E.~T.}\ \bibnamefont {Campbell}},\ }\bibfield
  {title} {\bibinfo {title} {Parallel window decoding enables scalable fault
  tolerant quantum computation},\ }\href
  {https://www.nature.com/articles/s41467-023-42482-1} {\bibfield  {journal}
  {\bibinfo  {journal} {Nature Communications}\ }\textbf {\bibinfo {volume}
  {14}},\ \bibinfo {pages} {7040} (\bibinfo {year} {2023})}\BibitemShut
  {NoStop}%
\bibitem [{\citenamefont {Tan}\ \emph {et~al.}(2023)\citenamefont {Tan},
  \citenamefont {Zhang}, \citenamefont {Chao}, \citenamefont {Shi},\ and\
  \citenamefont {Chen}}]{https://doi.org/10.48550/arxiv.2209.09219}%
  \BibitemOpen
  \bibfield  {author} {\bibinfo {author} {\bibfnamefont {X.}~\bibnamefont
  {Tan}}, \bibinfo {author} {\bibfnamefont {F.}~\bibnamefont {Zhang}}, \bibinfo
  {author} {\bibfnamefont {R.}~\bibnamefont {Chao}}, \bibinfo {author}
  {\bibfnamefont {Y.}~\bibnamefont {Shi}},\ and\ \bibinfo {author}
  {\bibfnamefont {J.}~\bibnamefont {Chen}},\ }\bibfield  {title} {\bibinfo
  {title} {Scalable surface-code decoders with parallelization in time},\
  }\href {https://doi.org/10.1103/PRXQuantum.4.040344} {\bibfield  {journal}
  {\bibinfo  {journal} {PRX Quantum}\ }\textbf {\bibinfo {volume} {4}},\
  \bibinfo {pages} {040344} (\bibinfo {year} {2023})}\BibitemShut {NoStop}%
\bibitem [{\citenamefont {Bombín}\ \emph {et~al.}(2023)\citenamefont
  {Bombín}, \citenamefont {Dawson}, \citenamefont {Liu}, \citenamefont
  {Nickerson}, \citenamefont {Pastawski},\ and\ \citenamefont
  {Roberts}}]{bombin2023modular}%
  \BibitemOpen
  \bibfield  {author} {\bibinfo {author} {\bibfnamefont {H.}~\bibnamefont
  {Bombín}}, \bibinfo {author} {\bibfnamefont {C.}~\bibnamefont {Dawson}},
  \bibinfo {author} {\bibfnamefont {Y.-H.}\ \bibnamefont {Liu}}, \bibinfo
  {author} {\bibfnamefont {N.}~\bibnamefont {Nickerson}}, \bibinfo {author}
  {\bibfnamefont {F.}~\bibnamefont {Pastawski}},\ and\ \bibinfo {author}
  {\bibfnamefont {S.}~\bibnamefont {Roberts}},\ }\href
  {https://arxiv.org/abs/2303.04846} {\bibinfo {title} {Modular decoding:
  parallelizable real-time decoding for quantum computers}} (\bibinfo {year}
  {2023})\BibitemShut {NoStop}%
\bibitem [{\citenamefont {Micali}\ and\ \citenamefont
  {Vazirani}(1980)}]{4567800}%
  \BibitemOpen
  \bibfield  {author} {\bibinfo {author} {\bibfnamefont {S.}~\bibnamefont
  {Micali}}\ and\ \bibinfo {author} {\bibfnamefont {V.~V.}\ \bibnamefont
  {Vazirani}},\ }\bibfield  {title} {\bibinfo {title} {An ${O}\left(\sqrt{|V|
  \cdot |E|}\right)$ algoithm for finding maximum matching in general graphs},\
  }in\ \href {https://doi.org/10.1109/SFCS.1980.12} {\emph {\bibinfo
  {booktitle} {21st Annual Symposium on Foundations of Computer Science (sfcs
  1980)}}}\ (\bibinfo {year} {1980})\ pp.\ \bibinfo {pages}
  {17--27}\BibitemShut {NoStop}%
\bibitem [{\citenamefont {Kolmogorov}(2009)}]{kolmogorov2009blossom}%
  \BibitemOpen
  \bibfield  {author} {\bibinfo {author} {\bibfnamefont {V.}~\bibnamefont
  {Kolmogorov}},\ }\bibfield  {title} {\bibinfo {title} {Blossom v: a new
  implementation of a minimum cost perfect matching algorithm},\ }\href
  {https://link.springer.com/article/10.1007/s12532-009-0002-8} {\bibfield
  {journal} {\bibinfo  {journal} {Mathematical Programming Computation}\
  }\textbf {\bibinfo {volume} {1}},\ \bibinfo {pages} {43} (\bibinfo {year}
  {2009})}\BibitemShut {NoStop}%
\bibitem [{\citenamefont {Dezső}\ \emph {et~al.}(2011)\citenamefont {Dezső},
  \citenamefont {Jüttner},\ and\ \citenamefont {Kovács}}]{DEZSO201123}%
  \BibitemOpen
  \bibfield  {author} {\bibinfo {author} {\bibfnamefont {B.}~\bibnamefont
  {Dezső}}, \bibinfo {author} {\bibfnamefont {A.}~\bibnamefont {Jüttner}},\
  and\ \bibinfo {author} {\bibfnamefont {P.}~\bibnamefont {Kovács}},\
  }\bibfield  {title} {\bibinfo {title} {Lemon – an open source c++ graph
  template library},\ }\href
  {https://doi.org/https://doi.org/10.1016/j.entcs.2011.06.003} {\bibfield
  {journal} {\bibinfo  {journal} {Electronic Notes in Theoretical Computer
  Science}\ }\textbf {\bibinfo {volume} {264}},\ \bibinfo {pages} {23}
  (\bibinfo {year} {2011})},\ \bibinfo {note} {proceedings of the Second
  Workshop on Generative Technologies (WGT) 2010}\BibitemShut {NoStop}%
\bibitem [{\citenamefont {Fowler}(2013)}]{fowler2013optimal}%
  \BibitemOpen
  \bibfield  {author} {\bibinfo {author} {\bibfnamefont {A.~G.}\ \bibnamefont
  {Fowler}},\ }\href@noop {} {\bibinfo {title} {Optimal complexity correction
  of correlated errors in the surface code}} (\bibinfo {year} {2013}),\ \Eprint
  {https://arxiv.org/abs/1310.0863} {arXiv:1310.0863 [quant-ph]} \BibitemShut
  {NoStop}%
\bibitem [{\citenamefont {Dijkstra}(1959)}]{dijkstra1959note}%
  \BibitemOpen
  \bibfield  {author} {\bibinfo {author} {\bibfnamefont {E.~W.}\ \bibnamefont
  {Dijkstra}},\ }\bibfield  {title} {\bibinfo {title} {A note on two problems
  in connexion with graphs.},\ }\href
  {https://doi.org/https://doi.org/10.1007/BF01386390} {\bibfield  {journal}
  {\bibinfo  {journal} {Numerische Mathematik}\ }\textbf {\bibinfo {volume}
  {1}},\ \bibinfo {pages} {269} (\bibinfo {year} {1959})}\BibitemShut {NoStop}%
\bibitem [{\citenamefont {Fredman}\ and\ \citenamefont
  {Tarjan}(1987)}]{10.1145/28869.28874}%
  \BibitemOpen
  \bibfield  {author} {\bibinfo {author} {\bibfnamefont {M.~L.}\ \bibnamefont
  {Fredman}}\ and\ \bibinfo {author} {\bibfnamefont {R.~E.}\ \bibnamefont
  {Tarjan}},\ }\bibfield  {title} {\bibinfo {title} {Fibonacci heaps and their
  uses in improved network optimization algorithms},\ }\href
  {https://doi.org/10.1145/28869.28874} {\bibfield  {journal} {\bibinfo
  {journal} {J. ACM}\ }\textbf {\bibinfo {volume} {34}},\ \bibinfo {pages}
  {596–615} (\bibinfo {year} {1987})}\BibitemShut {NoStop}%
\bibitem [{\citenamefont {Haeupler}\ \emph {et~al.}(2024)\citenamefont
  {Haeupler}, \citenamefont {Hladik}, \citenamefont {Rozhon}, \citenamefont
  {Tarjan},\ and\ \citenamefont {Tetek}}]{10756107}%
  \BibitemOpen
  \bibfield  {author} {\bibinfo {author} {\bibfnamefont {B.}~\bibnamefont
  {Haeupler}}, \bibinfo {author} {\bibfnamefont {R.}~\bibnamefont {Hladik}},
  \bibinfo {author} {\bibfnamefont {V.}~\bibnamefont {Rozhon}}, \bibinfo
  {author} {\bibfnamefont {R.~E.}\ \bibnamefont {Tarjan}},\ and\ \bibinfo
  {author} {\bibfnamefont {J.}~\bibnamefont {Tetek}},\ }\bibfield  {title}
  {\bibinfo {title} {{ Universal Optimality of Dijkstra Via Beyond-Worst-Case
  Heaps }},\ }in\ \href {https://doi.org/10.1109/FOCS61266.2024.00125} {\emph
  {\bibinfo {booktitle} {2024 IEEE 65th Annual Symposium on Foundations of
  Computer Science (FOCS)}}}\ (\bibinfo  {publisher} {IEEE Computer Society},\
  \bibinfo {address} {Los Alamitos, CA, USA},\ \bibinfo {year} {2024})\ pp.\
  \bibinfo {pages} {2099--2130}\BibitemShut {NoStop}%
\bibitem [{\citenamefont {Das}\ \emph {et~al.}(2022)\citenamefont {Das},
  \citenamefont {Locharla},\ and\ \citenamefont
  {Jones}}]{10.1145/3503222.3507707}%
  \BibitemOpen
  \bibfield  {author} {\bibinfo {author} {\bibfnamefont {P.}~\bibnamefont
  {Das}}, \bibinfo {author} {\bibfnamefont {A.}~\bibnamefont {Locharla}},\ and\
  \bibinfo {author} {\bibfnamefont {C.}~\bibnamefont {Jones}},\ }\bibfield
  {title} {\bibinfo {title} {Lilliput: a lightweight low-latency lookup-table
  decoder for near-term quantum error correction},\ }in\ \href
  {https://doi.org/10.1145/3503222.3507707} {\emph {\bibinfo {booktitle}
  {Proceedings of the 27th ACM International Conference on Architectural
  Support for Programming Languages and Operating Systems}}},\ \bibinfo {series
  and number} {ASPLOS '22}\ (\bibinfo  {publisher} {Association for Computing
  Machinery},\ \bibinfo {address} {New York, NY, USA},\ \bibinfo {year}
  {2022})\ p.\ \bibinfo {pages} {541–553}\BibitemShut {NoStop}%
\bibitem [{\citenamefont {Liao}\ \emph {et~al.}(2023)\citenamefont {Liao},
  \citenamefont {Suzuki}, \citenamefont {Tanimoto}, \citenamefont {Ueno},\ and\
  \citenamefont {Tokunaga}}]{liao2023wit}%
  \BibitemOpen
  \bibfield  {author} {\bibinfo {author} {\bibfnamefont {W.}~\bibnamefont
  {Liao}}, \bibinfo {author} {\bibfnamefont {Y.}~\bibnamefont {Suzuki}},
  \bibinfo {author} {\bibfnamefont {T.}~\bibnamefont {Tanimoto}}, \bibinfo
  {author} {\bibfnamefont {Y.}~\bibnamefont {Ueno}},\ and\ \bibinfo {author}
  {\bibfnamefont {Y.}~\bibnamefont {Tokunaga}},\ }\bibfield  {title} {\bibinfo
  {title} {Wit-greedy: hardware system design of weighted iterative greedy
  decoder for surface code},\ }in\ \href
  {https://doi.org/https://doi.org/10.1145/3566097.3567933} {\emph {\bibinfo
  {booktitle} {Proceedings of the 28th Asia and South Pacific Design Automation
  Conference}}}\ (\bibinfo {year} {2023})\ pp.\ \bibinfo {pages}
  {209--215}\BibitemShut {NoStop}%
\bibitem [{\citenamefont {Berkowitz}(1984)}]{BERKOWITZ1984147}%
  \BibitemOpen
  \bibfield  {author} {\bibinfo {author} {\bibfnamefont {S.~J.}\ \bibnamefont
  {Berkowitz}},\ }\bibfield  {title} {\bibinfo {title} {On computing the
  determinant in small parallel time using a small number of processors},\
  }\href {https://doi.org/https://doi.org/10.1016/0020-0190(84)90018-8}
  {\bibfield  {journal} {\bibinfo  {journal} {Information Processing Letters}\
  }\textbf {\bibinfo {volume} {18}},\ \bibinfo {pages} {147} (\bibinfo {year}
  {1984})}\BibitemShut {NoStop}%
\bibitem [{\citenamefont {Patterson}\ and\ \citenamefont
  {Hennessy}(2020)}]{patterson2020computer}%
  \BibitemOpen
  \bibfield  {author} {\bibinfo {author} {\bibfnamefont {D.}~\bibnamefont
  {Patterson}}\ and\ \bibinfo {author} {\bibfnamefont {J.}~\bibnamefont
  {Hennessy}},\ }\href
  {https://shop.elsevier.com/books/computer-organization-and-design-risc-v-edition/patterson/978-0-12-820331-6}
  {\emph {\bibinfo {title} {Computer Organization and Design RISC-V Edition:
  The Hardware Software Interface}}},\ The Morgan Kaufmann Series in Computer
  Architecture and Design\ (\bibinfo  {publisher} {Morgan Kaufmann},\ \bibinfo
  {year} {2020})\BibitemShut {NoStop}%
\bibitem [{\citenamefont {Bunimov}\ and\ \citenamefont
  {Schimmler}(2003)}]{10.1007/978-3-540-45209-6_127}%
  \BibitemOpen
  \bibfield  {author} {\bibinfo {author} {\bibfnamefont {V.}~\bibnamefont
  {Bunimov}}\ and\ \bibinfo {author} {\bibfnamefont {M.}~\bibnamefont
  {Schimmler}},\ }\bibfield  {title} {\bibinfo {title} {Efficient parallel
  multiplication algorithm for large integers},\ }in\ \href
  {https://doi.org/https://doi.org/10.1007/978-3-540-45209-6_127} {\emph
  {\bibinfo {booktitle} {Euro-Par 2003 Parallel Processing}}},\ \bibinfo
  {editor} {edited by\ \bibinfo {editor} {\bibfnamefont {H.}~\bibnamefont
  {Kosch}}, \bibinfo {editor} {\bibfnamefont {L.}~\bibnamefont
  {B{\"o}sz{\"o}rm{\'e}nyi}},\ and\ \bibinfo {editor} {\bibfnamefont
  {H.}~\bibnamefont {Hellwagner}}}\ (\bibinfo  {publisher} {Springer Berlin
  Heidelberg},\ \bibinfo {address} {Berlin, Heidelberg},\ \bibinfo {year}
  {2003})\ pp.\ \bibinfo {pages} {923--928}\BibitemShut {NoStop}%
\bibitem [{\citenamefont {Matsumoto}\ and\ \citenamefont
  {Nishimura}(1998)}]{10.1145/272991.272995}%
  \BibitemOpen
  \bibfield  {author} {\bibinfo {author} {\bibfnamefont {M.}~\bibnamefont
  {Matsumoto}}\ and\ \bibinfo {author} {\bibfnamefont {T.}~\bibnamefont
  {Nishimura}},\ }\bibfield  {title} {\bibinfo {title} {Mersenne twister: a
  623-dimensionally equidistributed uniform pseudo-random number generator},\
  }\href {https://doi.org/10.1145/272991.272995} {\bibfield  {journal}
  {\bibinfo  {journal} {ACM Trans. Model. Comput. Simul.}\ }\textbf {\bibinfo
  {volume} {8}},\ \bibinfo {pages} {3–30} (\bibinfo {year}
  {1998})}\BibitemShut {NoStop}%
\bibitem [{\citenamefont {Wills}\ \emph {et~al.}(2024)\citenamefont {Wills},
  \citenamefont {Hsieh},\ and\ \citenamefont
  {Yamasaki}}]{wills2024constantoverheadmagicstatedistillation}%
  \BibitemOpen
  \bibfield  {author} {\bibinfo {author} {\bibfnamefont {A.}~\bibnamefont
  {Wills}}, \bibinfo {author} {\bibfnamefont {M.-H.}\ \bibnamefont {Hsieh}},\
  and\ \bibinfo {author} {\bibfnamefont {H.}~\bibnamefont {Yamasaki}},\ }\href
  {https://arxiv.org/abs/2408.07764} {\bibinfo {title} {Constant-overhead magic
  state distillation}} (\bibinfo {year} {2024}),\ \Eprint
  {https://arxiv.org/abs/2408.07764} {arXiv:2408.07764 [quant-ph]} \BibitemShut
  {NoStop}%
\bibitem [{\citenamefont {Beverland}\ \emph {et~al.}(2021)\citenamefont
  {Beverland}, \citenamefont {Kubica},\ and\ \citenamefont
  {Svore}}]{PRXQuantum.2.020341}%
  \BibitemOpen
  \bibfield  {author} {\bibinfo {author} {\bibfnamefont {M.~E.}\ \bibnamefont
  {Beverland}}, \bibinfo {author} {\bibfnamefont {A.}~\bibnamefont {Kubica}},\
  and\ \bibinfo {author} {\bibfnamefont {K.~M.}\ \bibnamefont {Svore}},\
  }\bibfield  {title} {\bibinfo {title} {Cost of universality: A comparative
  study of the overhead of state distillation and code switching with color
  codes},\ }\href {https://doi.org/10.1103/PRXQuantum.2.020341} {\bibfield
  {journal} {\bibinfo  {journal} {PRX Quantum}\ }\textbf {\bibinfo {volume}
  {2}},\ \bibinfo {pages} {020341} (\bibinfo {year} {2021})}\BibitemShut
  {NoStop}%
\bibitem [{\citenamefont {Harrington}(2004)}]{Harrington_phd}%
  \BibitemOpen
  \bibfield  {author} {\bibinfo {author} {\bibfnamefont {J.~W.}\ \bibnamefont
  {Harrington}},\ }\emph {\bibinfo {title} {Analysis of quantum
  error-correcting codes: symplectic lattice codes and toric codes}},\ \href
  {https://thesis.library.caltech.edu/1747/} {Ph.D. thesis},\ \bibinfo
  {school} {California Institute of Technology} (\bibinfo {year}
  {2004})\BibitemShut {NoStop}%
\bibitem [{\citenamefont {Kubica}\ and\ \citenamefont
  {Preskill}(2019)}]{PhysRevLett.123.020501}%
  \BibitemOpen
  \bibfield  {author} {\bibinfo {author} {\bibfnamefont {A.}~\bibnamefont
  {Kubica}}\ and\ \bibinfo {author} {\bibfnamefont {J.}~\bibnamefont
  {Preskill}},\ }\bibfield  {title} {\bibinfo {title} {Cellular-automaton
  decoders with provable thresholds for topological codes},\ }\href
  {https://doi.org/10.1103/PhysRevLett.123.020501} {\bibfield  {journal}
  {\bibinfo  {journal} {Phys. Rev. Lett.}\ }\textbf {\bibinfo {volume} {123}},\
  \bibinfo {pages} {020501} (\bibinfo {year} {2019})}\BibitemShut {NoStop}%
\bibitem [{\citenamefont {Vasmer}\ \emph {et~al.}(2021)\citenamefont {Vasmer},
  \citenamefont {Browne},\ and\ \citenamefont {Kubica}}]{vasmer2021cellular}%
  \BibitemOpen
  \bibfield  {author} {\bibinfo {author} {\bibfnamefont {M.}~\bibnamefont
  {Vasmer}}, \bibinfo {author} {\bibfnamefont {D.~E.}\ \bibnamefont {Browne}},\
  and\ \bibinfo {author} {\bibfnamefont {A.}~\bibnamefont {Kubica}},\
  }\bibfield  {title} {\bibinfo {title} {Cellular automaton decoders for
  topological quantum codes with noisy measurements and beyond},\ }\href
  {https://www.nature.com/articles/s41598-021-81138-2} {\bibfield  {journal}
  {\bibinfo  {journal} {Scientific reports}\ }\textbf {\bibinfo {volume}
  {11}},\ \bibinfo {pages} {2027} (\bibinfo {year} {2021})}\BibitemShut
  {NoStop}%
\bibitem [{\citenamefont {Bombin}\ \emph {et~al.}(2013)\citenamefont {Bombin},
  \citenamefont {Chhajlany}, \citenamefont {Horodecki},\ and\ \citenamefont
  {Martin-Delgado}}]{Bombin_2013_self_correcting}%
  \BibitemOpen
  \bibfield  {author} {\bibinfo {author} {\bibfnamefont {H.}~\bibnamefont
  {Bombin}}, \bibinfo {author} {\bibfnamefont {R.~W.}\ \bibnamefont
  {Chhajlany}}, \bibinfo {author} {\bibfnamefont {M.}~\bibnamefont
  {Horodecki}},\ and\ \bibinfo {author} {\bibfnamefont {M.~A.}\ \bibnamefont
  {Martin-Delgado}},\ }\bibfield  {title} {\bibinfo {title} {Self-correcting
  quantum computers},\ }\href {https://doi.org/10.1088/1367-2630/15/5/055023}
  {\bibfield  {journal} {\bibinfo  {journal} {New Journal of Physics}\ }\textbf
  {\bibinfo {volume} {15}},\ \bibinfo {pages} {055023} (\bibinfo {year}
  {2013})}\BibitemShut {NoStop}%
\bibitem [{\citenamefont {Battistel}\ \emph {et~al.}(2023)\citenamefont
  {Battistel}, \citenamefont {Chamberland}, \citenamefont {Johar},
  \citenamefont {Overwater}, \citenamefont {Sebastiano}, \citenamefont
  {Skoric}, \citenamefont {Ueno},\ and\ \citenamefont
  {Usman}}]{battistel2023real}%
  \BibitemOpen
  \bibfield  {author} {\bibinfo {author} {\bibfnamefont {F.}~\bibnamefont
  {Battistel}}, \bibinfo {author} {\bibfnamefont {C.}~\bibnamefont
  {Chamberland}}, \bibinfo {author} {\bibfnamefont {K.}~\bibnamefont {Johar}},
  \bibinfo {author} {\bibfnamefont {R.~W.}\ \bibnamefont {Overwater}}, \bibinfo
  {author} {\bibfnamefont {F.}~\bibnamefont {Sebastiano}}, \bibinfo {author}
  {\bibfnamefont {L.}~\bibnamefont {Skoric}}, \bibinfo {author} {\bibfnamefont
  {Y.}~\bibnamefont {Ueno}},\ and\ \bibinfo {author} {\bibfnamefont
  {M.}~\bibnamefont {Usman}},\ }\bibfield  {title} {\bibinfo {title} {Real-time
  decoding for fault-tolerant quantum computing: Progress, challenges and
  outlook},\ }\href
  {https://iopscience.iop.org/article/10.1088/2399-1984/aceba6} {\bibfield
  {journal} {\bibinfo  {journal} {Nano Futures}\ }\textbf {\bibinfo {volume}
  {7}},\ \bibinfo {pages} {032003} (\bibinfo {year} {2023})}\BibitemShut
  {NoStop}%
\bibitem [{\citenamefont {Aliferis}\ \emph {et~al.}(2006)\citenamefont
  {Aliferis}, \citenamefont {Gottesman},\ and\ \citenamefont
  {Preskill}}]{10.5555/2011665.2011666}%
  \BibitemOpen
  \bibfield  {author} {\bibinfo {author} {\bibfnamefont {P.}~\bibnamefont
  {Aliferis}}, \bibinfo {author} {\bibfnamefont {D.}~\bibnamefont
  {Gottesman}},\ and\ \bibinfo {author} {\bibfnamefont {J.}~\bibnamefont
  {Preskill}},\ }\bibfield  {title} {\bibinfo {title} {Quantum accuracy
  threshold for concatenated distance-3 codes},\ }\href
  {https://doi.org/10.26421/QIC6.2-1} {\bibfield  {journal} {\bibinfo
  {journal} {Quantum Info. Comput.}\ }\textbf {\bibinfo {volume} {6}},\
  \bibinfo {pages} {97–165} (\bibinfo {year} {2006})}\BibitemShut {NoStop}%
\bibitem [{\citenamefont {Horsman}\ \emph {et~al.}(2012)\citenamefont
  {Horsman}, \citenamefont {Fowler}, \citenamefont {Devitt},\ and\
  \citenamefont {Van~Meter}}]{horsman2012surface}%
  \BibitemOpen
  \bibfield  {author} {\bibinfo {author} {\bibfnamefont {D.}~\bibnamefont
  {Horsman}}, \bibinfo {author} {\bibfnamefont {A.~G.}\ \bibnamefont {Fowler}},
  \bibinfo {author} {\bibfnamefont {S.}~\bibnamefont {Devitt}},\ and\ \bibinfo
  {author} {\bibfnamefont {R.}~\bibnamefont {Van~Meter}},\ }\bibfield  {title}
  {\bibinfo {title} {Surface code quantum computing by lattice surgery},\
  }\href {https://iopscience.iop.org/article/10.1088/1367-2630/14/12/123011}
  {\bibfield  {journal} {\bibinfo  {journal} {New Journal of Physics}\ }\textbf
  {\bibinfo {volume} {14}},\ \bibinfo {pages} {123011} (\bibinfo {year}
  {2012})}\BibitemShut {NoStop}%
\bibitem [{\citenamefont {Litinski}(2019)}]{Litinski2019gameofsurfacecodes}%
  \BibitemOpen
  \bibfield  {author} {\bibinfo {author} {\bibfnamefont {D.}~\bibnamefont
  {Litinski}},\ }\bibfield  {title} {\bibinfo {title} {A {G}ame of {S}urface
  {C}odes: {L}arge-{S}cale {Q}uantum {C}omputing with {L}attice {S}urgery},\
  }\href {https://doi.org/10.22331/q-2019-03-05-128} {\bibfield  {journal}
  {\bibinfo  {journal} {{Quantum}}\ }\textbf {\bibinfo {volume} {3}},\ \bibinfo
  {pages} {128} (\bibinfo {year} {2019})}\BibitemShut {NoStop}%
\bibitem [{\citenamefont {Gottesman}\ and\ \citenamefont
  {Chuang}(1999)}]{gottesmanchuang1999}%
  \BibitemOpen
  \bibfield  {author} {\bibinfo {author} {\bibfnamefont {D.}~\bibnamefont
  {Gottesman}}\ and\ \bibinfo {author} {\bibfnamefont {I.~L.}\ \bibnamefont
  {Chuang}},\ }\bibfield  {title} {\bibinfo {title} {Demonstrating the
  viability of universal quantum computation using teleportation and
  single-qubit operations},\ }\href {https://doi.org/10.1038/46503} {\bibfield
  {journal} {\bibinfo  {journal} {Nature}\ }\textbf {\bibinfo {volume} {402}},\
  \bibinfo {pages} {390 } (\bibinfo {year} {1999})}\BibitemShut {NoStop}%
\bibitem [{\citenamefont {Zhou}\ \emph {et~al.}(2000)\citenamefont {Zhou},
  \citenamefont {Leung},\ and\ \citenamefont {Chuang}}]{PhysRevA.62.052316}%
  \BibitemOpen
  \bibfield  {author} {\bibinfo {author} {\bibfnamefont {X.}~\bibnamefont
  {Zhou}}, \bibinfo {author} {\bibfnamefont {D.~W.}\ \bibnamefont {Leung}},\
  and\ \bibinfo {author} {\bibfnamefont {I.~L.}\ \bibnamefont {Chuang}},\
  }\bibfield  {title} {\bibinfo {title} {Methodology for quantum logic gate
  construction},\ }\href {https://doi.org/10.1103/PhysRevA.62.052316}
  {\bibfield  {journal} {\bibinfo  {journal} {Phys. Rev. A}\ }\textbf {\bibinfo
  {volume} {62}},\ \bibinfo {pages} {052316} (\bibinfo {year}
  {2000})}\BibitemShut {NoStop}%
\bibitem [{\citenamefont {Paetznick}\ and\ \citenamefont
  {Reichardt}(2013)}]{PhysRevLett.111.090505}%
  \BibitemOpen
  \bibfield  {author} {\bibinfo {author} {\bibfnamefont {A.}~\bibnamefont
  {Paetznick}}\ and\ \bibinfo {author} {\bibfnamefont {B.~W.}\ \bibnamefont
  {Reichardt}},\ }\bibfield  {title} {\bibinfo {title} {Universal
  fault-tolerant quantum computation with only transversal gates and error
  correction},\ }\href {https://doi.org/10.1103/PhysRevLett.111.090505}
  {\bibfield  {journal} {\bibinfo  {journal} {Phys. Rev. Lett.}\ }\textbf
  {\bibinfo {volume} {111}},\ \bibinfo {pages} {090505} (\bibinfo {year}
  {2013})}\BibitemShut {NoStop}%
\bibitem [{\citenamefont {Bravyi}\ \emph {et~al.}(2014)\citenamefont {Bravyi},
  \citenamefont {Suchara},\ and\ \citenamefont {Vargo}}]{PhysRevA.90.032326}%
  \BibitemOpen
  \bibfield  {author} {\bibinfo {author} {\bibfnamefont {S.}~\bibnamefont
  {Bravyi}}, \bibinfo {author} {\bibfnamefont {M.}~\bibnamefont {Suchara}},\
  and\ \bibinfo {author} {\bibfnamefont {A.}~\bibnamefont {Vargo}},\ }\bibfield
   {title} {\bibinfo {title} {Efficient algorithms for maximum likelihood
  decoding in the surface code},\ }\href
  {https://doi.org/10.1103/PhysRevA.90.032326} {\bibfield  {journal} {\bibinfo
  {journal} {Phys. Rev. A}\ }\textbf {\bibinfo {volume} {90}},\ \bibinfo
  {pages} {032326} (\bibinfo {year} {2014})}\BibitemShut {NoStop}%
\bibitem [{\citenamefont {Darmawan}\ \emph {et~al.}(2024)\citenamefont
  {Darmawan}, \citenamefont {Nakata}, \citenamefont {Tamiya},\ and\
  \citenamefont {Yamasaki}}]{PhysRevResearch.6.023055}%
  \BibitemOpen
  \bibfield  {author} {\bibinfo {author} {\bibfnamefont {A.~S.}\ \bibnamefont
  {Darmawan}}, \bibinfo {author} {\bibfnamefont {Y.}~\bibnamefont {Nakata}},
  \bibinfo {author} {\bibfnamefont {S.}~\bibnamefont {Tamiya}},\ and\ \bibinfo
  {author} {\bibfnamefont {H.}~\bibnamefont {Yamasaki}},\ }\bibfield  {title}
  {\bibinfo {title} {Low-depth random clifford circuits for quantum coding
  against pauli noise using a tensor-network decoder},\ }\href
  {https://doi.org/10.1103/PhysRevResearch.6.023055} {\bibfield  {journal}
  {\bibinfo  {journal} {Phys. Rev. Res.}\ }\textbf {\bibinfo {volume} {6}},\
  \bibinfo {pages} {023055} (\bibinfo {year} {2024})}\BibitemShut {NoStop}%
\bibitem [{\citenamefont {Leverrier}\ \emph {et~al.}(2015)\citenamefont
  {Leverrier}, \citenamefont {Tillich},\ and\ \citenamefont {Zemor}}]{7354429}%
  \BibitemOpen
  \bibfield  {author} {\bibinfo {author} {\bibfnamefont {A.}~\bibnamefont
  {Leverrier}}, \bibinfo {author} {\bibfnamefont {J.-P.}\ \bibnamefont
  {Tillich}},\ and\ \bibinfo {author} {\bibfnamefont {G.}~\bibnamefont
  {Zemor}},\ }\bibfield  {title} {\bibinfo {title} {Quantum expander codes},\
  }in\ \href {https://doi.org/10.1109/FOCS.2015.55} {\emph {\bibinfo
  {booktitle} {Proceedings of the 2015 IEEE 56th Annual Symposium on
  Foundations of Computer Science (FOCS)}}},\ \bibinfo {series and number}
  {FOCS '15}\ (\bibinfo  {publisher} {IEEE Computer Society},\ \bibinfo
  {address} {USA},\ \bibinfo {year} {2015})\ p.\ \bibinfo {pages}
  {810–824}\BibitemShut {NoStop}%
\bibitem [{\citenamefont {Fawzi}\ \emph
  {et~al.}(2018{\natexlab{b}})\citenamefont {Fawzi}, \citenamefont
  {Grospellier},\ and\ \citenamefont {Leverrier}}]{10.1145/3188745.3188886}%
  \BibitemOpen
  \bibfield  {author} {\bibinfo {author} {\bibfnamefont {O.}~\bibnamefont
  {Fawzi}}, \bibinfo {author} {\bibfnamefont {A.}~\bibnamefont {Grospellier}},\
  and\ \bibinfo {author} {\bibfnamefont {A.}~\bibnamefont {Leverrier}},\
  }\bibfield  {title} {\bibinfo {title} {Efficient decoding of random errors
  for quantum expander codes},\ }in\ \href
  {https://doi.org/10.1145/3188745.3188886} {\emph {\bibinfo {booktitle}
  {Proceedings of the 50th Annual ACM SIGACT Symposium on Theory of
  Computing}}},\ \bibinfo {series and number} {STOC 2018}\ (\bibinfo
  {publisher} {Association for Computing Machinery},\ \bibinfo {address} {New
  York, NY, USA},\ \bibinfo {year} {2018})\ p.\ \bibinfo {pages}
  {521–534}\BibitemShut {NoStop}%
\bibitem [{\citenamefont {Panteleev}\ and\ \citenamefont
  {Kalachev}(2022)}]{10.1145/3519935.3520017}%
  \BibitemOpen
  \bibfield  {author} {\bibinfo {author} {\bibfnamefont {P.}~\bibnamefont
  {Panteleev}}\ and\ \bibinfo {author} {\bibfnamefont {G.}~\bibnamefont
  {Kalachev}},\ }\bibfield  {title} {\bibinfo {title} {Asymptotically good
  quantum and locally testable classical ldpc codes},\ }in\ \href
  {https://doi.org/10.1145/3519935.3520017} {\emph {\bibinfo {booktitle}
  {Proceedings of the 54th Annual ACM SIGACT Symposium on Theory of
  Computing}}},\ \bibinfo {series and number} {STOC 2022}\ (\bibinfo
  {publisher} {Association for Computing Machinery},\ \bibinfo {address} {New
  York, NY, USA},\ \bibinfo {year} {2022})\ p.\ \bibinfo {pages}
  {375–388}\BibitemShut {NoStop}%
\bibitem [{\citenamefont {Leverrier}\ and\ \citenamefont
  {Zemor}(2022)}]{9996782}%
  \BibitemOpen
  \bibfield  {author} {\bibinfo {author} {\bibfnamefont {A.}~\bibnamefont
  {Leverrier}}\ and\ \bibinfo {author} {\bibfnamefont {G.}~\bibnamefont
  {Zemor}},\ }\bibfield  {title} {\bibinfo {title} {Quantum tanner codes},\
  }in\ \href {https://doi.org/10.1109/FOCS54457.2022.00117} {\emph {\bibinfo
  {booktitle} {2022 IEEE 63rd Annual Symposium on Foundations of Computer
  Science (FOCS)}}}\ (\bibinfo  {publisher} {IEEE Computer Society},\ \bibinfo
  {address} {Los Alamitos, CA, USA},\ \bibinfo {year} {2022})\ pp.\ \bibinfo
  {pages} {872--883}\BibitemShut {NoStop}%
\bibitem [{\citenamefont {Dinur}\ \emph {et~al.}(2023)\citenamefont {Dinur},
  \citenamefont {Hsieh}, \citenamefont {Lin},\ and\ \citenamefont
  {Vidick}}]{10.1145/3564246.3585101}%
  \BibitemOpen
  \bibfield  {author} {\bibinfo {author} {\bibfnamefont {I.}~\bibnamefont
  {Dinur}}, \bibinfo {author} {\bibfnamefont {M.-H.}\ \bibnamefont {Hsieh}},
  \bibinfo {author} {\bibfnamefont {T.-C.}\ \bibnamefont {Lin}},\ and\ \bibinfo
  {author} {\bibfnamefont {T.}~\bibnamefont {Vidick}},\ }\bibfield  {title}
  {\bibinfo {title} {Good quantum ldpc codes with linear time decoders},\ }in\
  \href {https://doi.org/10.1145/3564246.3585101} {\emph {\bibinfo {booktitle}
  {Proceedings of the 55th Annual ACM Symposium on Theory of Computing}}},\
  \bibinfo {series and number} {STOC 2023}\ (\bibinfo  {publisher} {Association
  for Computing Machinery},\ \bibinfo {address} {New York, NY, USA},\ \bibinfo
  {year} {2023})\ p.\ \bibinfo {pages} {905–918}\BibitemShut {NoStop}%
\bibitem [{\citenamefont {Gu}\ \emph {et~al.}(2024)\citenamefont {Gu},
  \citenamefont {Tang}, \citenamefont {Caha}, \citenamefont {Choe},
  \citenamefont {He},\ and\ \citenamefont {Kubica}}]{gu2024single}%
  \BibitemOpen
  \bibfield  {author} {\bibinfo {author} {\bibfnamefont {S.}~\bibnamefont
  {Gu}}, \bibinfo {author} {\bibfnamefont {E.}~\bibnamefont {Tang}}, \bibinfo
  {author} {\bibfnamefont {L.}~\bibnamefont {Caha}}, \bibinfo {author}
  {\bibfnamefont {S.~H.}\ \bibnamefont {Choe}}, \bibinfo {author}
  {\bibfnamefont {Z.}~\bibnamefont {He}},\ and\ \bibinfo {author}
  {\bibfnamefont {A.}~\bibnamefont {Kubica}},\ }\bibfield  {title} {\bibinfo
  {title} {Single-shot decoding of good quantum ldpc codes},\ }\href
  {https://link.springer.com/article/10.1007/s00220-024-04951-6} {\bibfield
  {journal} {\bibinfo  {journal} {Communications in Mathematical Physics}\
  }\textbf {\bibinfo {volume} {405}},\ \bibinfo {pages} {85} (\bibinfo {year}
  {2024})}\BibitemShut {NoStop}%
\bibitem [{\citenamefont {Delfosse}\ and\ \citenamefont
  {Nickerson}(2021)}]{Delfosse2021almostlineartime}%
  \BibitemOpen
  \bibfield  {author} {\bibinfo {author} {\bibfnamefont {N.}~\bibnamefont
  {Delfosse}}\ and\ \bibinfo {author} {\bibfnamefont {N.~H.}\ \bibnamefont
  {Nickerson}},\ }\bibfield  {title} {\bibinfo {title} {Almost-linear time
  decoding algorithm for topological codes},\ }\href
  {https://doi.org/10.22331/q-2021-12-02-595} {\bibfield  {journal} {\bibinfo
  {journal} {{Quantum}}\ }\textbf {\bibinfo {volume} {5}},\ \bibinfo {pages}
  {595} (\bibinfo {year} {2021})}\BibitemShut {NoStop}%
\bibitem [{\citenamefont {Panteleev}\ and\ \citenamefont
  {Kalachev}(2021)}]{Panteleev2021degeneratequantum}%
  \BibitemOpen
  \bibfield  {author} {\bibinfo {author} {\bibfnamefont {P.}~\bibnamefont
  {Panteleev}}\ and\ \bibinfo {author} {\bibfnamefont {G.}~\bibnamefont
  {Kalachev}},\ }\bibfield  {title} {\bibinfo {title} {Degenerate {Q}uantum
  {LDPC} {C}odes {W}ith {G}ood {F}inite {L}ength {P}erformance},\ }\href
  {https://doi.org/10.22331/q-2021-11-22-585} {\bibfield  {journal} {\bibinfo
  {journal} {{Quantum}}\ }\textbf {\bibinfo {volume} {5}},\ \bibinfo {pages}
  {585} (\bibinfo {year} {2021})}\BibitemShut {NoStop}%
\bibitem [{\citenamefont {Duclos-Cianci}\ and\ \citenamefont
  {Poulin}(2010)}]{PhysRevLett.104.050504}%
  \BibitemOpen
  \bibfield  {author} {\bibinfo {author} {\bibfnamefont {G.}~\bibnamefont
  {Duclos-Cianci}}\ and\ \bibinfo {author} {\bibfnamefont {D.}~\bibnamefont
  {Poulin}},\ }\bibfield  {title} {\bibinfo {title} {Fast decoders for
  topological quantum codes},\ }\href
  {https://doi.org/10.1103/PhysRevLett.104.050504} {\bibfield  {journal}
  {\bibinfo  {journal} {Phys. Rev. Lett.}\ }\textbf {\bibinfo {volume} {104}},\
  \bibinfo {pages} {050504} (\bibinfo {year} {2010})}\BibitemShut {NoStop}%
\bibitem [{\citenamefont {Steane}(1996)}]{S3}%
  \BibitemOpen
  \bibfield  {author} {\bibinfo {author} {\bibfnamefont {A.}~\bibnamefont
  {Steane}},\ }\bibfield  {title} {\bibinfo {title} {Multiple-particle
  interference and quantum error correction},\ }\href
  {https://doi.org/10.1098/rspa.1996.0136} {\bibfield  {journal} {\bibinfo
  {journal} {Proceedings of the Royal Society of London. Series A:
  Mathematical, Physical and Engineering Sciences}\ }\textbf {\bibinfo {volume}
  {452}},\ \bibinfo {pages} {2551} (\bibinfo {year} {1996})}\BibitemShut
  {NoStop}%
\bibitem [{\citenamefont {Moussa}(2016)}]{PhysRevA.94.042316}%
  \BibitemOpen
  \bibfield  {author} {\bibinfo {author} {\bibfnamefont {J.~E.}\ \bibnamefont
  {Moussa}},\ }\bibfield  {title} {\bibinfo {title} {Transversal clifford gates
  on folded surface codes},\ }\href
  {https://doi.org/10.1103/PhysRevA.94.042316} {\bibfield  {journal} {\bibinfo
  {journal} {Phys. Rev. A}\ }\textbf {\bibinfo {volume} {94}},\ \bibinfo
  {pages} {042316} (\bibinfo {year} {2016})}\BibitemShut {NoStop}%
\bibitem [{\citenamefont {Vasmer}\ and\ \citenamefont
  {Browne}(2019)}]{PhysRevA.100.012312}%
  \BibitemOpen
  \bibfield  {author} {\bibinfo {author} {\bibfnamefont {M.}~\bibnamefont
  {Vasmer}}\ and\ \bibinfo {author} {\bibfnamefont {D.~E.}\ \bibnamefont
  {Browne}},\ }\bibfield  {title} {\bibinfo {title} {Three-dimensional surface
  codes: Transversal gates and fault-tolerant architectures},\ }\href
  {https://doi.org/10.1103/PhysRevA.100.012312} {\bibfield  {journal} {\bibinfo
   {journal} {Phys. Rev. A}\ }\textbf {\bibinfo {volume} {100}},\ \bibinfo
  {pages} {012312} (\bibinfo {year} {2019})}\BibitemShut {NoStop}%
\bibitem [{\citenamefont {Iverson}(2020)}]{iverson2020aspects}%
  \BibitemOpen
  \bibfield  {author} {\bibinfo {author} {\bibfnamefont {J.~K.}\ \bibnamefont
  {Iverson}},\ }\emph {\bibinfo {title} {Aspects of Fault-Tolerant Quantum
  Computation}},\ \href {https://thesis.library.caltech.edu/13729/} {Ph.D.
  thesis},\ \bibinfo  {school} {California Institute of Technology} (\bibinfo
  {year} {2020})\BibitemShut {NoStop}%
\bibitem [{\citenamefont {Higgott}(2022)}]{higgott2022pymatching}%
  \BibitemOpen
  \bibfield  {author} {\bibinfo {author} {\bibfnamefont {O.}~\bibnamefont
  {Higgott}},\ }\bibfield  {title} {\bibinfo {title} {Pymatching: A python
  package for decoding quantum codes with minimum-weight perfect matching},\
  }\href {https://doi.org/10.1145/3505637} {\bibfield  {journal} {\bibinfo
  {journal} {ACM Transactions on Quantum Computing}\ }\textbf {\bibinfo
  {volume} {3}},\ \bibinfo {pages} {1} (\bibinfo {year} {2022})}\BibitemShut
  {NoStop}%
\bibitem [{\citenamefont {Siek}\ \emph {et~al.}(2001)\citenamefont {Siek},
  \citenamefont {Lee},\ and\ \citenamefont {Lumsdaine}}]{siek2001boost}%
  \BibitemOpen
  \bibfield  {author} {\bibinfo {author} {\bibfnamefont {J.~G.}\ \bibnamefont
  {Siek}}, \bibinfo {author} {\bibfnamefont {L.-Q.}\ \bibnamefont {Lee}},\ and\
  \bibinfo {author} {\bibfnamefont {A.}~\bibnamefont {Lumsdaine}},\ }\href@noop
  {} {\emph {\bibinfo {title} {The Boost Graph Library: User Guide and
  Reference Manual. Addison-Wesley.}}}\ (\bibinfo  {publisher} {Pearson
  Education},\ \bibinfo {year} {2001})\BibitemShut {NoStop}%
\bibitem [{\citenamefont {Duan}\ \emph {et~al.}(2023)\citenamefont {Duan},
  \citenamefont {Mao}, \citenamefont {Shu},\ and\ \citenamefont
  {Yin}}]{10353179}%
  \BibitemOpen
  \bibfield  {author} {\bibinfo {author} {\bibfnamefont {R.}~\bibnamefont
  {Duan}}, \bibinfo {author} {\bibfnamefont {J.}~\bibnamefont {Mao}}, \bibinfo
  {author} {\bibfnamefont {X.}~\bibnamefont {Shu}},\ and\ \bibinfo {author}
  {\bibfnamefont {L.}~\bibnamefont {Yin}},\ }\bibfield  {title} {\bibinfo
  {title} {{ A Randomized Algorithm for Single-Source Shortest Path on
  Undirected Real-Weighted Graphs }},\ }in\ \href
  {https://doi.org/10.1109/FOCS57990.2023.00035} {\emph {\bibinfo {booktitle}
  {2023 IEEE 64th Annual Symposium on Foundations of Computer Science
  (FOCS)}}}\ (\bibinfo  {publisher} {IEEE Computer Society},\ \bibinfo
  {address} {Los Alamitos, CA, USA},\ \bibinfo {year} {2023})\ pp.\ \bibinfo
  {pages} {484--492}\BibitemShut {NoStop}%
\bibitem [{\citenamefont {Csanky}(1976)}]{doi:10.1137/0205040}%
  \BibitemOpen
  \bibfield  {author} {\bibinfo {author} {\bibfnamefont {L.}~\bibnamefont
  {Csanky}},\ }\bibfield  {title} {\bibinfo {title} {Fast parallel matrix
  inversion algorithms},\ }\href {https://doi.org/10.1137/0205040} {\bibfield
  {journal} {\bibinfo  {journal} {SIAM Journal on Computing}\ }\textbf
  {\bibinfo {volume} {5}},\ \bibinfo {pages} {618} (\bibinfo {year}
  {1976})}\BibitemShut {NoStop}%
\bibitem [{\citenamefont {Preparata}\ and\ \citenamefont
  {Sarwate}(1978)}]{PREPARATA1978148}%
  \BibitemOpen
  \bibfield  {author} {\bibinfo {author} {\bibfnamefont {F.}~\bibnamefont
  {Preparata}}\ and\ \bibinfo {author} {\bibfnamefont {D.}~\bibnamefont
  {Sarwate}},\ }\bibfield  {title} {\bibinfo {title} {An improved parallel
  processor bound in fast matrix inversion},\ }\href
  {https://doi.org/https://doi.org/10.1016/0020-0190(78)90079-0} {\bibfield
  {journal} {\bibinfo  {journal} {Information Processing Letters}\ }\textbf
  {\bibinfo {volume} {7}},\ \bibinfo {pages} {148} (\bibinfo {year}
  {1978})}\BibitemShut {NoStop}%
\bibitem [{\citenamefont {Mahajan}\ and\ \citenamefont
  {Vinay}(1997)}]{10.5555/866057}%
  \BibitemOpen
  \bibfield  {author} {\bibinfo {author} {\bibfnamefont {M.}~\bibnamefont
  {Mahajan}}\ and\ \bibinfo {author} {\bibfnamefont {V.}~\bibnamefont
  {Vinay}},\ }\href {https://dl.acm.org/doi/10.5555/866057} {\emph {\bibinfo
  {title} {Determinant: Combinatorics, Algorithms, and Complexity}}},\ \bibinfo
  {type} {Tech. Rep.}\ (\bibinfo {year} {1997})\BibitemShut {NoStop}%
\bibitem [{\citenamefont {Johansson}(2020)}]{johansson2020fast}%
  \BibitemOpen
  \bibfield  {author} {\bibinfo {author} {\bibfnamefont {F.}~\bibnamefont
  {Johansson}},\ }\href@noop {} {\bibinfo {title} {On a fast and nearly
  division-free algorithm for the characteristic polynomial}} (\bibinfo {year}
  {2020}),\ \Eprint {https://arxiv.org/abs/2011.12573} {arXiv:2011.12573
  [math.NA]} \BibitemShut {NoStop}%
\bibitem [{\citenamefont {Aaronson}\ and\ \citenamefont
  {Gottesman}(2004)}]{PhysRevA.70.052328}%
  \BibitemOpen
  \bibfield  {author} {\bibinfo {author} {\bibfnamefont {S.}~\bibnamefont
  {Aaronson}}\ and\ \bibinfo {author} {\bibfnamefont {D.}~\bibnamefont
  {Gottesman}},\ }\bibfield  {title} {\bibinfo {title} {Improved simulation of
  stabilizer circuits},\ }\href {https://doi.org/10.1103/PhysRevA.70.052328}
  {\bibfield  {journal} {\bibinfo  {journal} {Phys. Rev. A}\ }\textbf {\bibinfo
  {volume} {70}},\ \bibinfo {pages} {052328} (\bibinfo {year}
  {2004})}\BibitemShut {NoStop}%
\bibitem [{\citenamefont {Tomita}\ and\ \citenamefont
  {Svore}(2014)}]{PhysRevA.90.062320}%
  \BibitemOpen
  \bibfield  {author} {\bibinfo {author} {\bibfnamefont {Y.}~\bibnamefont
  {Tomita}}\ and\ \bibinfo {author} {\bibfnamefont {K.~M.}\ \bibnamefont
  {Svore}},\ }\bibfield  {title} {\bibinfo {title} {Low-distance surface codes
  under realistic quantum noise},\ }\href
  {https://doi.org/10.1103/PhysRevA.90.062320} {\bibfield  {journal} {\bibinfo
  {journal} {Phys. Rev. A}\ }\textbf {\bibinfo {volume} {90}},\ \bibinfo
  {pages} {062320} (\bibinfo {year} {2014})}\BibitemShut {NoStop}%
\bibitem [{\citenamefont {Gidney}(2021)}]{gidney2021stim}%
  \BibitemOpen
  \bibfield  {author} {\bibinfo {author} {\bibfnamefont {C.}~\bibnamefont
  {Gidney}},\ }\bibfield  {title} {\bibinfo {title} {Stim: a fast stabilizer
  circuit simulator},\ }\href {https://doi.org/10.22331/q-2021-07-06-497}
  {\bibfield  {journal} {\bibinfo  {journal} {{Quantum}}\ }\textbf {\bibinfo
  {volume} {5}},\ \bibinfo {pages} {497} (\bibinfo {year} {2021})}\BibitemShut
  {NoStop}%
\bibitem [{\citenamefont {Hagberg}\ \emph {et~al.}(2008)\citenamefont
  {Hagberg}, \citenamefont {Swart},\ and\ \citenamefont
  {Schult}}]{hagberg2008exploring}%
  \BibitemOpen
  \bibfield  {author} {\bibinfo {author} {\bibfnamefont {A.}~\bibnamefont
  {Hagberg}}, \bibinfo {author} {\bibfnamefont {P.~J.}\ \bibnamefont {Swart}},\
  and\ \bibinfo {author} {\bibfnamefont {D.~A.}\ \bibnamefont {Schult}},\
  }\href {https://www.osti.gov/biblio/960616} {\emph {\bibinfo {title}
  {Exploring network structure, dynamics, and function using NetworkX}}},\
  \bibinfo {type} {Tech. Rep.}\ (\bibinfo  {institution} {Los Alamos National
  Laboratory (LANL), Los Alamos, NM (United States)},\ \bibinfo {year}
  {2008})\BibitemShut {NoStop}%
\bibitem [{\citenamefont {Pattison}\ and\ \citenamefont
  {Nguyen}(2025)}]{ChrisPersonal}%
  \BibitemOpen
  \bibfield  {author} {\bibinfo {author} {\bibfnamefont {C.~A.}\ \bibnamefont
  {Pattison}}\ and\ \bibinfo {author} {\bibfnamefont {Q.}~\bibnamefont
  {Nguyen}},\ }\href@noop {} {}\bibinfo {howpublished} {personal communication}
  (\bibinfo {year} {2025})\BibitemShut {NoStop}%
\bibitem [{\citenamefont {Tremblay}\ \emph {et~al.}(2022)\citenamefont
  {Tremblay}, \citenamefont {Delfosse},\ and\ \citenamefont
  {Beverland}}]{PhysRevLett.129.050504}%
  \BibitemOpen
  \bibfield  {author} {\bibinfo {author} {\bibfnamefont {M.~A.}\ \bibnamefont
  {Tremblay}}, \bibinfo {author} {\bibfnamefont {N.}~\bibnamefont {Delfosse}},\
  and\ \bibinfo {author} {\bibfnamefont {M.~E.}\ \bibnamefont {Beverland}},\
  }\bibfield  {title} {\bibinfo {title} {Constant-overhead quantum error
  correction with thin planar connectivity},\ }\href
  {https://doi.org/10.1103/PhysRevLett.129.050504} {\bibfield  {journal}
  {\bibinfo  {journal} {Phys. Rev. Lett.}\ }\textbf {\bibinfo {volume} {129}},\
  \bibinfo {pages} {050504} (\bibinfo {year} {2022})}\BibitemShut {NoStop}%
\bibitem [{\citenamefont {Xu}\ \emph {et~al.}(2024)\citenamefont {Xu},
  \citenamefont {Bonilla~Ataides}, \citenamefont {Pattison}, \citenamefont
  {Raveendran}, \citenamefont {Bluvstein}, \citenamefont {Wurtz}, \citenamefont
  {Vasi{\'c}}, \citenamefont {Lukin}, \citenamefont {Jiang},\ and\
  \citenamefont {Zhou}}]{xu2024constant}%
  \BibitemOpen
  \bibfield  {author} {\bibinfo {author} {\bibfnamefont {Q.}~\bibnamefont
  {Xu}}, \bibinfo {author} {\bibfnamefont {J.~P.}\ \bibnamefont
  {Bonilla~Ataides}}, \bibinfo {author} {\bibfnamefont {C.~A.}\ \bibnamefont
  {Pattison}}, \bibinfo {author} {\bibfnamefont {N.}~\bibnamefont
  {Raveendran}}, \bibinfo {author} {\bibfnamefont {D.}~\bibnamefont
  {Bluvstein}}, \bibinfo {author} {\bibfnamefont {J.}~\bibnamefont {Wurtz}},
  \bibinfo {author} {\bibfnamefont {B.}~\bibnamefont {Vasi{\'c}}}, \bibinfo
  {author} {\bibfnamefont {M.~D.}\ \bibnamefont {Lukin}}, \bibinfo {author}
  {\bibfnamefont {L.}~\bibnamefont {Jiang}},\ and\ \bibinfo {author}
  {\bibfnamefont {H.}~\bibnamefont {Zhou}},\ }\bibfield  {title} {\bibinfo
  {title} {Constant-overhead fault-tolerant quantum computation with
  reconfigurable atom arrays},\ }\href
  {https://doi.org/https://doi.org/10.1038/s41567-024-02479-z} {\bibfield
  {journal} {\bibinfo  {journal} {Nature Physics}\ }\textbf {\bibinfo {volume}
  {20}},\ \bibinfo {pages} {1084} (\bibinfo {year} {2024})}\BibitemShut
  {NoStop}%
\bibitem [{\citenamefont {Razborov}\ and\ \citenamefont
  {Rudich}(1997)}]{RAZBOROV199724}%
  \BibitemOpen
  \bibfield  {author} {\bibinfo {author} {\bibfnamefont {A.~A.}\ \bibnamefont
  {Razborov}}\ and\ \bibinfo {author} {\bibfnamefont {S.}~\bibnamefont
  {Rudich}},\ }\bibfield  {title} {\bibinfo {title} {Natural proofs},\ }\href
  {https://doi.org/https://doi.org/10.1006/jcss.1997.1494} {\bibfield
  {journal} {\bibinfo  {journal} {Journal of Computer and System Sciences}\
  }\textbf {\bibinfo {volume} {55}},\ \bibinfo {pages} {24} (\bibinfo {year}
  {1997})}\BibitemShut {NoStop}%
\bibitem [{\citenamefont {Furst}\ \emph {et~al.}(1984)\citenamefont {Furst},
  \citenamefont {Saxe},\ and\ \citenamefont {Sipser}}]{furst1984parity}%
  \BibitemOpen
  \bibfield  {author} {\bibinfo {author} {\bibfnamefont {M.}~\bibnamefont
  {Furst}}, \bibinfo {author} {\bibfnamefont {J.~B.}\ \bibnamefont {Saxe}},\
  and\ \bibinfo {author} {\bibfnamefont {M.}~\bibnamefont {Sipser}},\
  }\bibfield  {title} {\bibinfo {title} {Parity, circuits, and the
  polynomial-time hierarchy},\ }\href
  {https://link.springer.com/article/10.1007/BF01744431} {\bibfield  {journal}
  {\bibinfo  {journal} {Mathematical systems theory}\ }\textbf {\bibinfo
  {volume} {17}},\ \bibinfo {pages} {13} (\bibinfo {year} {1984})}\BibitemShut
  {NoStop}%
\bibitem [{\citenamefont {Yao}(1985)}]{4568121}%
  \BibitemOpen
  \bibfield  {author} {\bibinfo {author} {\bibfnamefont {A.~C.-C.}\
  \bibnamefont {Yao}},\ }\bibfield  {title} {\bibinfo {title} {Separating the
  polynomial-time hierarchy by oracles},\ }in\ \href
  {https://doi.org/10.1109/SFCS.1985.49} {\emph {\bibinfo {booktitle} {26th
  Annual Symposium on Foundations of Computer Science (sfcs 1985)}}}\ (\bibinfo
  {year} {1985})\ pp.\ \bibinfo {pages} {1--10}\BibitemShut {NoStop}%
\bibitem [{\citenamefont {Hastad}(1986)}]{10.1145/12130.12132}%
  \BibitemOpen
  \bibfield  {author} {\bibinfo {author} {\bibfnamefont {J.}~\bibnamefont
  {Hastad}},\ }\bibfield  {title} {\bibinfo {title} {Almost optimal lower
  bounds for small depth circuits},\ }in\ \href
  {https://doi.org/10.1145/12130.12132} {\emph {\bibinfo {booktitle}
  {Proceedings of the Eighteenth Annual ACM Symposium on Theory of
  Computing}}},\ \bibinfo {series and number} {STOC '86}\ (\bibinfo
  {publisher} {Association for Computing Machinery},\ \bibinfo {address} {New
  York, NY, USA},\ \bibinfo {year} {1986})\ p.\ \bibinfo {pages}
  {6–20}\BibitemShut {NoStop}%
\bibitem [{\citenamefont {Razborov}(1987)}]{razborov1987lower}%
  \BibitemOpen
  \bibfield  {author} {\bibinfo {author} {\bibfnamefont {A.~A.}\ \bibnamefont
  {Razborov}},\ }\bibfield  {title} {\bibinfo {title} {Lower bounds on the size
  of bounded depth circuits over a complete basis with logical addition},\
  }\href {https://link.springer.com/article/10.1007/BF01137685} {\bibfield
  {journal} {\bibinfo  {journal} {Mat. Zametki}\ }\textbf {\bibinfo {volume}
  {41}},\ \bibinfo {pages} {598} (\bibinfo {year} {1987})}\BibitemShut
  {NoStop}%
\bibitem [{\citenamefont {Smolensky}(1987)}]{10.1145/28395.28404}%
  \BibitemOpen
  \bibfield  {author} {\bibinfo {author} {\bibfnamefont {R.}~\bibnamefont
  {Smolensky}},\ }\bibfield  {title} {\bibinfo {title} {Algebraic methods in
  the theory of lower bounds for boolean circuit complexity},\ }in\ \href
  {https://doi.org/10.1145/28395.28404} {\emph {\bibinfo {booktitle}
  {Proceedings of the Nineteenth Annual ACM Symposium on Theory of
  Computing}}},\ \bibinfo {series and number} {STOC '87}\ (\bibinfo
  {publisher} {Association for Computing Machinery},\ \bibinfo {address} {New
  York, NY, USA},\ \bibinfo {year} {1987})\ p.\ \bibinfo {pages}
  {77–82}\BibitemShut {NoStop}%
\bibitem [{\citenamefont {Sahay}\ \emph {et~al.}(2024)\citenamefont {Sahay},
  \citenamefont {Lin}, \citenamefont {Huang}, \citenamefont {Brown},\ and\
  \citenamefont {Puri}}]{sahay2024errorcorrectiontransversalcnot}%
  \BibitemOpen
  \bibfield  {author} {\bibinfo {author} {\bibfnamefont {K.}~\bibnamefont
  {Sahay}}, \bibinfo {author} {\bibfnamefont {Y.}~\bibnamefont {Lin}}, \bibinfo
  {author} {\bibfnamefont {S.}~\bibnamefont {Huang}}, \bibinfo {author}
  {\bibfnamefont {K.~R.}\ \bibnamefont {Brown}},\ and\ \bibinfo {author}
  {\bibfnamefont {S.}~\bibnamefont {Puri}},\ }\href
  {https://arxiv.org/abs/2408.01393} {\bibinfo {title} {Error correction of
  transversal cnot gates for scalable surface code computation}} (\bibinfo
  {year} {2024}),\ \Eprint {https://arxiv.org/abs/2408.01393} {arXiv:2408.01393
  [quant-ph]} \BibitemShut {NoStop}%
\bibitem [{\citenamefont {Strassen}(1969)}]{strassen1969gaussian}%
  \BibitemOpen
  \bibfield  {author} {\bibinfo {author} {\bibfnamefont {V.}~\bibnamefont
  {Strassen}},\ }\bibfield  {title} {\bibinfo {title} {Gaussian elimination is
  not optimal},\ }\href {https://link.springer.com/article/10.1007/BF02165411}
  {\bibfield  {journal} {\bibinfo  {journal} {Numerische mathematik}\ }\textbf
  {\bibinfo {volume} {13}},\ \bibinfo {pages} {354} (\bibinfo {year}
  {1969})}\BibitemShut {NoStop}%
\bibitem [{\citenamefont {Huang}\ \emph {et~al.}(2016)\citenamefont {Huang},
  \citenamefont {Smith}, \citenamefont {Henry},\ and\ \citenamefont {Van
  De~Geijn}}]{7877137}%
  \BibitemOpen
  \bibfield  {author} {\bibinfo {author} {\bibfnamefont {J.}~\bibnamefont
  {Huang}}, \bibinfo {author} {\bibfnamefont {T.~M.}\ \bibnamefont {Smith}},
  \bibinfo {author} {\bibfnamefont {G.~M.}\ \bibnamefont {Henry}},\ and\
  \bibinfo {author} {\bibfnamefont {R.~A.}\ \bibnamefont {Van De~Geijn}},\
  }\bibfield  {title} {\bibinfo {title} {Strassen's algorithm reloaded},\ }in\
  \href {https://doi.org/10.1109/SC.2016.58} {\emph {\bibinfo {booktitle} {SC
  '16: Proceedings of the International Conference for High Performance
  Computing, Networking, Storage and Analysis}}}\ (\bibinfo {year} {2016})\
  pp.\ \bibinfo {pages} {690--701}\BibitemShut {NoStop}%
\end{thebibliography}%

\section*{Acknowledgements}
The authors acknowledge discussions with Oscar Higgott, Christopher A.\ Pattison, Thomas R. Scruby, Shiro Tamiya, and Michael Vasmer.

\section*{Funding}
This work was supported by JST PRESTO Grant Number JPMJPR201A, JPMJPR23FC, JSPS KAKENHI Grant Number JP23K19970, JST COI-NEXT Grant Number JPMJPF2014, JST Moonshot R\&D Grant No. JPMJMS2061, and MEXT Quantum Leap Flagship Program (MEXT QLEAP) JPMXS0118069605, JPMXS0120351339, JPMXS0120319794\@.

\section*{Author contributions}

The authors contributed equally to this work.
Both authors contributed to the conception of the work, the analysis and interpretation of the work, and the preparation of the manuscript.

\section*{Competing interests}

The authors declare no competing interests.

\section*{Data and materials availability}

No data is used in this study.
The code used in this work is available from the corresponding author upon reasonable request.

\clearpage

\renewcommand{\theequation}{S\arabic{equation}}
\renewcommand{\thetheorem}{S\arabic{theorem}}
\renewcommand{\theproposition}{S\arabic{proposition}}
\renewcommand{\thelemma}{S\arabic{theorem}}
\renewcommand{\thecorollary}{S\arabic{theorem}}
\renewcommand{\thedefinition}{S\arabic{definition}}
\renewcommand{\theremark}{S\arabic{remark}}
\renewcommand{\theexample}{S\arabic{example}}
\renewcommand{\thesection}{S\arabic{section}}
\renewcommand{\thetable}{S\arabic{table}}
\renewcommand{\thefigure}{S\arabic{figure}}
\renewcommand{\theHfigure}{S\arabic{figure}}
\setcounter{equation}{0}
\setcounter{theorem}{0}
\setcounter{figure}{0}
\setcounter{section}{0}

\section*{Supplementary Materials}

\section{Avoiding the backlog problem with the polylog-time MWPM decoder for 2D surface codes}
\label{supple:backlog}
\begin{figure}
\includegraphics[width=0.8\linewidth]{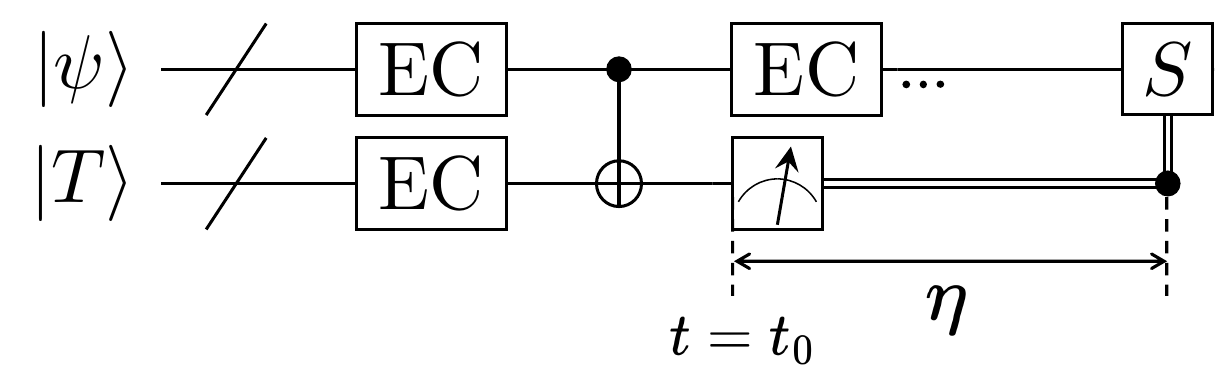}
\caption{\label{fig:teleportation_se}
Reaction time in the $T$-gate teleportation circuit. An EC gadget consists of $d$ rounds of syndrome measurements without including wait operations for decoding, unlike the definition of EC gadgets for 3D subsystem surface codes in Methods, where wait operations are performed during decoding. The time $t=t_0$ marks the start of the measurement of the auxiliary logical qubit, and the duration $\eta$ is called reaction time. During the reaction time, EC gadgets on the data block are repeatedly applied.
}
\end{figure}

Here we explain how to use our polylog-time MWPM decoder for 2D surface codes, with which we can avoid a backlog problem raised in Ref.~\cite{T10} while still being able to perform the MWPM decoding in chronological order.
If an original circuit consists only of Clifford gates, we do not have to decode syndrome data online, and there is no need to actively apply recovery operations.
This is because we can keep track of Pauli recovery operations and update the Pauli frame in classical software; thus, in this case, we can decode offline, i.e., delay the decoding until the end of the circuit.
However, if the circuit contains non-Clifford gates, we must perform the decoding online.
For instance, when we apply the non-Clifford $T$ gate using a magic state and gate teleportation as shown in Fig.~\ref{fig:teleportation_se}, the correct logical measurement outcome must be obtained using a decoder to decide whether to apply the Clifford $S$ correction, and this process cannot be delayed.
In such a situation, if the decoder cannot keep up with the speed at which syndromes are generated, we experience exponential slowdown during the computation, which is called a backlog problem~\cite{T10}.

We briefly review why the backlog problem happens when we apply several $T$ gates (see Ref.~\cite{T10} for more details).
Let $r_{\mathrm{gen}}$ be the rate at which syndromes are generated, and $r_{\mathrm{proc}}$ be the rate at which syndromes are processed by the decoder.
Suppose we have to process the syndromes generated during the past time $\Delta_{\mathrm{gen}}$ to know the appropriate Clifford correction.
The volume of the syndromes generated during $\Delta_{\mathrm{gen}}$ is $r_{\mathrm{gen}}\Delta_{\mathrm{gen}} $, so we spend the time $\qty(r_{\mathrm{gen}}/r_{\mathrm{proc}})\Delta_{\mathrm{gen}}=f\Delta_{\mathrm{gen}}$ to decode them, where $f\coloneqq r_{\mathrm{gen}}/r_{\mathrm{proc}}$.
While the decoder is running, the data qubits are waiting, and syndrome measurements are repeated, so additional syndromes $r_{\mathrm{gen}}f\Delta_{\mathrm{gen}}$ are generated. 
If we want to apply another $T$ gate subsequently, we have to process, at least, these syndromes to determine the correct Clifford correction, taking the time $\qty(r_{\mathrm{gen}}/r_{\mathrm{proc}})f\Delta_{\mathrm{gen}}=f^2\Delta_{\mathrm{gen}}$.
Recursively applying this argument, the reaction time defined by $\eta$ in Fig.~\ref{fig:teleportation_se} of the $k$th $T$ gate is at least $f^k \Delta_{\mathrm{gen}}$, which grows exponentially if $f>1$.

\begin{figure}
\includegraphics[width=0.7\linewidth]{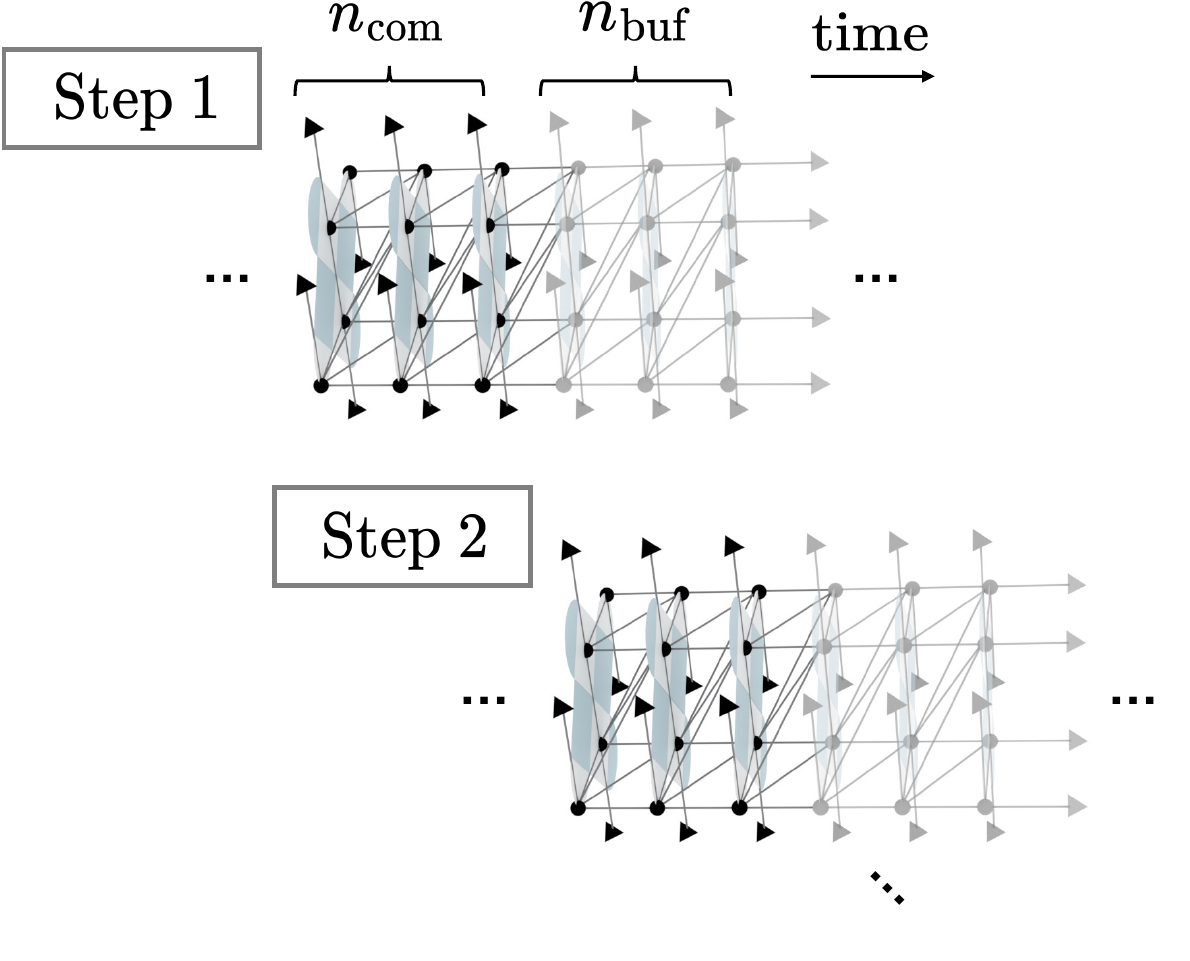}
\caption{\label{fig:sliding_window}
The sliding window decoding for the $d=3$ 2D rotated surface code. In this figure, we set $n_\mathrm{com}=n_\mathrm{buf}=d$, following Ref.~\cite{https://doi.org/10.48550/arxiv.2209.08552}. In Step 1, a window of length $n_\mathrm{com}+n_\mathrm{buf}$ is decoded, and then the corrections for the commit region (dark color) are accepted. The corrections for the buffer region (light color) will be discarded. Based on the corrections in Step 1, the syndromes of the past time boundary in Step 2 are modified. The decoding procedure is then repeated for the next window.
}
\end{figure}

Sliding window decoding is an approach for online decoding, originally proposed under the name of overlapping recovery method~\cite{doi:10.1063/1.1499754}.
We show a schematic of the sliding window decoding for the $d=3$ 2D rotated surface code in Fig.~\ref{fig:sliding_window}.
In this approach, instead of decoding the whole syndrome history at once, the global decoding problem is divided into multiple windows, i.e., subsets of consecutive syndrome data along the time direction. 
The decoder sequentially processes these windows in chronological order, using the result of the previous window for the decoding of the next window.
Each window consists of an $n_\mathrm{com}$-round commit region followed by an $n_\mathrm{buf}$-round buffer region, where we will accept the decoding result of a commit region as a final correction, while that of a buffer region will be discarded.
A sufficiently large buffer region is necessary to guarantee that the decoding result of the sliding window scheme remains the same as that of the global decoding.
In particular, it is known that it suffices to set the number of syndrome measurement rounds in a buffer region to be $n_\mathrm{buf}=d$~\cite{doi:10.1063/1.1499754}.
With this, error chains starting in the commit region are prevented from extending beyond the window more frequently than they would in global decoding, thereby maintaining nearly identical logical error rates in the leading order.
Once the decoding result of the current window is obtained, the decoder will move to the next window; at this point, error chains that cross between the commit and buffer regions of the current window create additional detection events at the first round of the next window, which are referred to as artificial detection events.
The detectors in the first round of the next window become closed boundaries.\footnote{We define a detector to be open if there is a fault that flips only this single detector; otherwise, it is closed. For example, detectors in the final round of each window in the sliding window decoding are open, as shown in Fig.~\ref{fig:sliding_window}.}
Then, decoding of the next window starts with these artificial detection events.

The sliding window decoding is still sequential, and hence, the computation will experience the backlog problem if the decoding time of a window is larger than the time to generate syndromes of the window.
Let $\tau_{\mathrm{sg}}$ be the syndrome generation time per round and $T_\mathrm{w}$ be the decoding time of a window.
If each window is responsible for committing $n_\mathrm{com}$ rounds, to meet the throughput requirement for avoiding the backlog problem, we require
\begin{equation}
\label{eq:backlog}
    T_\mathrm{w}<\tau_{\mathrm{sg}} n_\mathrm{com}.
\end{equation}
In most of the analyses of the backlog problem, it is assumed that the decoding complexity of an inner decoder, i.e., the one that solves the decoding problem of a window, is superlinear in the volume of the syndrome to be decoded.
In this case, \eqref{eq:backlog} becomes 
\begin{equation}
\label{eq:backlog_super}
    \Omega((n_{\mathrm{com}}+n_{\mathrm{buf}})d^2)<\tau_{\mathrm{sg}} n_\mathrm{com}.
\end{equation}
It is obvious that as $d$ grows larger, there will be a point at which \eqref{eq:backlog_super} is no longer satisfied, causing a backlog problem given that $\tau_{\mathrm{sg}}$ is fixed.

By contrast, our decoder achieves a polylog parallel runtime, which fundamentally changes the situation:
\begin{equation}
\label{eq:backlog_polylog}
    O(\polylog ((n_{\mathrm{com}}+n_{\mathrm{buf}})d^2))<\tau_{\mathrm{sg}} n_\mathrm{com}.
\end{equation}
If we adopt a typical choice of $n_{\mathrm{com}}=n_{\mathrm{buf}}=d$, then the requirement~\eqref{eq:backlog_polylog} becomes 
\begin{equation}
\label{eq:backlog_polylog_d}
    O(\polylog\qty(d))< \tau_{\mathrm{sg}}d.
\end{equation}
Since the right-hand side of \eqref{eq:backlog_polylog_d} grows faster than the left-hand side, there exists a sufficiently large $d$ beyond which~\eqref{eq:backlog} will always be satisfied. 
Therefore, with our polylog-time parallel MWPM decoder, the sliding window decoding of the 2D surface codes no longer suffers from the backlog problem in the asymptotic regime.

We discuss the reaction time $\eta$ of gate teleportation in Fig.~\ref{fig:teleportation_se} achieved by the sliding window decoding in the regime where \eqref{eq:backlog} holds.
In this case, the reaction time $\eta$ is the duration from the start of the measurement of the auxiliary code block ($t=t_0$) until the decoder determines the logical measurement outcome---in other words, until it is decided whether to apply the $S$ gate for correction.
We let $\tau_l$ denote the latency from the completion of the measurement until its outcome becomes available to a classical computer, which may be nonzero due to finite input-output (IO) speed.
Here we consider the case where $n_\mathrm{com}=n_\mathrm{buf}=d$.
According to the standard fault-tolerant protocol in which a gate gadget is followed by an EC gadget~\cite{G}, there should be an EC gadget between the \textsc{CNOT} gate and the destructive measurement of the logical auxiliary qubit in Fig.~\ref{fig:teleportation_se}; however, we omit it because it is not necessarily required for fault tolerance, as discussed in Ref.~\cite{sahay2024errorcorrectiontransversalcnot}. 
This omission optimizes the total depth of the circuit and helps in avoiding generating hyperedges in the detector graph when performing correlated decoding~\cite{sahay2024errorcorrectiontransversalcnot}.

In our analysis, for the decoding scheme of the transversal \textsc{CNOT} gate, we decode data and auxiliary block independently.
However, if the sliding window decoding were not employed, one could instead perform correlated decoding by an MWPM decoder without requiring a hypergraph-matching decoder.
This would be feasible owing to the aforementioned omission of the EC gadget; in particular, if one decodes $d$ rounds of syndromes on the data block right before the \textsc{CNOT} gate---without a buffer region---and applies the correction before proceeding with the subsequent $d$ rounds of syndrome measurements, then correlated decoding can be used.
For more details on the correlated decoding schemes for transversal \textsc{CNOT} gates on 2D surface codes, see Ref.~\cite{sahay2024errorcorrectiontransversalcnot}.

To determine the correct logical measurement outcome, all syndromes on the data block preceding the \textsc{CNOT} gate must be decoded.
To decode the $d$ rounds of syndromes right before the \textsc{CNOT} gate, the $d$ rounds of syndromes after the timing of applying the \textsc{CNOT} gate must also be generated, as shown in Fig.~\ref{fig:teleportation_se}; after all, in the sliding window decoding, the decoder is invoked after generating windows that include the commit region and also additional $d$ rounds of syndrome measurements as a buffer region.
The window that includes the syndromes of $d$ rounds preceding the \textsc{CNOT} gate and $d$ rounds following the \textsc{CNOT} gate is generated at $t=t_0+d\tau_{\mathrm{sg}}$, and the decoder for this window can begin processing at $t=t_0+d\tau_{\mathrm{sg}}+\tau_l$.
First, let us consider the case of $\tau_l<d\tau_{\mathrm{sg}}$ for simplicity.
In the regime where $T_\mathrm{w}$ is smaller than the time to generate a new window as in~\eqref{eq:backlog}, i.e., $T_\mathrm{w}<d\tau_{\mathrm{sg}}$, if $T_\mathrm{w}$ is even smaller than $d\tau_{\mathrm{sg}}-\tau_l$, the decoding finishes by $t=t_0+2d\tau_{\mathrm{sg}}$.
Otherwise, i.e., if $T_\mathrm{w}$ is still larger than $d\tau_{\mathrm{sg}}-\tau_l$, then the decoding may not finish by $t=t_0+2d\tau_{\mathrm{sg}}$; however, due to~\eqref{eq:backlog}, it finishes by $t=t_0+2d\tau_{\mathrm{sg}}+\tau_l $.
Note that decoding of syndromes on the auxiliary block has already been finished at this point, which is performed as a decoding of $(d+1)$-round window composed solely of a commit region that consists of $d$ rounds preceding the \textsc{CNOT} gate and one round constructed from the destructive measurement outcomes.
To avoid storing an excessive number of lookup tables for constructing path graphs in our decoding strategy, logical gates must always be performed after a multiple of $d$ rounds of syndrome measurements; with this protocol design, we can avoid the need to precompute lookup tables that depend on the communication latency $\tau_l$, by fixing the position of gate gadgets within the windows.
In other words, even if decoding is finished, we wait before applying the next logical gates until a multiple of $d$ rounds of syndrome measurements have been completed.
Thus, the reaction time is given by
\begin{equation}
\label{eq:reaction_time}
\eta= \begin{cases}
2d\tau_{\mathrm{sg}}, & \text{if } 0<T_\mathrm{w} \leq d\tau_{\mathrm{sg}}-\tau_l, \\[1ex]
3d\tau_{\mathrm{sg}}, & \text{if } d\tau_{\mathrm{sg}}-\tau_l < T_\mathrm{w} < d\tau_{\mathrm{sg}}.
\end{cases}
\end{equation}
More generally, for any $\tau_l$, the following equation holds:
\begin{equation}
\label{eq:reaction_time_general}
\eta= \begin{cases}
2d\tau_{\mathrm{sg}}+d\tau_{\mathrm{sg}} \cdot \left\lceil \frac{\tau_l}{d\tau_{\mathrm{sg}}} -1\right\rceil, & \text{if } 0<T_\mathrm{w} \leq \Delta, \\[1ex]
2d\tau_{\mathrm{sg}}+d\tau_{\mathrm{sg}} \cdot \left\lceil \frac{\tau_l}{d\tau_{\mathrm{sg}}} \right\rceil, & \text{if } \Delta < T_\mathrm{w} < d\tau_{\mathrm{sg}},
\end{cases}
\end{equation}
where $\Delta =d\tau_{\mathrm{sg}}(1+\left\lceil \tau_l/d\tau_{\mathrm{sg}} -1\right\rceil )-\tau_l$.
Equation~\eqref{eq:reaction_time_general} is equivalent to 
\begin{equation}
\label{eq:reaction_time_2}
\eta=2d\tau_{\mathrm{sg}}+ d\tau_{\mathrm{sg}} \cdot\left\lceil \frac{\tau_l+T_\mathrm{w}}{d\tau_{\mathrm{sg}}} -1\right\rceil,
\end{equation}
where $0<T_\mathrm{w} < d\tau_{\mathrm{sg}}$.

\begin{figure}
\includegraphics[width=\linewidth]{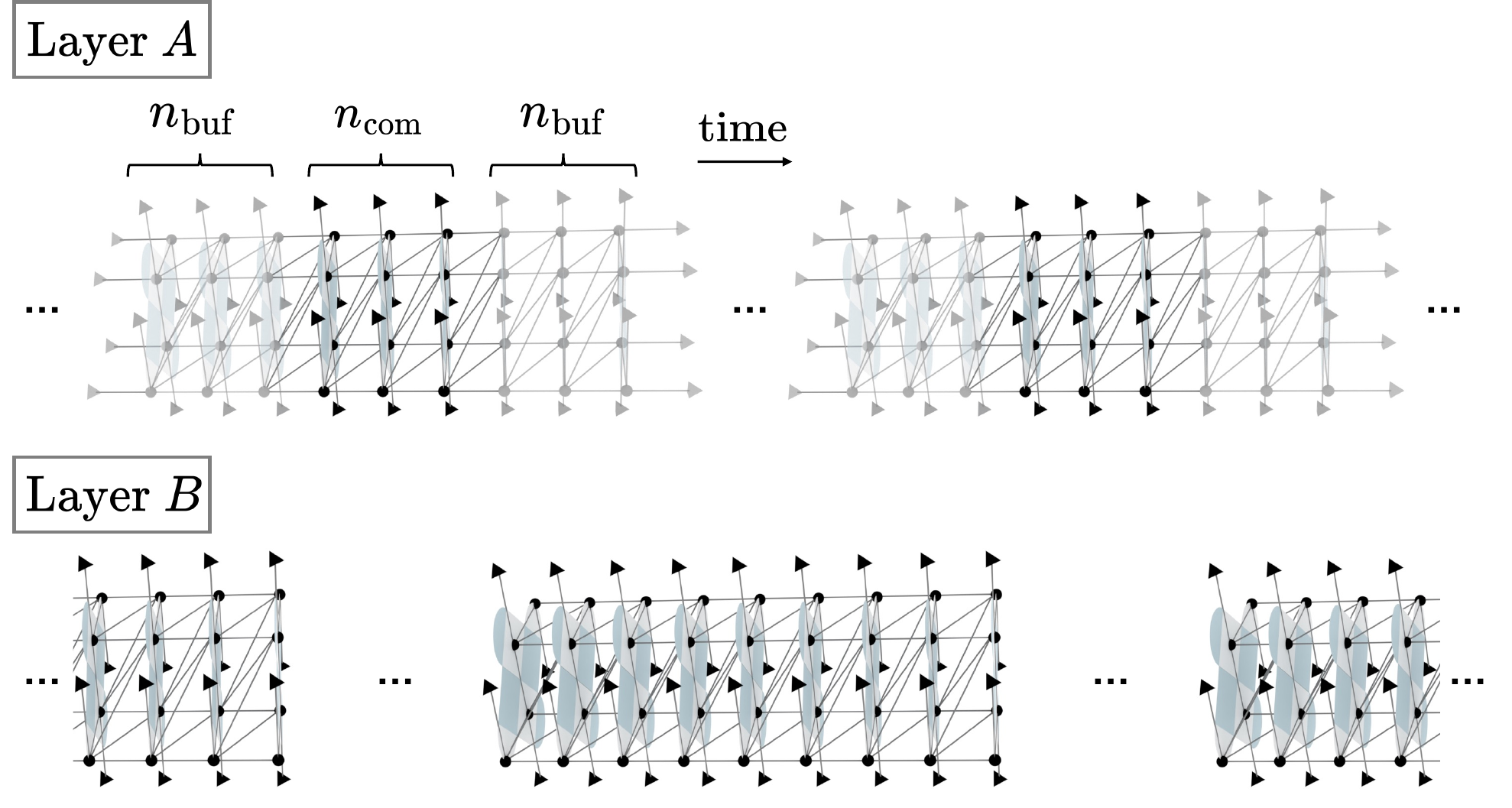}
\caption{\label{fig:parallel_window} The parallel window decoding for the $d=3$ 2D rotated surface code. We decode all windows in Layer $A$ and then the ones in Layer $B$, both in parallel. The initial and final time-like boundaries of a window are open in Layer $A$, whereas they are closed in Layer $B$. Here, we set $n_\mathrm{com}=n_\mathrm{buf}=d$, and $n_\mathrm{off}=3d$, following Ref.~\cite{https://doi.org/10.48550/arxiv.2209.08552}. With these parameters, the size of each window in both Layers $A$ and $B$ is $3d$.
}
\end{figure}

As we noted above, if we employ sliding window decoding using decoders with superlinear complexity, we inevitably encounter the backlog problem at some $d$; then, a leading proposal to avoid the backlog problem even using decoders with superlinear complexity has been a parallel window decoding~\cite{https://doi.org/10.48550/arxiv.2209.08552,https://doi.org/10.48550/arxiv.2209.09219,bombin2023modular}.
A schematic of the parallel window decoding using the same notations is shown in Fig.~\ref{fig:parallel_window}.
In the parallel window decoding, we have two types of windows: those in Layer $A$, and those in Layer $B$, as illustrated in Fig.~\ref{fig:parallel_window}.
First, windows in Layer $A$ are decoded in parallel, and then the rest of the windows in Layer $B$ are decoded depending on the decoding results of the Layer $A$\@.
Decoding in the Layer $B$ can also be parallelized.
In this approach, each window in the Layer $A$ consists of a commit region and two buffer regions for both past and future directions.
A window in the Layer $B$ consists of only a commit region.
In Ref.~\cite{https://doi.org/10.48550/arxiv.2209.08552}, it is proposed to set $n_\mathrm{buf}=d$, $n_\mathrm{com}=d$, and $n_\mathrm{off}=3d$ in Layer $A$, where $n_\mathrm{off}$ is the number of rounds between the end of the commit region and the start of the commit region for the next window, and we use these parameters in the following discussion.
More details on the different choices of these parameters are examined in Ref.~\cite{https://doi.org/10.48550/arxiv.2209.09219}.
In this scheme, independent of the number of the whole syndrome rounds, the decoding time is  $2T_\mathrm{w}$, accounting for the decoding of Layers $A$ and $B$.
Thus, given enough classical resources, the decoding throughput can be made arbitrarily high, effectively solving the backlog problem.
By a similar argument as above, the reaction time $\eta$ in the parallel window decoding scheme is
\begin{equation}
\label{eq:reaction_time_parallel}
\eta=2d\tau_{\mathrm{sg}}+ d\tau_{\mathrm{sg}} \cdot\left\lceil \frac{\tau_l+2T_\mathrm{w}}{d\tau_{\mathrm{sg}}} -1\right\rceil
\end{equation}
for any $T_\mathrm{w}$, indicating that the reaction time of the sliding window decoding and the parallel window decoding with our polylog-time MWPM decoder are indeed comparable in the asymptotic regime.

To summarize, our polylog-time MWPM decoder makes it possible to perform decoding while avoiding the backlog problem, even when employing the sliding window decoding in chronological order.
 In this work, we focus on the asymptotic scaling, leaving the estimation of practical decoding time for future investigations; however, it is important to note that whether the backlog problem can be avoided at a finite code distance depends on the actual decoding time, which is strongly influenced by constant factors in the decoding complexity as well as communication latencies.
Although asymptotically it suffices to utilize the sliding window scheme to avoid the backlog problem, there may be a finite $d$ such that the decoding time of a window is longer than the time to generate the window.
In such cases, one can still use the parallel window decoding.
In the asymptotic regime where $2T_\mathrm{w}$ becomes smaller than $d\tau_{\mathrm{sg}}$ and the effect of finite $\tau_l$ is also sufficiently small, both the sliding window and parallel window decoding yield the same reaction time of $2d\tau_{\mathrm{sg}}$, as shown in~\eqref{eq:reaction_time_2} and~\eqref{eq:reaction_time_parallel}.
However, the sliding window decoding takes $T_\mathrm{w}$ to decode a window, while the parallel window decoding takes $2T_\mathrm{w}$ for decoding Layers $A$ and $B$, as indicated in~\eqref{eq:reaction_time_2} and~\eqref{eq:reaction_time_parallel}.
Also, considering that the size of each window in the parallel window decoding is typically larger than that of the sliding window decoding, one can see that the decoding time $T_\mathrm{w}$ of a window for the sliding window decoding can be shorter than that of the parallel window decoding.
These suggest that in some cases with finite $d$, sliding window decoding can achieve a shorter reaction time of $2d\tau_\mathrm{sg}$ while the parallel window decoding may require a longer reaction time of $3d\tau_\mathrm{sg}$.
Whereas further study is needed to thoroughly compare the practical performances of the sliding window decoding and parallel window decoding with our polylog-time MWPM decoder, our key contribution is to demonstrate the feasibility of employing MWPM decoding for the sliding window decoding of 2D surface codes with arbitrarily large code sizes, which is fundamental for proving the threshold existence and overhead bounds and also potentially enables simpler implementations of the fault-tolerant protocols with the topological codes.

\section{The Samuelson-Berkowitz algorithm for parallel computation of the matrix determinant}
\label{sec:determinant}

For completeness, we here review the Samuelson-Berkowitz algorithm~\cite{BERKOWITZ1984147} as a suitable method for computing the determinant in our decoder.
In our description, for matrix multiplication of $N\times N$ matrices $A$ and $B$, we use a conventional parallel method using $O(N^3)$ processors running in $O(\log(N))$ time; that is, for each $i,j\in\{1,\ldots,N\}$ in parallel, the $(i,j)$ element of $C=AB$ is obtained by computing the addition
\begin{equation}
\label{eq:matrix_multiplication}
    C_{i,j}=\sum_{k=1}^{N}A_{i,k}B_{k,j}
\end{equation}
in parallel.
In particular, the addition of $N$ numbers $x_1,x_2,\ldots,x_N$ is performed by first computing $x_i+x_{i+1}$ in parallel for all $i\in\{1,3,5\ldots\}$, then computing $(x_i+x_{i+1})+(x_{i+2}+x_{i+3})$ in parallel, and iteratively repeating this until obtaining $x_1+\cdots+x_N$ within $O(\log(N))$ steps.
The same parallel procedure is applied for the multiplication of $N$ numbers (and those of matrices).
Note that the Faddeev-LeVerrier algorithm in Ref.~\cite{PREPARATA1978148} may also be applicable to our decoder if we appropriately rescale the matrix elements of $B$ and $B_\mathrm{sub}^{(i,j)}$ to convert them into integer matrices, which may lead to a shorter runtime than the Samuelson-Berkowitz algorithm in some cases~\cite{johansson2020fast}.
We may use asymptotically faster algorithms for matrix multiplication than the conventional $O(N^3)$ method, such as Strassen's algorithm~\cite{strassen1969gaussian}, which have larger constant factors but may be faster on large scales~\cite{7877137}.
We leave a detailed comparison of the implementations of our decoder using these different algorithms in a practical regime for future work.

\begin{widetext}
The Samuelson-Berkowitz algorithm~\cite{BERKOWITZ1984147} computes $\det(A)$ of an $N\times N$ matrix $A$ as follows.
Starting from $M_0\coloneqq A$, recursively for each $t\in\{1,\ldots,N\}$, we define a scalar $A_{t,t}$ representing $A$'s $(t,t)$ element, an $(N-t)\times 1$ matrix $S_t$, a $1\times(N-t)$ matrix $R_t$, and an $(N-t)\times (N-t)$ matrix $M_{t}$ as
\begin{align}
   \left(\begin{matrix}
       A_{t,t} & R_t\\
       S_t & M_t
   \end{matrix}\right)\coloneqq M_{t-1}.
\end{align}
Using the matrix elements of the given $N\times N$ matrix $A$, we can directly obtain $S_t$, $R_t$, and $M_t$ by
\begin{align}
\label{eq:B_t}
A=\begin{pNiceArray}{c|c|c|c|cw{c}{1cm}c}[margin]
    \Block{1-1}{A_{1,1}} & \Block{1-6}{R_1}  & & & & & \\
    \Hline
    \Block{6-1}{S_1} & \Block{1-1}{A_{2,2}} & \Block{1-5}{R_2} & & & & \\
    \Hline
    & \Block{5-1}{S_2} & \Block{1-1}{\ddots} & \Block{1-4}{\vdots} & & & \\
    \Hline
    & & \Block{4-1}{\cdots} & A_{t,t} & \Block{1-3}{R_t} & & \\
    \Hline
    & & & \Block{3-1}{S_t} & \Block{3-3}{M_t} & & \\
    & & & & & &\\
    & & & & & &
\end{pNiceArray}.
\end{align}
Then, for each $t\in\{1,2,\ldots,N\}$, we define an $(N+2-t)\times(N+1-t)$ matrix $C_t$ as a lower triangular Toplitz (diagonal-constant) matrix given by its $(j,1)$ element
\begin{align}
\label{eq:def_C_t}
    {\qty(C_t)}_{j,1}\coloneqq\begin{cases}
        -1&\text{if $j=1$},\\
        A_{t,t}&\text{if $j=2$},\\
        R_t M_t^{j-3} S_t&\text{otherwise};
    \end{cases}
\end{align}
that is, it holds that
\begin{align}
   C_t&=\begin{pNiceMatrix}[margin]
       -1 & 0 & 0 & 0 & 0 & 0 \\
       A_{t,t} & -1 & 0 & 0 & 0 & 0 \\
       R_t M_t^0 S_t & A_{t,t} & -1 & 0 & 0 & 0 \\
       R_t M_t^1 S_t & R_t M_t^0 S_t & A_{t,t} & -1 & 0 & 0 \\
       R_t M_t^2 S_t & R_t M_t^1 S_t & R_t M_t^0 S_t & A_{t,t} & \ddots & 0 \\
       \vdots & \ddots & \ddots & \ddots & \ddots & -1 \\
       R_t M_t^{N-1-t} S_t & \cdots & \cdots & \cdots & R_t M_t^0 S_t & A_{t,t} 
   \end{pNiceMatrix}\quad\text{for $t\in\{1,\ldots,N-1\}$},\\
   C_{N}&=\begin{pmatrix}
       -1\\
       A_{N,N}
   \end{pmatrix}.
\end{align}
For each $C_t$, we can compute each matrix element of $C_t$ by performing the matrix multiplication in parallel using $S_t$, $M_t$, and $R_t$ given by~\eqref{eq:B_t}.
To obtain these matrix elements efficiently, for each $t$ in parallel, we first compute a sequence
\begin{align}
    M_t^{2^0}&=M_t,\\
    M_t^{2^1}&=M_t^{2^0} M_t^{2^0},\\
    M_t^{2^2}&=M_t^{2^1} M_t^{2^1},\\
    M_t^{2^3}&=M_t^{2^2} M_t^{2^2},\\
    &\vdots 
\end{align}
of $O(\log(N))$ matrices; then, following the procedure described in Ref.~\cite{BERKOWITZ1984147}, we compute $R_t M_t^n S_t$ for each $n\in\{0,1,\ldots,N-1-t\}$ by  using these $O(\log(N))$ matrices in combination to perform parallel matrix multiplication
\begin{align}
    R_t M_t^n S_t=R_t\qty(\qty(M_t^{2^0})^{b_0}\qty(M_t^{2^1})^{b_1}\qty(M_t^{2^2})^{b_2}\cdots) S_t,
\end{align}
where $b_0,b_1,b_2,\ldots\in\qty{0,1}$ represent the binary expansion of $n$ given by $n=b_0 2^0+b_1 2^1+b_2 2^2+\cdots$.

To see how the matrix $C_t$ leads to the determinant, let $(p_0,p_1,\ldots,p_N)$ be coefficients of the characteristic polynomial of $A$, i.e.,
\begin{equation}
\label{eq:characteristic}
    p\qty(\lambda)=\sum_{n=0}^{N}p_{N-n}\lambda^n\coloneqq\det(A-\lambda\mathds{1}).
\end{equation}
Then, as shown in Theorem~5 of Ref.~\cite{BERKOWITZ1984147},
we can obtain these coefficients by
\begin{align}
\label{eq:p_N_C}
    \begin{pmatrix}
        p_0\\
        p_1\\
        \vdots\\
        p_N
    \end{pmatrix}=C_1 C_2 \cdots C_N.
\end{align}
We remark that Ref.~\cite{BERKOWITZ1984147} contains some typographical errors, but they can be corrected as follows.
Claim~1 of Ref.~\cite{BERKOWITZ1984147} should be corrected to
\begin{align}
    p(\lambda)=\qty(A_{1,1}-\lambda)\det\qty(M_1-\lambda\mathds{1})-R_1\qty[\adj\qty(M_1-\lambda\mathds{1})]S_1,
\end{align}
where $\adj\qty(M_1-\lambda\mathds{1})$ is the adjugate matrix of $M_1-\lambda\mathds{1}$.
Claim~2 of Ref.~\cite{BERKOWITZ1984147} is stated correctly and, in our notation, reads as
\begin{align}
    \adj\qty(M_1-\lambda\mathds{1})=-\sum_{k=2}^{N}\qty(q_0M_1^{k-2}+q_1M_1^{k-3}+\cdots+q_{k-2}\mathds{1})\lambda^{N-k},
\end{align}
where $q_0,\ldots,q_{N-1}$ are coefficients of the characteristic polynomial of $M_1$, given by
\begin{align}
\label{eq:q_lambda}
    q(\lambda)=\sum_{n=0}^{N-1}q_{N-n-1}\lambda^n\coloneqq\det\qty(M_1-\lambda\mathds{1}).
\end{align}
The proof of Claim~2 relies on the property of adjugate matrices
\begin{align}
    \qty[\adj\qty(M_1-\lambda\mathds{1})]\qty(M_1-\lambda\mathds{1})=q(\lambda)\mathds{1},
\end{align}
and also, in the corrected notation,
\begin{align}
    &\qty[-\sum_{k=2}^{N}\qty(q_0M_1^{k-2}+\cdots+q_{k-2}\mathds{1})\lambda^{N-k}]\qty(M_1-\lambda\mathds{1})\nonumber\\
    &=-\qty[\sum_{k=2}^{N}\qty(q_0M_1^{k-1}+\cdots+q_{k-2}M_1)\lambda^{N-k}]+\qty[\sum_{k=2}^{N}\qty(q_0M_1^{k-2}+\cdots+q_{k-2}\mathds{1})\lambda^{N-k+1}]\\
    &=-\qty[\sum_{k=2}^{N}\qty(q_0M_1^{k-1}+\cdots+q_{k-2}M_1)\lambda^{N-k}]+\qty[\sum_{l=1}^{N-1}\qty(q_0M_1^{l-1}+\cdots+q_{l-2}M_1+q_{l-1}\mathds{1})\lambda^{N-l}]\\
    &=-\qty(q_0M_1^{N-1}+\cdots+q_{N-2}M_1)+\qty[\sum_{l=1}^{N-1}\qty(q_{l-1}\mathds{1})\lambda^{N-l}]\\
    &=-\qty(q_0M_1^{N-1}+\cdots+q_{N-2}M_1)+\qty[\sum_{n=1}^{N-1}q_{N-n-1}\lambda^{n}\mathds{1}]\\
    &=q\qty(\lambda)\mathds{1},
\end{align}
where the last line follows from the Cayley-Hamilton theorem
\begin{align}
    \sum_{n=0}^{N-1}q_{N-n-1}M_1^n=0
\end{align}
applied to the first term and using the definition of $q(\lambda)$ in~\eqref{eq:q_lambda}.
Then, using the corrected versions of Claims~1 and~2, the key equation~$(\star)$ in Ref.~\cite{BERKOWITZ1984147} should be corrected to
\begin{align}
    p(\lambda)&=\qty(A_{1,1}-\lambda)\det\qty(M_1-\lambda\mathds{1})+R_1\qty[\sum_{k=2}^{N}\qty(q_0M_1^{k-2}+q_1M_1^{k-3}+\cdots+q_{k-2}\mathds{1})\lambda^{N-k}]S_1.
\end{align}
From this, with the corrected definition of $C_t$ as in~\eqref{eq:def_C_t}, the proof of Theorem~5 in Ref.~\cite{BERKOWITZ1984147} implies
\begin{align}
    \begin{pNiceMatrix}
    p_0\\
    p_1\\
    \vdots\\
    p_N
    \end{pNiceMatrix}=C_1\begin{pNiceMatrix}
    q_0\\
    q_1\\
    \vdots\\
    q_{N-1}
    \end{pNiceMatrix}.
\end{align}
Applying this relation recursively yields~\eqref{eq:p_N_C}.

Therefore, by performing the matrix multiplication on the right-hand side of~\eqref{eq:p_N_C} in parallel and adapting the last element of the vector on the left-hand side of~\eqref{eq:p_N_C}, we obtain the determinant by
\begin{equation}
    \det(A)=p_N,
\end{equation}
where we substitute $\lambda=0$ in~\eqref{eq:characteristic}.
Note that since each $C_t$ is a lower triangular Toplitz matrix, one may use a more space-efficient algorithm for parallel matrix multiplication than that for general matrices in~\eqref{eq:matrix_multiplication}~\cite{BERKOWITZ1984147}, while this improvement does not affect the runtime scaling.
Due to the parallelizability of matrix multiplication, the overall algorithm for computing $\det(A)$ can be parallelized to achieve an $O(\log^2(N))$ runtime, as shown in Ref.~\cite{BERKOWITZ1984147}.
\end{widetext}

\end{document}